\documentclass[twocolumn]{aastex63}
\usepackage{graphicx}
\usepackage{enumerate}
\usepackage{amssymb, amsmath}
\usepackage{natbib}
\usepackage{color}
\usepackage{ulem}
\usepackage{dblfnote}
\usepackage{appendix}
\usepackage{hyperref}
\usepackage[stable]{footmisc}

\definecolor{red}{rgb}{1.0,0.0,0.0}
\def\pasa{PASA}
\def\procspie{Proc.~SPIE}

\begin{document}

\author[0000-0002-6879-3030]{Taichi Uyama}
    \affiliation{Infrared Processing and Analysis Center, California Institute of Technology, Pasadena, CA 91125, USA}
    \affiliation{NASA Exoplanet Science Institute, Pasadena, CA 91125, USA}
    \affiliation{National Astronomical Observatory of Japan, 2-21-1 Osawa, Mitaka, Tokyo 181-8588, Japan}
\author[0000-0002-7405-3119]{Thayne Currie}
    \affiliation{NASA-Ames Research Center, Moffett Blvd., Moffett Field, CA 94035, USA}
    \affiliation{Subaru Telescope, National Astronomical Observatory of Japan, National Institutes of Natural Sciences, 650 North A`oh$\bar{o}$k$\bar{u}$ Place, Hilo, HI 96720, USA}
    \affiliation{Eureka Scientific, 2452 Delmer, Suite 100, Oakland, CA 96002, USA}
\author[0000-0002-0101-8814]{Valentin Christiaens}
    \affiliation{School of Physics and Astronomy, Monash University, 10 College Walk, Clayton Melbourne 3800, Vic, Australia}
\author[0000-0001-7258-770X]{Jaehan Bae}
    \affiliation{Earth and Planets Laboratory, Carnegie Institution for Science, 5241 Broad Branch Road NW, Washington, DC 20015, USA}
    \affiliation{NHFP Sagan Fellow}
\author{Takayuki Muto}
    \affiliation{Division of Liberal Arts, Kogakuin University
2665-1, Nakano-cho, Hachioji-chi, Tokyo, 192-0015, Japan}
\author[0000-0003-3038-364X]{Sanemichi Z. Takahashi}
    \affiliation{National Astronomical Observatory of Japan, 2-21-1 Osawa, Mitaka, Tokyo 181-8588, Japan}
\author[0000-0003-1451-6836]{Ryo Tazaki}
    \affiliation{Division of Liberal Arts, Kogakuin University 2665-1, Nakano-cho, Hachioji-chi, Tokyo, 192-0015, Japan}
\author[0000-0001-7591-2731]{Marie Ygouf}
    \affiliation{Infrared Processing and Analysis Center, California Institute of Technology, Pasadena, CA 91125, USA}
\author{Jeremy N. Kasdin}
    \affiliation{Department of Mechanical Engineering, Princeton University, Princeton, NJ 08544, USA}
\author[0000-0001-5978-3247]{Tyler Groff}
    \affiliation{NASA-Goddard Space Flight Center, Greenbelt, MD 20771, USA}
\author[0000-0003-2630-8073]{Timothy D. Brandt}
    \affiliation{Department of Physics, University of California-Santa Barbara, Santa Barbara, CA 93106, USA}
\author[0000-0001-6305-7272]{Jeffrey Chilcote}
    \affiliation{Department of Physics, University of Notre Dame, 225 Nieuwland Science Hall, Notre Dame, IN, 46556, USA}
\author{Masahiko Hayashi}
    \affiliation{National Astronomical Observatory of Japan, 2-21-1 Osawa, Mitaka, Tokyo 181-8588, Japan}
\author[0000-0003-0241-8956]{Michael W. McElwain}
    \affiliation{Exoplanets and Stellar Astrophysics Laboratory, Code 667, Goddard Space Flight Center, Greenbelt, MD 20771, USA}
\author[0000-0002-1097-9908]{Olivier Guyon}
    \affiliation{Subaru Telescope, National Astronomical Observatory of Japan, National Institutes of Natural Sciences, 650 North A`oh$\bar{o}$k$\bar{u}$ Place, Hilo, HI 96720, USA}
    \affiliation{Steward Observatory, University of Arizona, Tucson, AZ 85721, USA}
    \affiliation{Astrobiology Center, National Institutes of Natural Sciences, 2-21-1 Osawa, Mitaka, Tokyo, Japan}
\author{Julien Lozi}
    \affiliation{Subaru Telescope, National Astronomical Observatory of Japan, National Institutes of Natural Sciences, 650 North A`oh$\bar{o}$k$\bar{u}$ Place, Hilo, HI  96720, USA}
\author[0000-0001-5213-6207]{Nemanja Jovanovic}
    \affiliation{Department of Astronomy, California Institute of Technology, 1200 E. California Blvd., Pasadena, CA 91125, USA}
\author[0000-0003-1180-4138]{Frantz Martinache}
    \affiliation{Universit$\acute{e}$ C$\hat{o}$te d'Azur, Observatoire de la C$\hat{o}$te d'Azur, CNRS, Laboratoire Lagrange, France}
\author[0000-0002-9294-1793]{Tomoyuki Kudo}
    \affiliation{Subaru Telescope, National Astronomical Observatory of Japan, National Institutes of Natural Sciences, 650 North A`oh$\bar{o}$k$\bar{u}$ Place, Hilo, HI  96720, USA}
\author[0000-0002-6510-0681]{Motohide Tamura}
    \affiliation{Department of Astronomy, The University of Tokyo, 7-3-1, Hongo, Bunkyo-ku, Tokyo 113-0033, Japan}
    \affiliation{Astrobiology Center of NINS, 2-21-1 Osawa, Mitaka, Tokyo 181-8588, Japan}
    \affiliation{National Astronomical Observatory of Japan, 2-21-1 Osawa, Mitaka, Tokyo 181-8588, Japan}
\author[0000-0002-5082-8880]{Eiji Akiyama}
    \affiliation{Department of Engineering, Niigata Institute of Technology, 1719 Fujihashi, Kashiwazaki, Niigata 945-1195, Japan}
\author[0000-0002-5627-5471]{Charles A. Beichman}
    \affiliation{NASA Exoplanet Science Institute, Pasadena, CA 91125, USA}
    \affiliation{Infrared Processing and Analysis Center, California Institute of Technology, Pasadena, CA 91125, USA}
\author[0000-0001-5440-1879]{Carol A. Grady}
    \affiliation{Exoplanets and Stellar Astrophysics Laboratory, Code 667, Goddard Space Flight Center, Greenbelt, MD 20771, USA}
    \affiliation{Eureka Scientific, 2452 Delmer, Suite 100, Oakland, CA 96002, USA}
    \affiliation{Goddard Center for Astrobiology, 8800 Greenbelt Road, Greenbelt, MD 20771, USA}
\author[0000-0002-9259-1164]{Gillian R. Knapp}
    \affiliation{Department of Astrophysical Science, Princeton University, Peyton Hall, Ivy Lane, Princeton, NJ 08544, USA}
\author[0000-0003-2815-7774]{Jungmi Kwon}
    \affiliation{Department of Astronomy, The University of Tokyo, 7-3-1, Hongo, Bunkyo-ku, Tokyo 113-0033, Japan}
\author[0000-0003-1799-1755]{Michael Sitko}
    \affiliation{Space Science Institute, 4765 Walnut St, Suite B, Boulder, CO 80301, USA}
\author[0000-0001-9248-7546]{Michihiro Takami}
    \affiliation{Institute of Astronomy and Astrophysics, Academia Sinica, National Taiwan University, No.1, Sec. 4, Roosevelt Rd, Taipei 10617, Taiwan, R.O.C.}
\author[0000-0002-4309-6343]{Kevin R. Wagner}
    \affiliation{Steward Observatory, University of Arizona, Tucson, AZ 85721, USA}
    \affiliation{NASA NExSS Earths in Other Solar SystemsTeam, USA}
\author[0000-0001-9209-1808]{John P. Wisniewski}
    \affiliation{Homer L. Dodge Department of Physics and Astronomy, University of Oklahoma, 440 W. Brooks Street, Norman, OK 73019, US}
\author[0000-0002-9024-4150]{Yi Yang}
    \affiliation{Department of Astronomy, The University of Tokyo, 7-3-1, Hongo, Bunkyo-ku, Tokyo 113-0033, Japan}
\affiliation{National Astronomical Observatory of Japan, 2-21-1 Osawa, Mitaka, Tokyo 181-8588, Japan}

\shortauthors{Uyama et al.}

\title{SCExAO/CHARIS High-Contrast Imaging of Spirals and Darkening Features in the HD 34700 A Protoplanetary Disk}

\begin{abstract}
We present Subaru/SCExAO+CHARIS broadband ($JHK$-band) integral field spectroscopy of HD 34700 A.  CHARIS data recover HD 34700 A's disk ring and confirm multiple spirals discovered in \cite{Monnier2019}.  We set limits on substellar companions of $\sim12\ M_{\rm Jup}$ at $0\farcs3$ (in the ring gap) and $\sim5\ M_{\rm Jup}$ at $0\farcs75$ (outside the ring).
The data reveal darkening effects on the ring and spiral, although we do not identify the origin of each feature such as shadows or physical features related to the outer spirals. Geometric albedoes converted from the surface brightness suggests a higher scale height and/or prominently abundant sub-micron dust at position angle between $\sim45^\circ$ and $90^\circ$.
Spiral fitting resulted in very large pitch angles ($\sim30-50^\circ$) and a stellar flyby of HD 34700 B or infall from a possible envelope is perhaps a reasonable scenario to explain the large pitch angles.

\end{abstract}

\section{Introduction}

Protoplanetary disks around young ($\lesssim$ 10 Myr) stars are key laboratories for exploring planet formation.
Recent high angular resolution observations of these disks in scattered light through thermal emission in the submillimeter reveal a variety of asymmetric features -- e.g. gaps, rings, and spirals -- that may be traced to planet formation processes \citep[e.g.][]{Avenhaus2018,Andrews2018}.  
Theoretical studies have predicted part of such asymmetric features are related to planet formation \citep[e.g.][]{Zhu2011, Dodson-Robinson2011} and recently VLT and MagAO high-contrast imaging observations reported the first convincing protoplanets within a gap of the PDS 70's protoplanetary disk \citep[][]{Keppler2018, Wagner2018, Haffert2019}.
High-contrast imaging opened a new window of investigating planet formation mechanism but the occurrence rate of detected young planets is much smaller \citep[$\sim$1-3\% at 10-300 au; e.g.][]{Bowler2016, Nielsen2019} than the occurrence rate of asymmetric disks that are favorable for planet formation \citep[e.g.][]{Dong2018-spiral}.

Recently HD 34700 A became one of the most intriguing young system with a large gap and multiple spirals in its disk \citep{Monnier2019}.
Previously HD 34700 A was known to be a binary (HD 34700 Aab) with a significant far-infrared (IR) excess that had been regarded as debris disk \citep[$\gtrsim$10 Myr;][]{Torres2004}. This system has two other companions (HD 34700~BC) located at $\sim5\farcs2$ and $\sim9\farcs2$ respectively \citep{Sterzik2005}.
A precise measurement of the parallax with {\it Gaia} ( $356.5^{+6.3}_{-6.0}$ pc) showed a larger distance than the previous assumption, which made one infer a younger age. \cite{Monnier2019} implemented radiative transfer modeling along with Gemini/GPI $JH$-band observations and proved that HD 34700 A is a young system ($\sim$5 Myr) surrounded by a protoplanetary disk.
Although their model showed good agreement with polarimetric data in $J$ band, it had differences somewhat between GPI-based $JH$-band total intensity and $H$-band polarized intensity.
Another intriguing feature in the HD 34700 A disk is its spiral features: previous high angular resolution observations have reported a variety of morphology in disks at various evolutionary stages \citep[e.g. AB Aur, SAO 206462, MWC 758, HD100453, HD 100546, HD 142527, Elias 2-27, CQ Tau, GG Tau;][]{Hashimoto2011, Muto2012, Grady2013, Wagner2015, Currie2015, Avenhaus2014, Perez2016, Uyama2020, Keppler2020}. Among these disks this object has the largest number of spirals in a disk, the mechanism of which is still unclear.

In this study we present integral field spectroscopy results of HD 34700 A taken with the Coronagraphic High Angular Resolution Imaging Spectrograph (CHARIS) and the Subaru Coronagraphic Extreme Adaptive Optics (SCExAO).
Our observation and several differential-imaging reductions detected the ring and multiple spirals. 
We also newly detected darkening features on the ring and one of the spirals. 
Section \ref{sec: Data} describes our observation, data reduction, and results.
We then implemented radiative transfer modeling from $J$ to $K$ band and investigated scattering profiles.
Our spiral fitting shows very large pitch angles ($\sim30^\circ-50^\circ$) and we discuss possible scenarios that can induce multiple spirals with such large pitch angles. Details of each topic are investigated in Section \ref{sec: Analysis and Discussion}.
Finally we summarize our work in Section \ref{sec: Summary}.

\section{Data} \label{sec: Data}
\subsection{Observations} \label{sec: Observations}
We used Subaru/SCExAO+CHARIS in broadband integral field spectroscopy (IFS) mode (1.16-2.37 $\mu$m, spectral resolution of $\mathcal{R}\sim$19, pixel scale = $0\farcs0162$ pixel$^{-1}$). 
In this paper we collapse the reduced IFS data cube into $JHK$-band images to discuss simultaneous multi-band imaging results.
HD 34700 A \citep[$J$=8.04, $H$=7.71, $K$=7.48;][]{Cutri2003-2MASS} was observed on 2019 January 12 UT with a Lyot coronagraph mask to suppress the starlight and a fixed pupil so that angular differential imaging \citep[ADI;][]{Marois2006} could be applied after. HR 2466 \citep[$J$=5.03, $H$=5.07, $K$=5.11;][]{Cutri2003-2MASS} was also observed for a PSF reference of reference-star differential imaging \citep[RDI;][]{Lafreniere2009}. Details about the data reduction are explained in Section \ref{sec: Data Reduction and Results}.
Astrogrids made from the star's PSF were added in the field of view (FoV) with 25nm amplitude modulation in the deformable mirror \citep{Jovanovic2015-astrogrids, Sahoo2020}, which provides accurate measurements of the central star's location and photometry.
The data were taken under very good seeing conditions ($\theta_{\rm V}$ $\sim$ 0$\farcs{}$4) and a typical FWHM was $\sim$30 (2 pix), 45, 55 mas in $JHK$ bands, respectively.
The total exposure time was 2168.6 seconds (1.475-sec single exposure $\times21$ coadds $\times 70$ cubes) for HD 34700 A and 2952.95 seconds (1.475-sec single exposure $\times 14$ coadds $\times 143$ cubes) for HR 2466. The HD 34700 A observation obtained $\sim28^\circ$ of parallactic angle change for ADI.

\begin{figure}
\begin{tabular}{ll}
\begin{minipage}{0.45\hsize}
    \flushleft
    \includegraphics[width=0.9\textwidth]{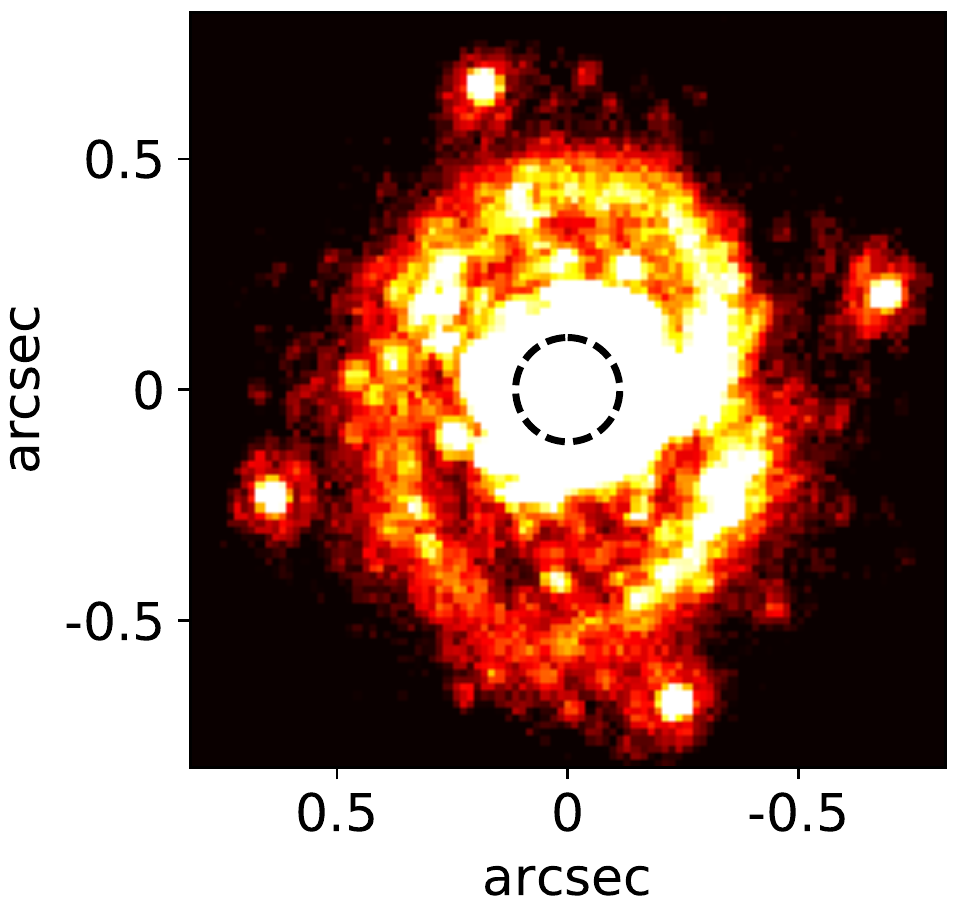}
\end{minipage}
\begin{minipage}{0.45\hsize}
    \flushleft
    \includegraphics[width=0.9\textwidth]{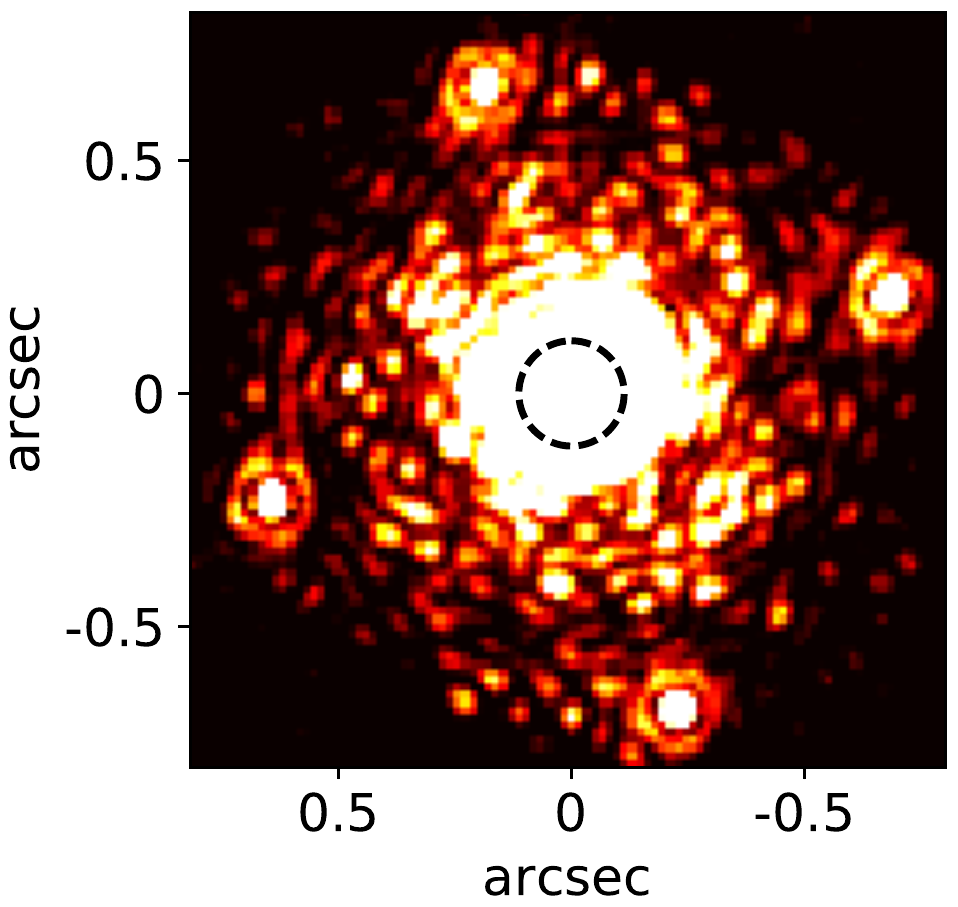}
\end{minipage}
\end{tabular}
\caption{Comparison of a single exposure for HD 34700 A (left) and HR 2466 (right) at channel 11 ($\lambda$=1.6296 $\mu$m). Color scale is arbitrary and these images are not rotated to North up. Astrogrids are located by the four edges in each FoV. Dashed black circle in each image indicates the coronagraph mask (113 mas in radius).}
\label{fig: raw images}
\end{figure}

\subsection{Data Reduction and Results} \label{sec: Data Reduction and Results}
We used CHARIS data reduction pipeline with the $\chi^2$ extraction algorithm \citep[][]{Brandt2017} to extract dark-subtracted, flat-fielded, and wavelength-calibrated data cubes with 22 uniform spectral channels from the CHARIS raw files for both HD 34700 A and HR 2466.
For spectrophotometric calibration we used appropriate Kuruz model atmospheres \citep[][]{Castelli2003} adopting G0V and A2V for spectral types of HD 34700 A and HR 2466 respectively.  
Single extracted data cubes show the ring feature of HD 34700 A without any post-processing (Figure \ref{fig: raw images}).

For post-processing PSF subtraction we implemented two reduction techniques: (1) RDI by following the way of \cite{Currie2019} to capture the ring morphology without self-subtraction (2) combination of ADI and spectral differential imaging \citep[SDI;][]{Vigan2015} by following the way of \cite{Currie2018} to get high contrast enough to investigate outer spirals and potential planetary-mass companions. In both data reductions we used the same data reduction pipelines as \cite{Currie2018, Currie2019}.
Our methods are described in more detail below.

\subsubsection{RDI} \label{sec: RDI}
First, we performed RDI by utilizing Karhunen-Lo$\grave{e}$ve Image Projection algorithms \citep[KLIP;][]{Soummer2012-KLIP}, where we adopted a ``full-frame subtraction'' on the CHARIS FoV ($r_{\rm min}=3$ pix for the inner working angle, $r_{\rm max}=65$ pix (1\farcs{}05) for the outer working angle, and $\Delta r=62$ pix for the subtraction separation).

\begin{figure*}
\begin{tabular}{ccc}
\begin{minipage}{0.3\hsize}
    \centering
    \includegraphics[width=\textwidth]{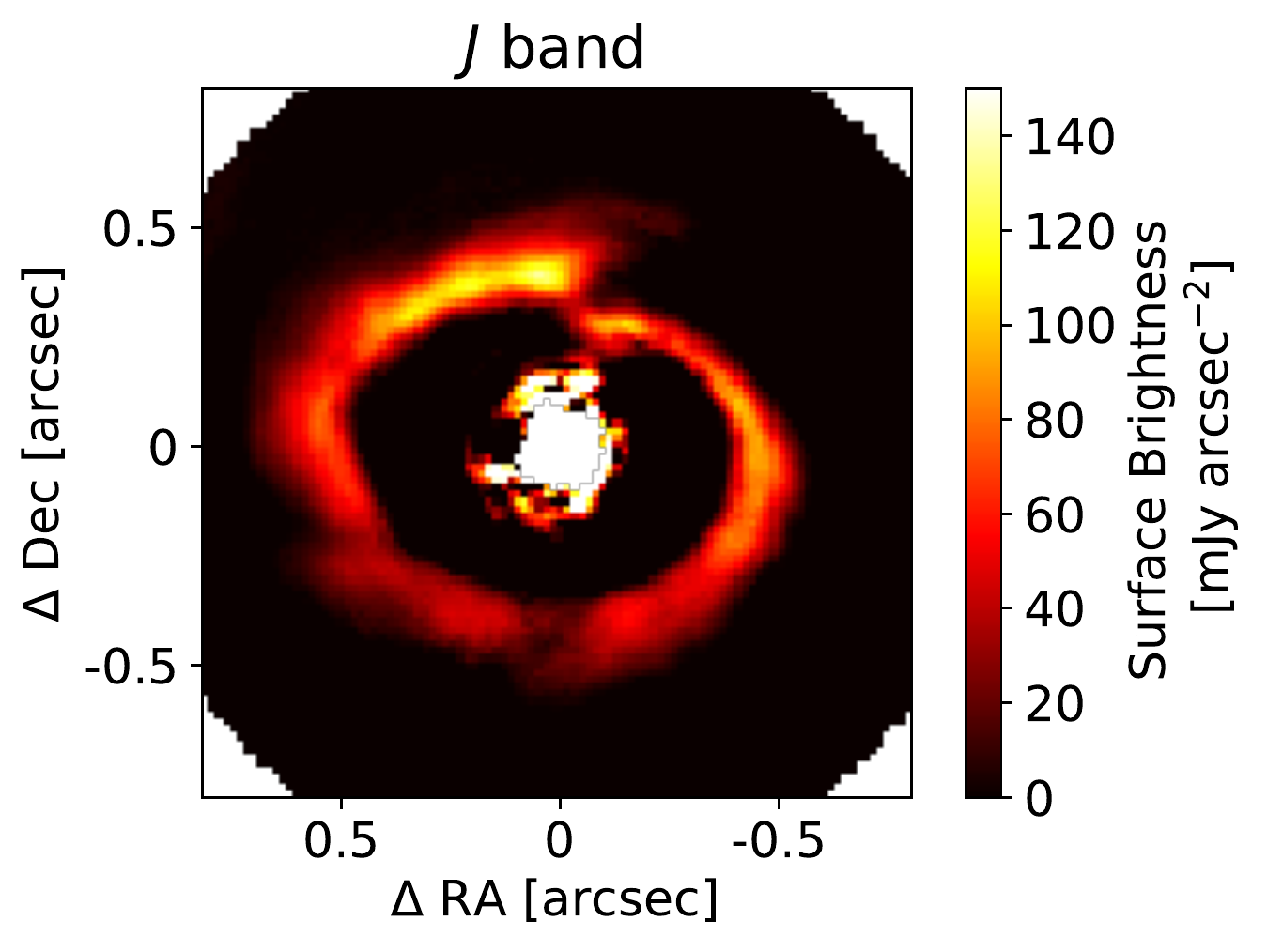}
\end{minipage}
\begin{minipage}{0.3\hsize}
    \centering
    \includegraphics[width=\textwidth]{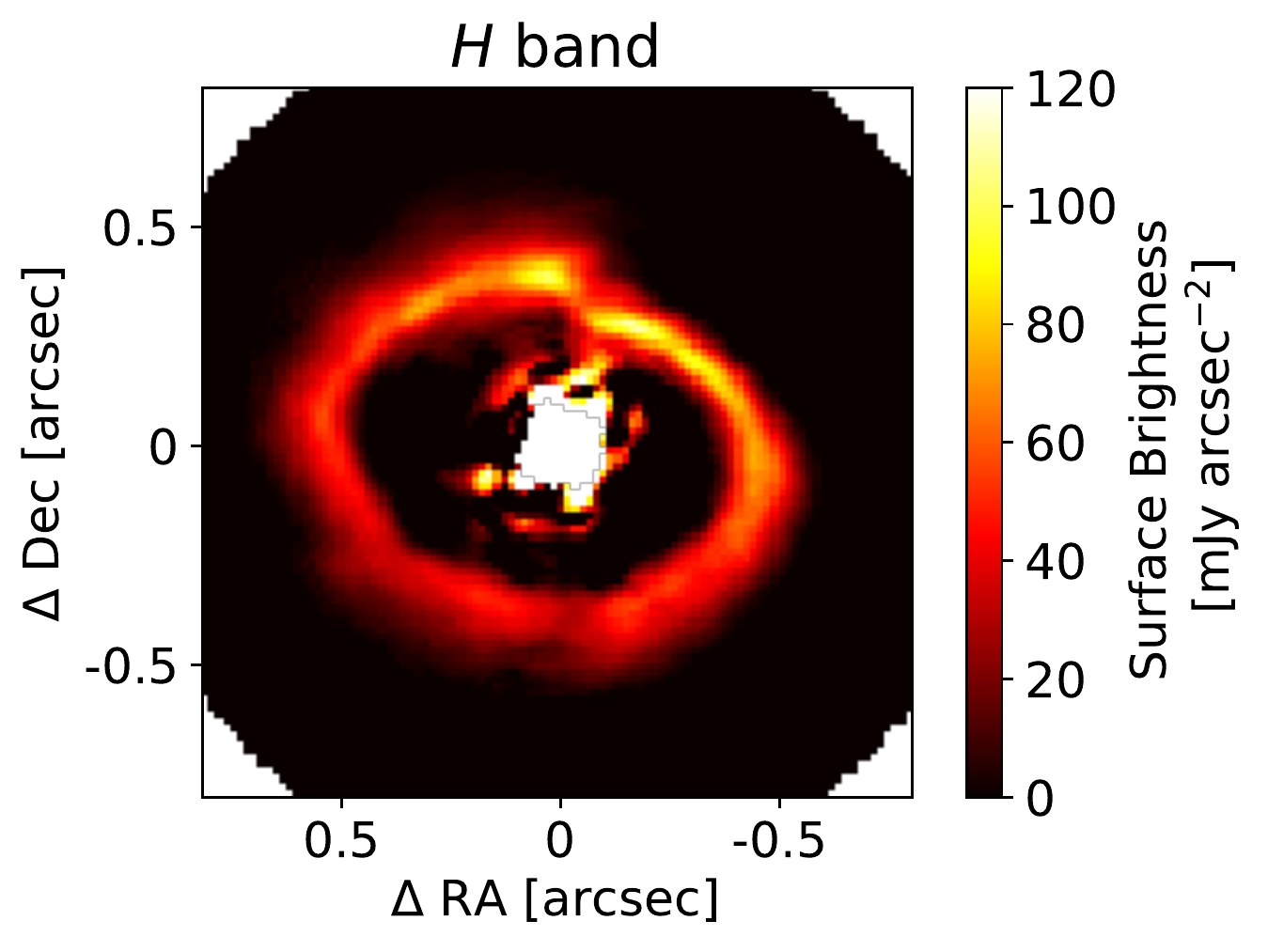}
\end{minipage}
\begin{minipage}{0.3\hsize}
    \centering
    \includegraphics[width=\textwidth]{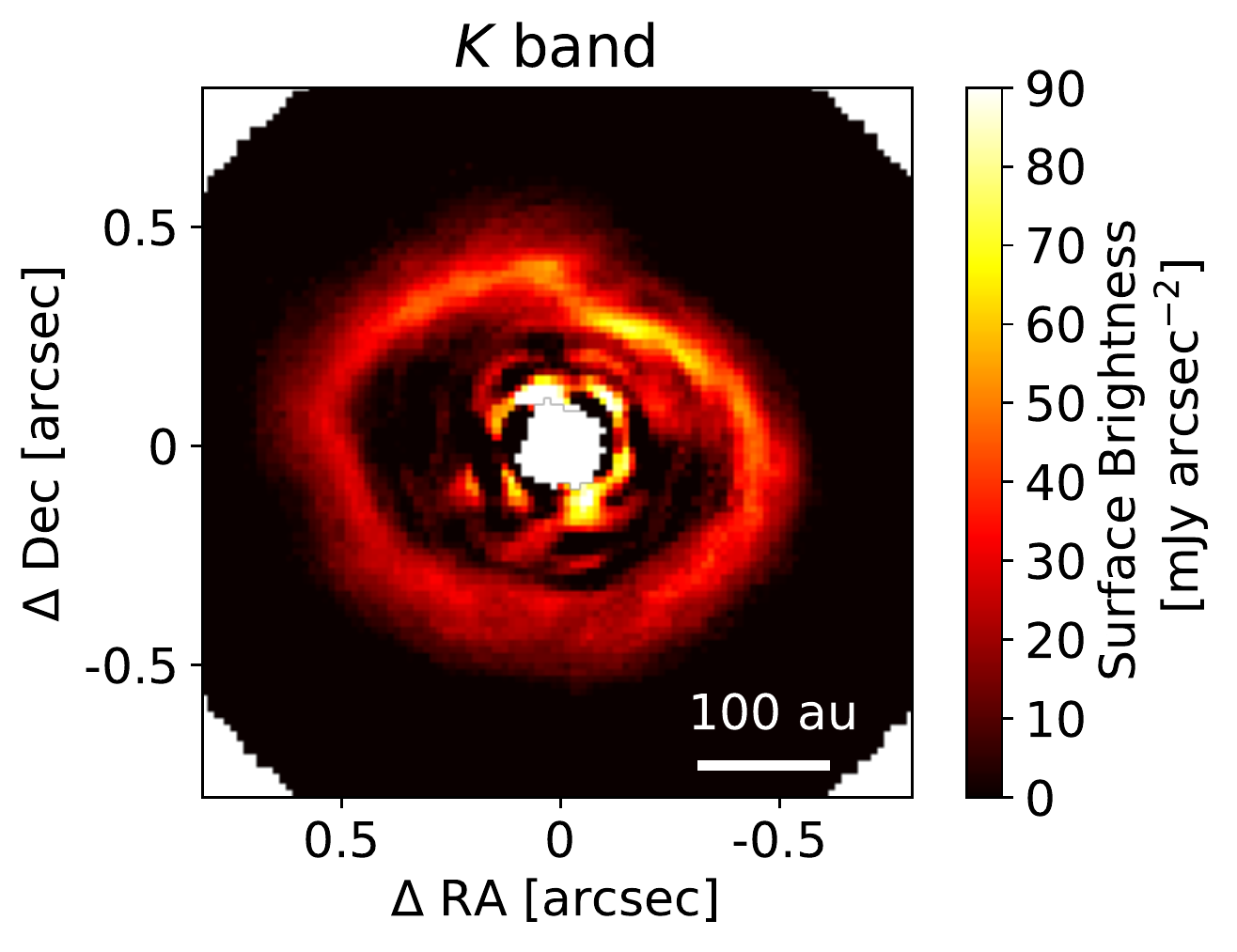}
\end{minipage}
\end{tabular}
\caption{RDI-KLIP (KL=5) reduction results at $J$ (left), $H$ (center), and $K$ (right) bands. The central unresolved binary (HD 34700 Aab) is masked by the reduction algorithm. North is up and East is left.}
\label{fig: RDI images}
\end{figure*}

\begin{figure}
    \centering
    \includegraphics[width=0.45\textwidth]{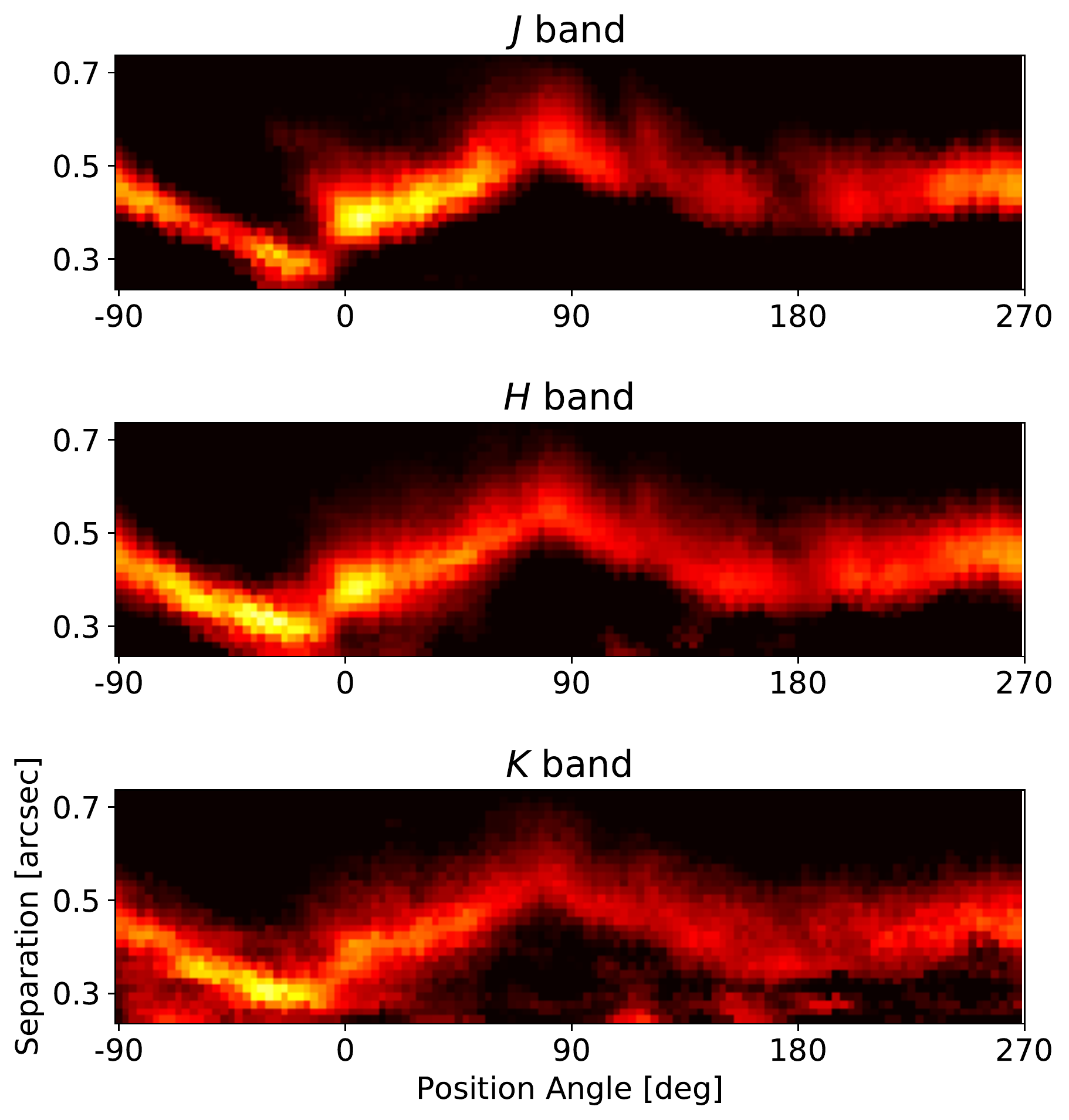}
    \caption{Polar-projected (East of North) images of Figure \ref{fig: RDI images} at the ring area. Color scale is set the same as Figure \ref{fig: RDI images}.}
    \label{fig: r-theta RDI}
\end{figure}

Figure \ref{fig: RDI images} compares collapsed $JHK$-band (1.154--2.387 $\mu$m) images of RDI-reduced (Karhunen-Lo$\grave{e}$ve - the number of basis vector; KL=5) HD 34700 A data and Figure \ref{fig: r-theta RDI} shows polar-projected images of Figure \ref{fig: RDI images}.
Here we excluded channels (channel No. 6-8: 1.3746-1.4714 $\mu$m, No. 15-17: 1.8672-1.9987 $\mu$m) that have stronger telluric absorption and lie either in the wings or outside of the nominal $JHK$ bandpasses.
We were able to resolve scattered light from the ring surface, but did not confirm an inner arc in the gap \cite{Monnier2019} reported. Regions interior to $\sim$240, 280, and 300 mas in $J$, $H$, and $K$ bands are dominated by residual speckle noise in our RDI reduction. Thus, we focus on characterizing disk features at wider separations (Section \ref{sec: Analysis and Discussion}).
Details about the ring feature are discussed in Section \ref{sec: Analysis and Discussion}.

The ring extends along the whole azimuthal direction and shows some asymmetric features such as darkening, which makes it difficult to calculate a radial noise profile to define error bars of the surface brightness. 
Therefore, we calculate standard deviations at the interior ($0\farcs3$) and exterior ($0\farcs7$) of the ring in each collapsed image and defined noise at the ring separation as interpolation of the standard deviations between the two separations. Here we adopted 3$\sigma$ clipping to mitigate effects of the presence of the ring at $\sim0\farcs3$ and the spirals at $\sim0\farcs7$. The scattering properties of the ring are discussed in detail in Section \ref{sec: Scattering Profiles}.

\subsubsection{ADI+SDI} \label{sec: ADI+SDI}
After the basic reductions as mentioned at the beginning of Section \ref{sec: Data Reduction and Results} we performed ADI reduction utilizing Locally Optimized Combination of Images \citep[LOCI;][]{Lafreniere2007} and Adaptive-LOCI \citep[A-LOCI;][]{Currie2012} algorithms.
A smaller separation of subtraction zones ($\Delta r=5$ pix) than the RDI reduction, a singular value decomposition (SVD) cutoff to truncate the diagonal terms of the covariance matrix of $SVD_{\rm lim}$ = $10^{-6}$ \citep[see also][]{Currie2013,Currie2018}, a rotation gap of $\delta=0.75$ to limit signal loss/biasing due to azimuthally displaced copies of the astrophysical signal, and a pixel mask over the subtraction zone \citep[e.g.][]{Currie2012} were adopted to generate weighed reference PSFs at different separations.
To further suppress residual speckles and achieve higher contrast we then performed SDI reduction on the ADI residuals. 

\begin{figure*}
\begin{tabular}{ccc}
\begin{minipage}{0.3\hsize}
    \centering
    \includegraphics[width=0.95\textwidth]{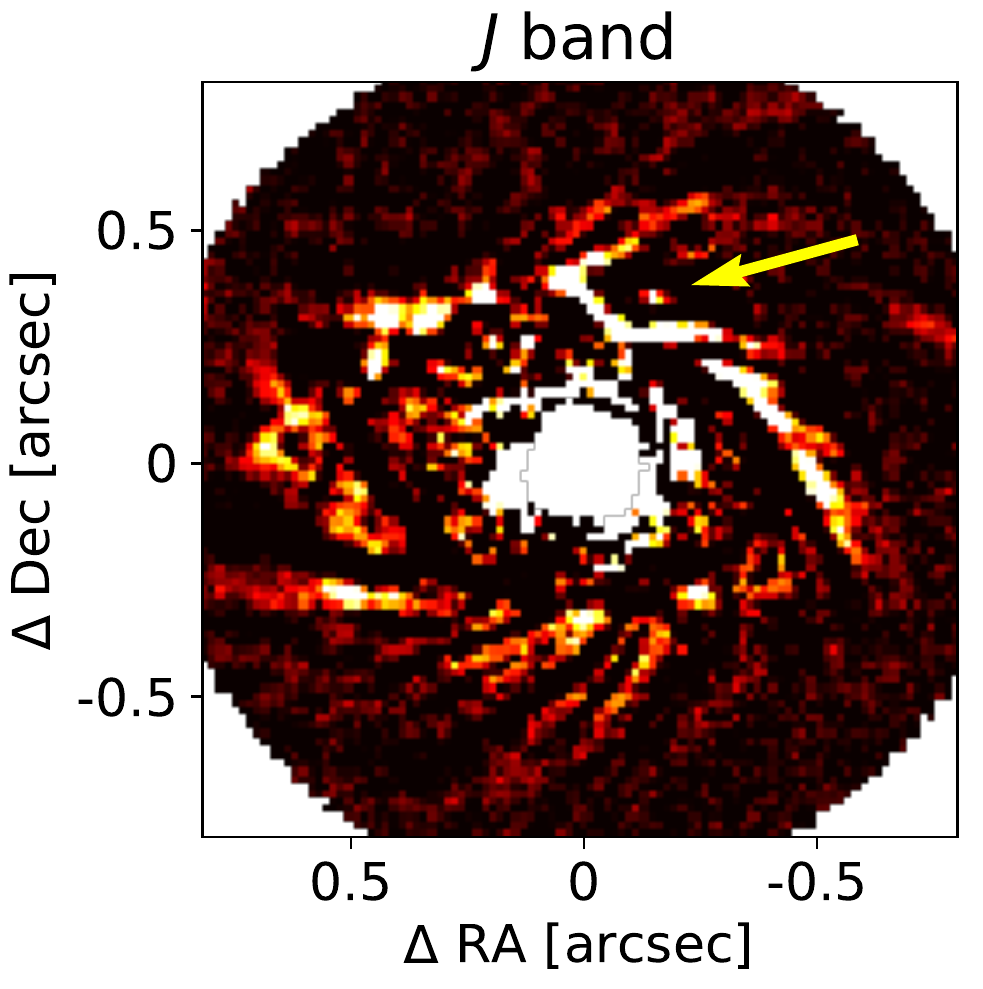}
\end{minipage}
\begin{minipage}{0.3\hsize}
    \centering
    \includegraphics[width=0.95\textwidth]{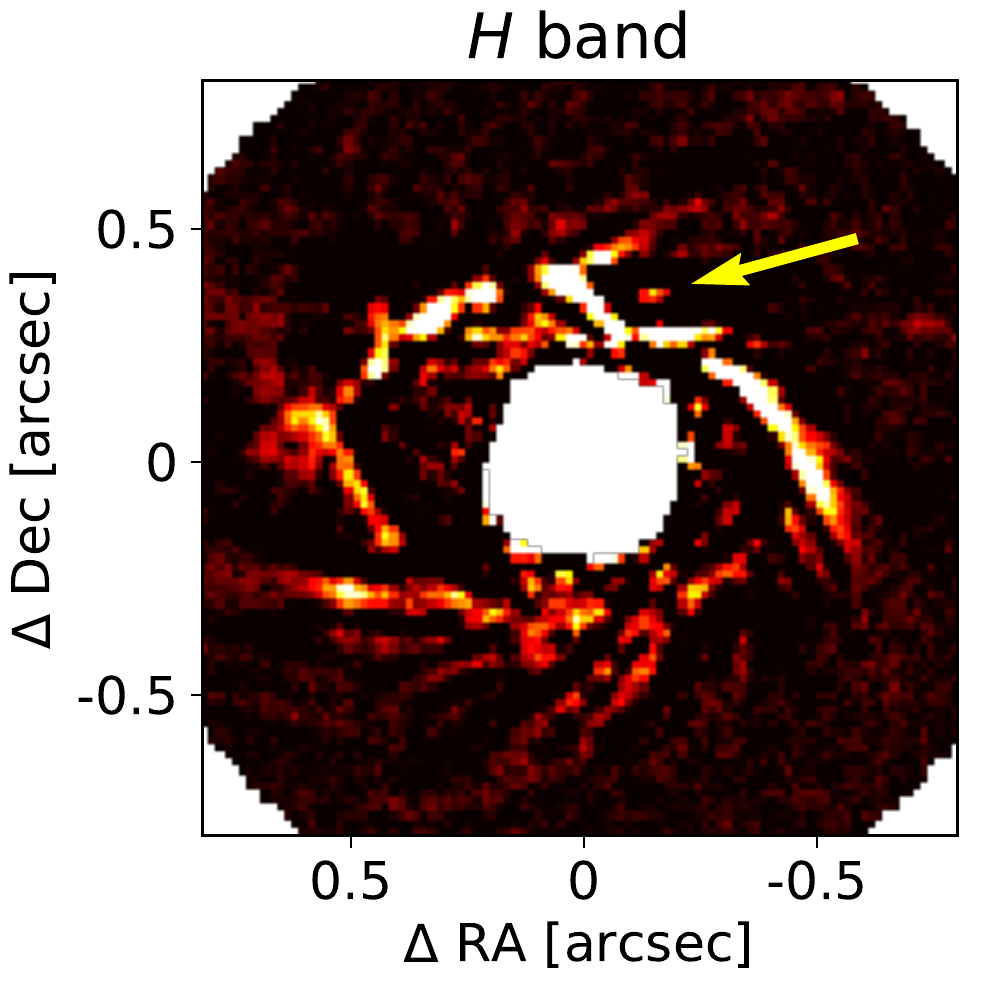}
\end{minipage}
\begin{minipage}{0.3\hsize}
    \centering
    \includegraphics[width=0.95\textwidth]{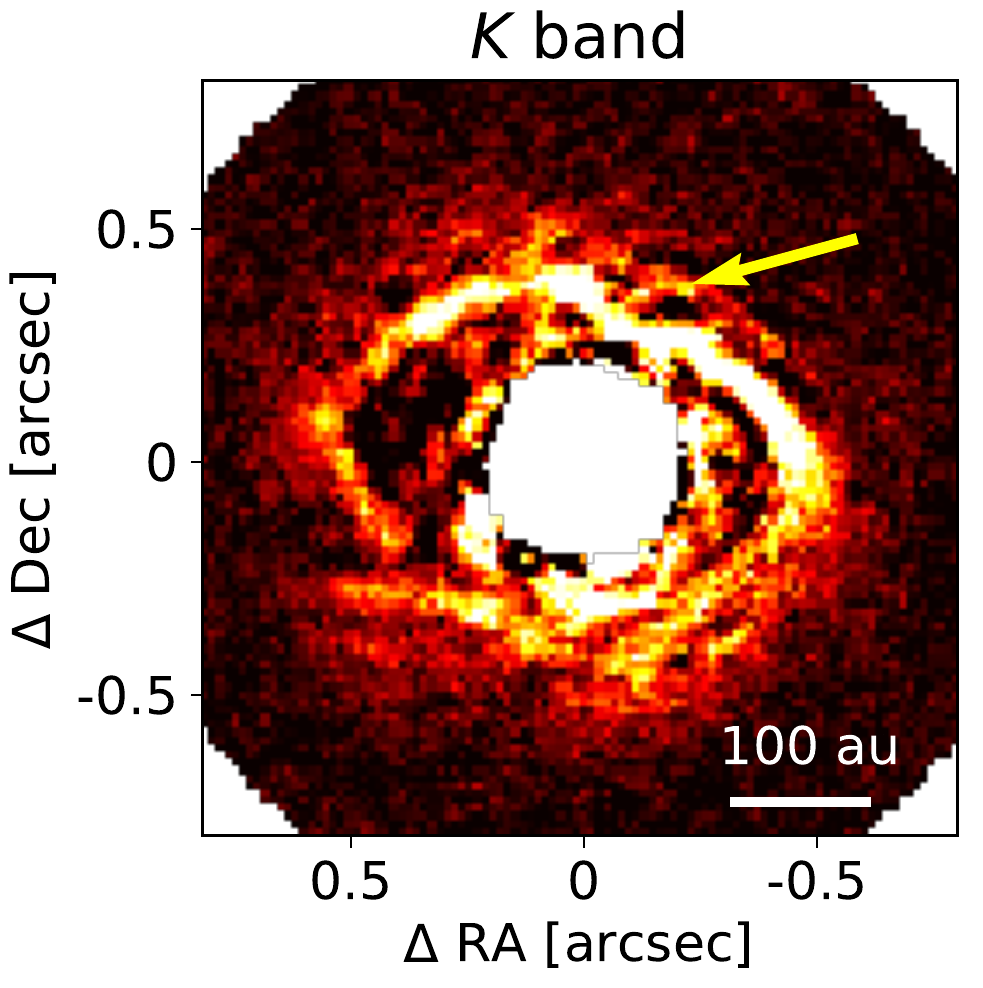}
\end{minipage}
\end{tabular}
\caption{Same comparison of the reduced images as Figure \ref{fig: RDI images} with ADI+SDI-ALOCI reduction. Color scale is arbitrary. A positive signal at a similar location to where \cite{Monnier2019} predicted a substellar-mass companion, which is likely a part of disk distorted or an artifact by the ADI+SDI reduction, is indicated by yellow arrow in each image (see Section \ref{sec: Constraints on Potential Companions} for details).}
\label{fig: ADISDI images}
\end{figure*}

\begin{figure}
    \centering
    \includegraphics[width=0.45\textwidth]{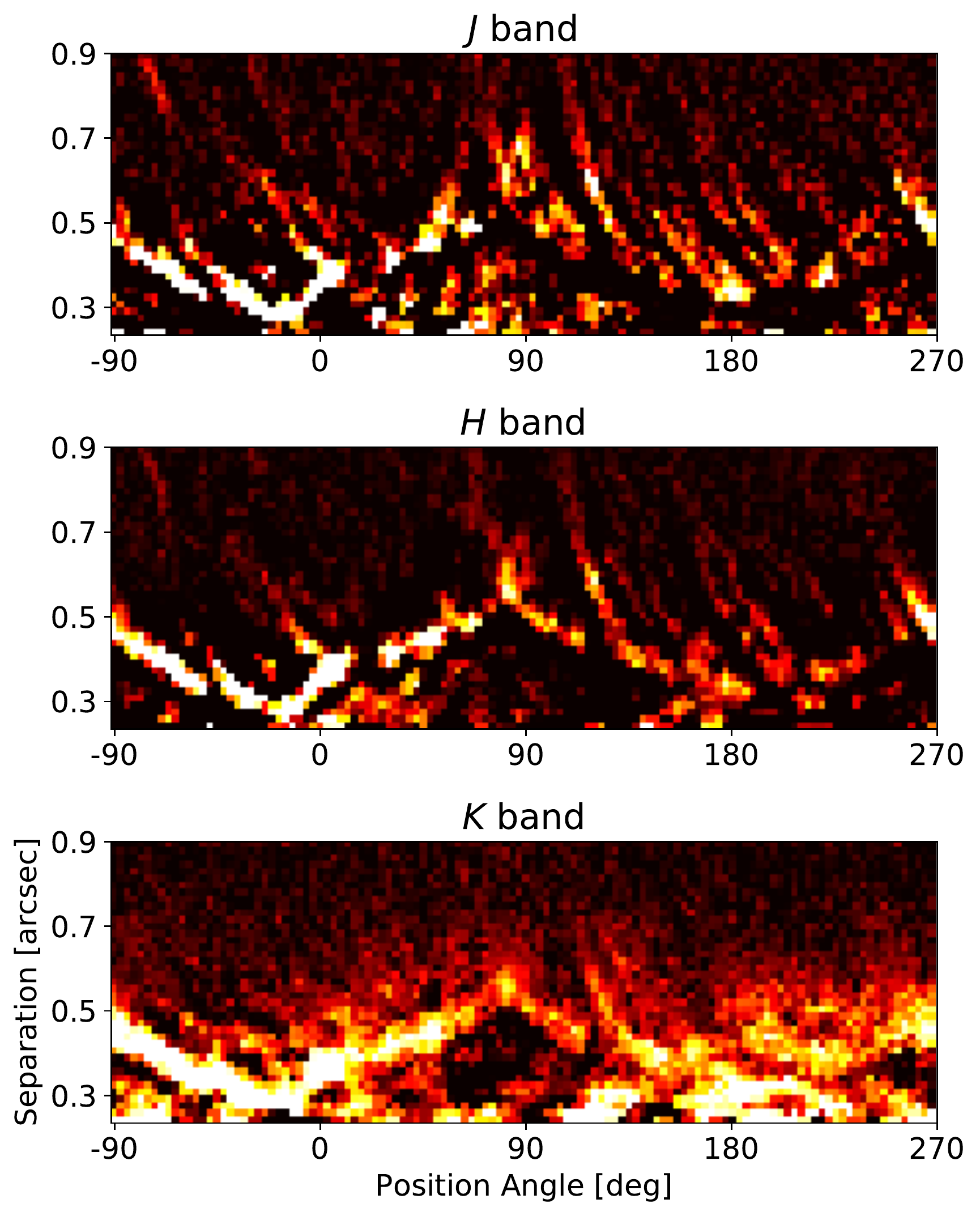}
    \caption{Same as Figure \ref{fig: r-theta RDI} for Figure \ref{fig: ADISDI images} at the ring+spiral area.}
    \label{fig: r-theta ADISDI}
\end{figure}

Figures \ref{fig: ADISDI images} and \ref{fig: r-theta ADISDI} show $J$, $H$, and $K$-band images reduced using ADI+SDI instead of RDI (see Figures \ref{fig: RDI images} and \ref{fig: r-theta RDI}).
We were able to detect several spiral features that are not detected by the RDI reduction (signal-to-noise ratios (SNRs) $\geq$4 along the spines of the spirals)\footnote{The noise is defined as standard deviation at separations between $0\farcs75-{\rm FWHM}/2$ and $0\farcs75+{\rm FWHM}/2$ in each ADI+SDI-reduced image.}.
Details of the spiral fitting are described in Section \ref{sec: Spiral Characterization}.

\subsection{Constraints on Potential Companions} \label{sec: Constraints on Potential Companions}
\begin{figure}
\centering
\includegraphics[width=0.48\textwidth]{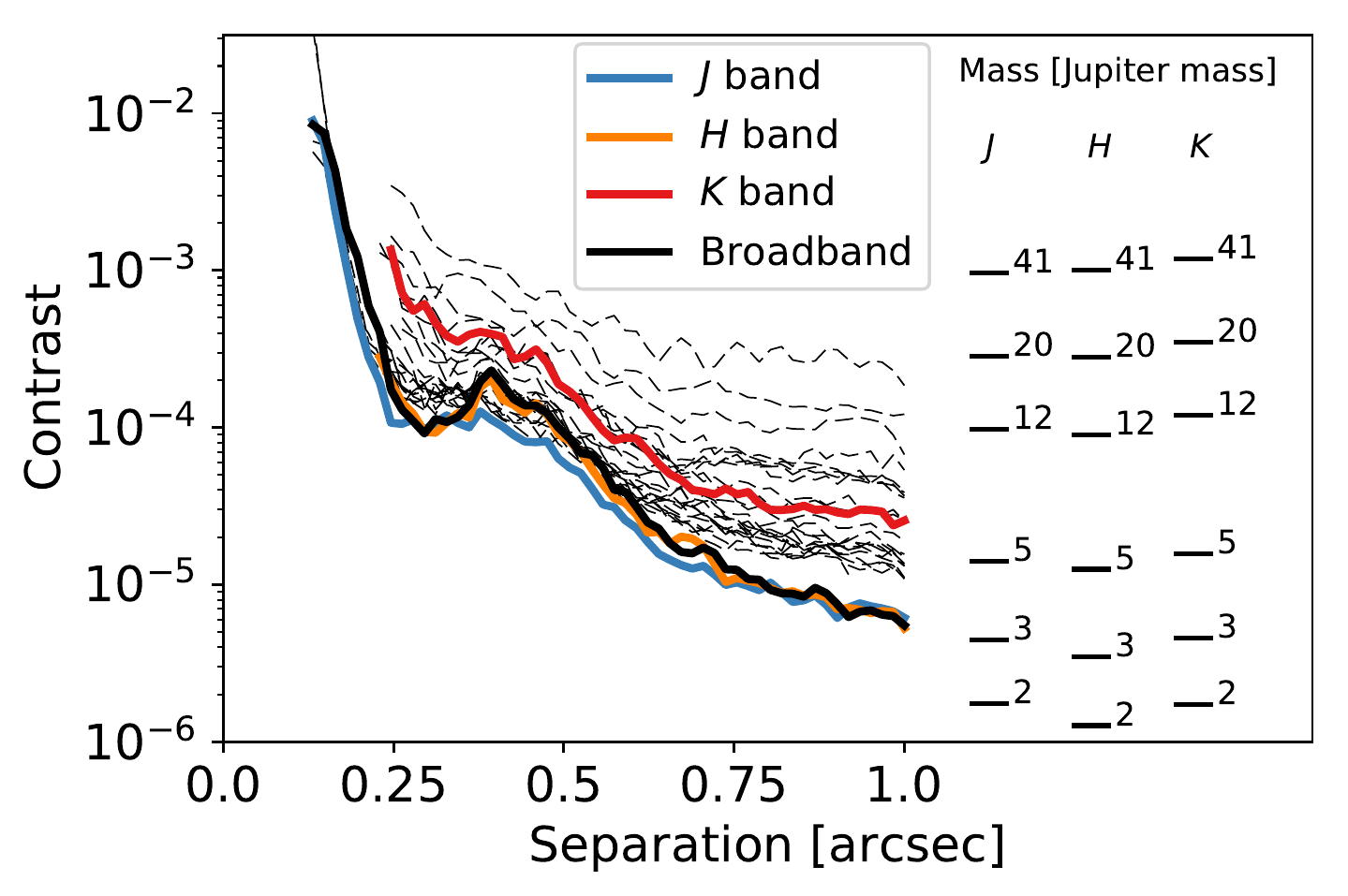}
\caption{5$\sigma$ contrast limits of our ADI+SDI result. Dashed lines correspond to that at each slice and $J$, $H$, $K$, and Broadband ($JHK$) correspond to those of collapsed images at each wavelength, respectively. We also plot mass as a function of contrast at three wavelengths assuming COND03 and 5 Myr.}
\label{fig: contrast}
\end{figure}

Our data did not reveal any substellar-mass companion candidates. 
We determined contrast limits by calculating radial noise profiles at each spectral channel, as in prior studies \citep{Currie2011}, including a {small sample statistics} correction \citep{Mawet2014}. 
We took account of throughput correction by estimating flux loss of injected fake point sources made by the ADI+SDI reduction.
We note that noise in this section is different from the noise of surface brightness used in Sections \ref{sec: RDI} and \ref{sec: ADI+SDI} because we aim at constraining point sources and thus used convolved images with aperture radii=FWHM/2.
Figure \ref{fig: contrast} shows 5$\sigma$ contrast limits of our ADI+SDI reduction results and comparison with mass unit at each band assuming a hot-start model \citep[COND03;][]{Baraffe2003-COND03} and 5 Myr. The broadband contrast achieved $10^{-4}$ ($\sim12\ M_{\rm Jup}$) at $0\farcs3$ and $10^{-5}$ ($\sim5\ M_{\rm Jup}$) at $0\farcs75$.
The detection limits are strongly affected by the bright ring and spirals at separations $\gtrsim 0\farcs4$, which bias an estimate of the noise. $K$-band contrast limits are poorer than $JH$-band limits because of the thermal background at channels of longer wavelength.
With a cold-start model \citep{Spiegel2012} a 10 $M_{\rm Jup}$ object corresponds to $\sim10^{-7}$ contrast at each band and we do not compare our detection limits with the cold-start model.

\begin{figure*}
\begin{tabular}{ccc}
\begin{minipage}{0.3\hsize}
    \centering
    \includegraphics[width=0.95\textwidth]{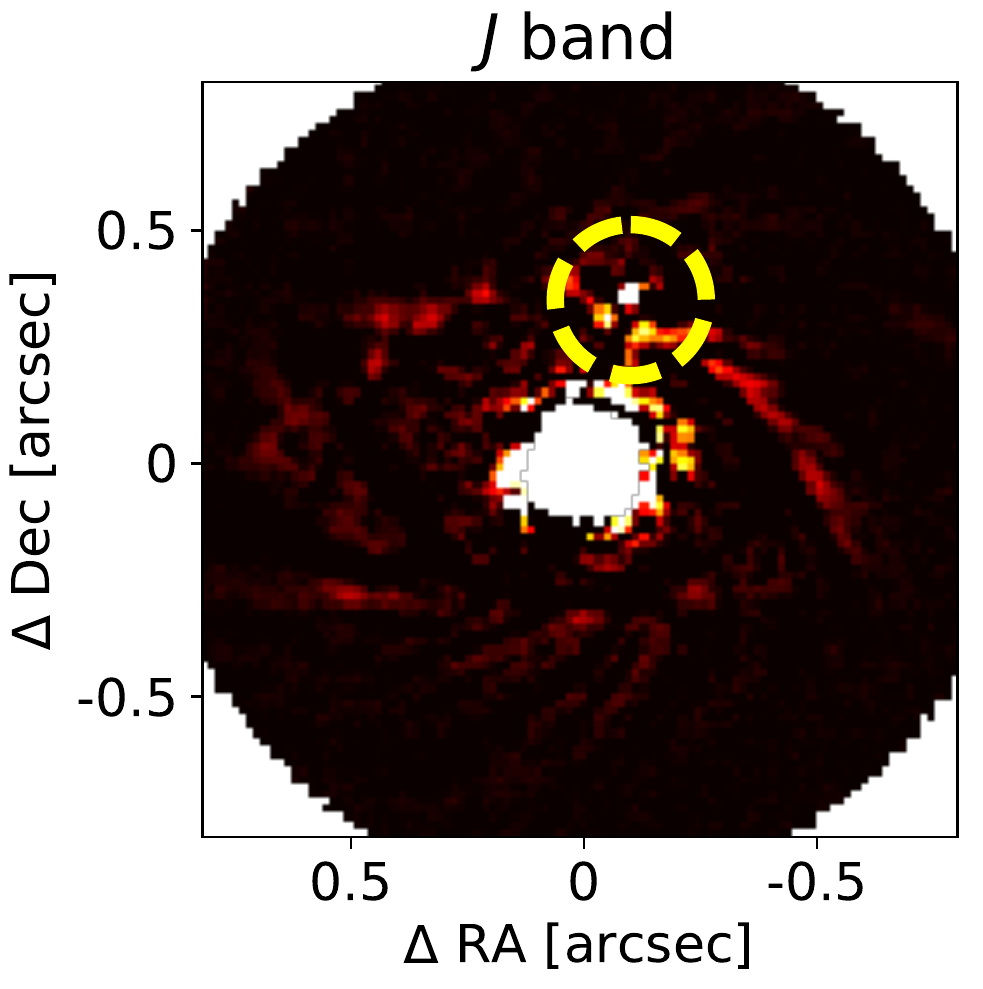}
\end{minipage}
\begin{minipage}{0.3\hsize}
    \centering
    \includegraphics[width=0.95\textwidth]{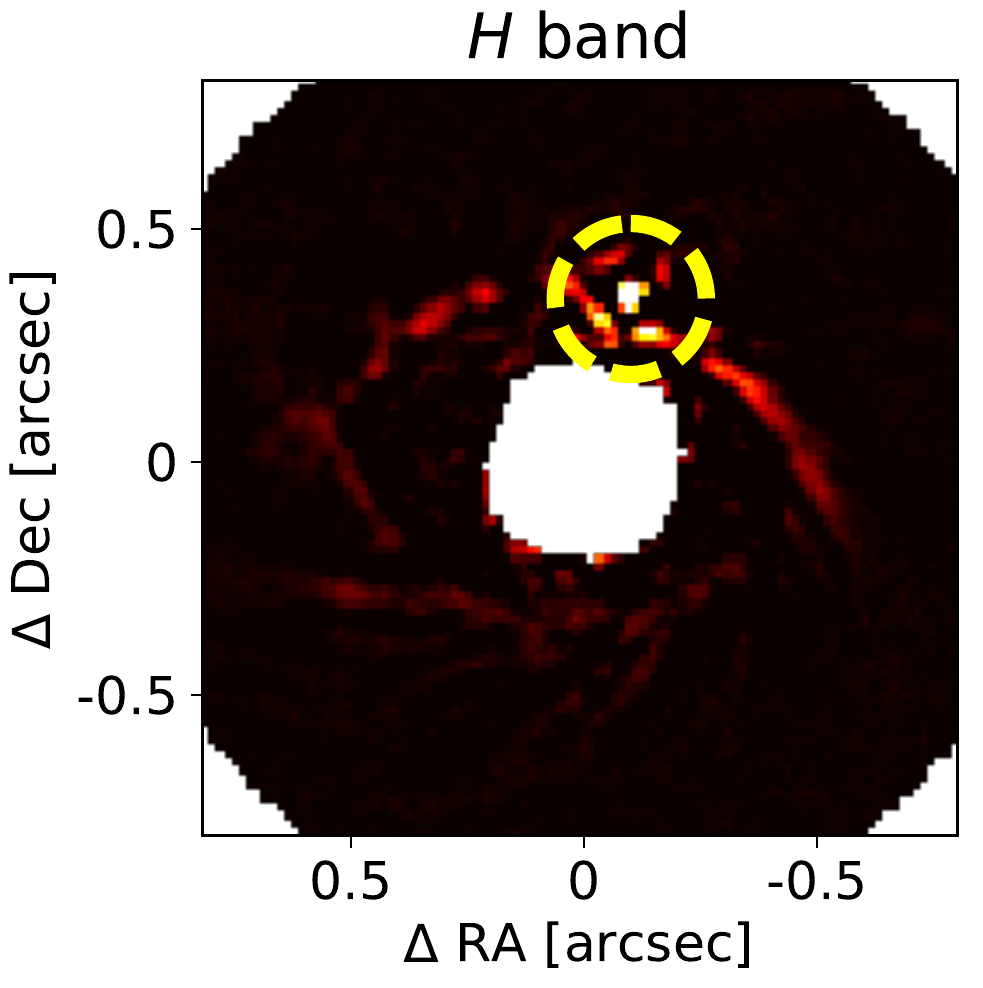}
\end{minipage}
\begin{minipage}{0.3\hsize}
    \centering
    \includegraphics[width=0.95\textwidth]{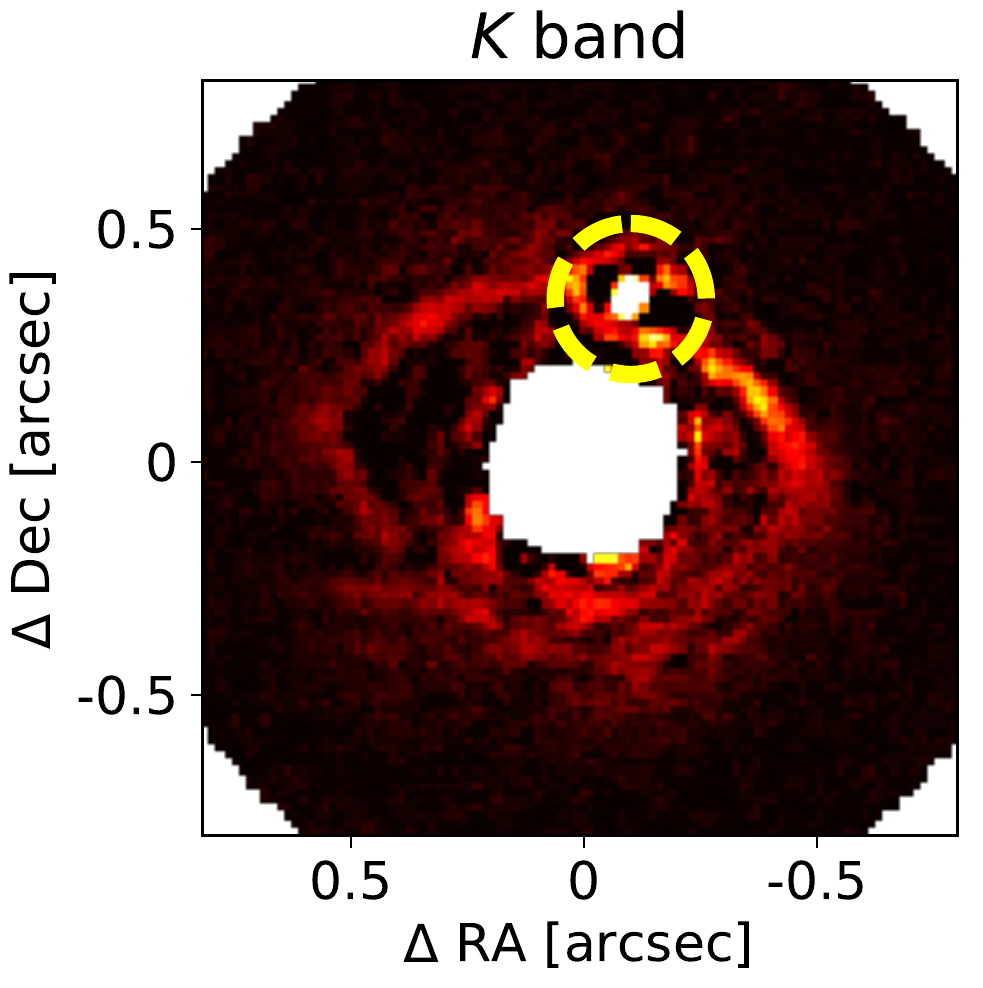}
\end{minipage}
\end{tabular}
\caption{As Figure \ref{fig: ADISDI images} with an injected fake source (indicated by dashed yellow circles) to test the hypothesis of a 50-$M_{\rm Jup}$ companion. We changed the color scale from Figure \ref{fig: ADISDI images} to clearly show the injected source.}
\label{fig: ADISDI with fake source}
\end{figure*}

To test a hypothesis of an eccentric ($e=0.2$) 50 $M_{\rm Jup}$ companion embedded in the disk \citep{Monnier2019}, we injected a fake source in the CHARIS data set and reran the ADI+SDI reduction. 
For a spectrum we made a planet model among $JHK$ bands by assuming $H$-band contrast of $10^{-2.8}$, which corresponds to 0.05 $M_\odot$ and $\sim$2800 K at 5 Myr in COND03 model, using DH Tau B's spectrum based on the VLT/SINFONI spectral library \citep[][]{Bonnefoy2014}.
For a location we injected the fake source at $0\farcs35$ North and $0\farcs1$ West from the center \cite[see also Figure 14 in][]{Monnier2019}.
Figure \ref{fig: ADISDI with fake source} shows the ADI+SDI images with the injected fake source indicated by the dashed yellow circle. Compared with the actual ADI+SDI result the fake source can clearly be seen and self-subtraction by this source distorts the nearby ring shape. Therefore we conclude that our observation could set a robust constraint on the potential substellar-mass companion \cite{Monnier2019} predicted.
We note that there is indeed a positive, albeit much fainter, signal at a similar location (indicated by yellow arrows in Figure \ref{fig: ADISDI images}). These signals are elongated and not significant among all the CHARIS channels. As the RDI result did not detect any counterpart this is likely part of the disk feature distorted or an artifact by the ADI+SDI reduction.

\section{Analysis and Discussion} \label{sec: Analysis and Discussion}
\subsection{Disk Morphology} \label{sec: Disk Morphology}
\begin{figure*}
    \centering
    \includegraphics[width=\textwidth]{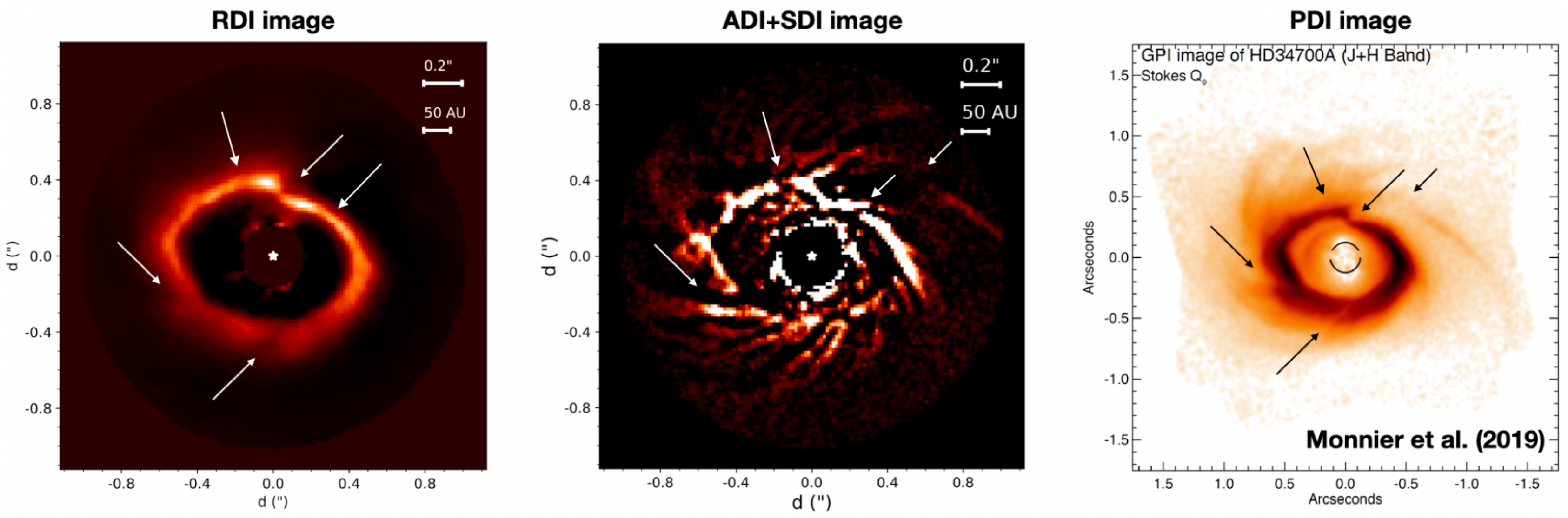}
    \caption{Comparison of the collapsed RDI and ADI+SDI images to the GPI-PDI image \citet{Monnier2019}, reproduced by permission of J. Monnier. 
    The arrows indicate darkening features (see Section \ref{sec: Shadows}). The central star is indicated by a white star in the masked region.}
    \label{fig:comparison_ADISDI_RDI_PDI}
\end{figure*}

Figure~\ref{fig:comparison_ADISDI_RDI_PDI} compares our new ADI+SDI and RDI images of HD 34700 A to the GPI-polarimetric differential imaging (PDI) result shown in \citet{Monnier2019}.
In this subsection we describe the ring, darkening features, and spirals in detail.

\subsubsection{Ring} \label{sec: Ring}

\begin{figure*}
    \centering
    \includegraphics[width=\textwidth]{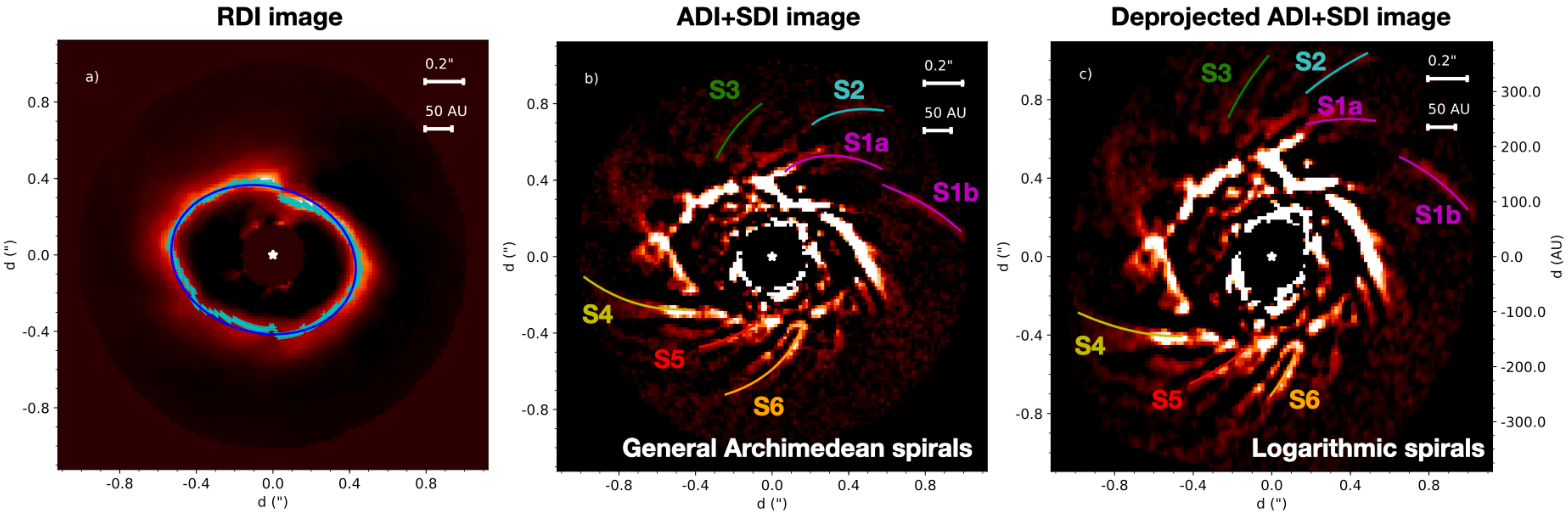}
    \caption{{\bf a)} Fit of the ring to an ellipse (blue curve) overlaid on the $JHK$band-collapsed RDI image. Cyan crosses show local radial maxima used for the fit. {\bf b)} Fit of the spiral arms seen in the collapsed ADI+SDI image to the equation of a general Archimedean spiral. {\bf c)} Deprojected ADI+SDI disk image (assuming a thin-disk), where the spirals are fit to the equation of a logarithmic spiral, in order to estimate their pitch angle. In all images the central star is indicated by a white star.}
    \label{fig:spiral_ring_fits}
\end{figure*}

We estimate that the RDI process alters the signal less than $\sim$15\% (see Section \ref{sec: Forward Modeling}), and thus we use the RDI images for our analysis.
We fit the bright edge of the cavity to an ellipse using the python ellipse fitting tool described in \citet{Hammel2020}. We provided as input to the python routine the pixel coordinates of the local radial maxima in the surface brightness profiles in $1^\circ$-wide azimuthal sections (cyan points in Figure~\ref{fig:spiral_ring_fits}a).
We performed the fit separately on the $J$-, $H$- and $K$-band RDI images (Figure~\ref{fig: r-theta RDI}) and found consistent results.
The uncertainties on each parameter of the ellipse were obtained in each band using the standard deviation of the Gaussian fit to the distribution of fitting results for 10000 bootstraps. Our final results are an average of the best fits obtained in each band, with the uncertainties for each band combined in quadrature. 
We found a semi-major axis of 487.1 mas $\pm$ 2.7 mas (173.6 AU $\pm$ 1.0 AU) and a shift of the center of the ellipse with respect to the star of 52.7 mas $\pm$ 2.3 mas towards a position angle (PA) of $110.8^\circ \pm 2.4^\circ$.

Assuming that the actual shape of the cavity is circular, our best-fit ellipse suggests a disk inclination of $40.9^\circ \pm 0.8^\circ$ and PA of semi-major axis of $74.5^\circ \pm 1.0^\circ$.
Regarding the uncertainty of PA we include the fitting uncertainty and CHARIS uncertainty on true north \citep[$0.27^\circ$; see Appendix A of][]{Currie2018}.
Our estimates of the cavity parameters and the disk inclination are all consistent to those inferred in \citet{Monnier2019} using a similar method applied to the PDI image, apart from the value of the PA of the semi-major axis of the disk (for which they found $69.0^\circ \pm 2.3^\circ$).
The slight discrepancy might be due to the difference of the scattering phase function between polarized intensity and total intensity.  
However, such difference can stem from the requirement for the shift of the center of the ellipse with respect to the star to lie along the semi-minor axis in their procedure. 
Considering the uncertainties on the centering of the star and the assumption of a circular shape cavity, we did not force this condition in our procedure.

We note that the assumption of a circular cavity is not necessarily correct, given that several disks with large cavities show a non-null eccentricity, such as HD~142527 and MWC~758 \citep{Avenhaus2014,Dong2018}.
New ALMA data probing the kinematics of the disk would provide an independent estimate of the inclination of the disk. The difference of inclination, if any, estimated from the scattered light and ALMA will suggest different distribution of gas/small grains/large grains and therefore we can answer whether the assumption of circular cavity is reasonable.

\subsubsection{Darkening Effects} \label{sec: Shadows}

The RDI images show evidence for multiple darkening areas on the bright edge of the cavity and these areas, except for the Northwest one, coincide with the GPI-PDI image, which are indicated by arrows in Figure \ref{fig:comparison_ADISDI_RDI_PDI}.
We also indicate these regions by gray shades in plots of surface brightness and geometric albedo of the ring (see Figure \ref{fig: azimuthal profile ring} and Section \ref{sec: Scattering Profiles} for details).
We first note that the darkening features on the ring, except the Northwest one, are located by the roots of S1, possibly S3, S4, and S6 (see also Figure \ref{fig:spiral_ring_fits}). 
Shadows, actual geometric features, or other scattering characteristics due to heterogeneous dust distribution may give explanations of the darkening features.
We individually investigate the possibility of the shadowing effect for each darkening feature but do not rule out other possibilities. We also note that reproducing all the darkening features by only the shadowing effect likely requires multiple inner disks, which may be dynamically unstable. It is hard to identify which mechanism is the most favorable for reproducing each feature in this study.

Prominent roughly symmetric shadowing effects can be seen to the North and South, at PA spanning $\sim-25^\circ$ to $30^\circ$ and $\sim155^\circ$ to $200^\circ$, and to the Northwest and Southeast, at PA spanning $\sim105^\circ$ to $120^\circ$ and $\sim290^\circ$ to $325^\circ$ (-70$^\circ$ to -35$^\circ$), respectively.
There might be other possible darkening areas that are marginally seen in our reduced images and the surface brightness profile (e.g. PA $\sim210^\circ$ in $J$ band), but they are less convincing than those mentioned above and we do not conclude such possible features as shadows or spiral roots in this study. 
The effect of the North-South symmetric shadows is seen in all bands, albeit stronger at a shorter wavelength. The other pair is only seen in $J$ and $H$ band suggesting optically-thin at longer wavelength (Figures~\ref{fig: r-theta RDI} and \ref{fig: azimuthal profile ring}) or different scattering characteristics (in this case the darkening feature corresponds to a non-shadowing effect).
These darkening features can also be seen in our ADI+SDI image, although less conspicuously given the presence of radial post-processing artifacts (left panel of Figure~\ref{fig:comparison_ADISDI_RDI_PDI}).
Furthermore, our ADI+SDI image suggests shadowing of a part of the main NW spiral, which appears to lie in the continuity of the Northern part of the symmetric N-S shadow.
 
A comparison of our images to the PDI image shown in \citet{Monnier2019} confirms the presence of all darkening areas, except one in the Northwest direction, in their image too (right panel of Figure \ref{fig:comparison_ADISDI_RDI_PDI}) - albeit not reported as such. 
The symmetric shadows are reminiscent of polarimetric imaging or space-based coronagraphic imaging of the disks such as HD~142527 \citep{Avenhaus2014,Marino2015a}, HD~100453 \citep{Benisty2017}, HD~163296 \citep{Wisniewski2008,Rich2019}, SAO~206462 \citep{Stolker2017}, and DoAr 44 \citep{Casassus2018}, and suggest the presence of an inclined inner disk.
These shadow features may also be reproduced by a combination of single shadows. 
The single shadow is reminiscent of the ones observed in the circumbinary disk of GG~Tau~A, which includes a close central binary similar to HD~34700~A \citep{Itoh2002,Itoh2014}, and in the transition disk of HD~169142 \citep{Quanz2013,Bertrang2018}. For GG Tau, several explanations for the single shadow have been proposed, including a dense clump in an accretion stream onto one component of the central binary, or a circumplanetary disk surrounding a protoplanet located in the cavity \citep{Krist2002,Canovas2017} as well as circumstellar disks around GG Tau Aa/b \citep{Baruer2019,Keppler2020}.

Finally we note that our observation did not detect any further inner object(s) down to $\sim0\farcs2$.
ALMA continuum observation may help to investigate possible inner disk(s). 
Follow-up high-contrast observations are also useful to investigate time variation of the shadows and to constrain inner objects as previous observations reported (possible) changes of shadow features \citep[in a time scale of years;][]{Wisniewski2008,Debes2017,Stolker2017,Uyama2018,Rich2019,Lass2020}.
Assuming that an inner object at a radius of $0\farcs2$ (73 au, slightly interior to the physical inner working angle in our $J$-band result)) casts a shadow on the ring (0.5”) and that we can identify the time variation of the shadow if the shadow shifts by 30 mas (= $J$-band angular resolution in our observation), the inner object should move 12 mas (4.4 au). A period of Keplerian rotation at 73 au around HD 34700 A is about 313 years and the 4.4-au movement takes 3 years.
A color discussion at the darkening areas with the high-contrast imaging may also help to investigate whether possible inner object(s) are optically-think or thin.
If the darkening areas are accompanied with actual geometric features the scattered light there might include multiple scattering, a ratio of which depends on dust properties \citep[e.g.][]{Takami2013}, and then detailed discussions with radiative transfer simulations are required for a synthetic understanding of the HD 34700 A's disk.

\subsubsection{Spiral Characterization} \label{sec: Spiral Characterization}

To increase SNRs of the faint spirals we used a median-combined ADI+SDI image using all CHARIS spectral channels. 
Although the ADI+SDI reduction can cause self-subtraction of the spiral features, our data reduction adopts reasonable settings to avoid biasing the actual morphology (see Section \ref{sec: ADI+SDI} for the settings). The rotation gap ($\delta=0.75$) limits the self-subtraction of the astrophysical signal caused by rotating the field.
With a local pixel masking over the subtraction zone, the astrophysical signal contained within the subtraction zone does not bias the LOCI coefficients and self-subtraction is reduced \citep[for details see][]{Currie2018}.

We followed the same procedure as in \citet{Reggiani2018} and \citet{Price2018} to identify the trace of spiral arms as local maxima in the radial intensity profile of the disk, and fit them to the equation of using the equations of general Archimedian and logarithmic spiral arms, respectively. 
The fits to the equation of general Archimedean spirals systematically yield the best morphological match, while the fit to logarithmic spirals is used for pitch angle estimation. In polar coordinates, a general Archimedean spiral is given by the equation $r = a + b \theta^n$, and a logarithmic spiral by $ r = r_0 e^{k \theta}$, where the pitch angle ($\phi=\arctan(k)$) is constant and determines the spiral.
With this procedure, we fit the six brightest spirals outside the ring, including two arcs likely tracing the same spiral but truncated due to shadowing from the inner disk (referred to as S1a and S1b). All the identified spirals have SNRs $\gtrsim4$ at their spines, except S3 (SNR $\sim3-3.5$), in the collapsed ADI+SDI image.
The first column in Table~\ref{Spiral_fits} reports the pitch angle measured for each spiral arm labeled in Figure~\ref{fig:spiral_ring_fits}.

\begin{table}
 \caption{Pitch angle of the spirals}
 \centering
 \begin{tabular}{cccc}\\ \hline\hline
 Spiral & $\phi$ & $\phi_{\rm deproj, thin}$ & $\phi_{\rm deproj, h=0.3r}$\\
  & ($\deg$) & ($\deg$) & ($\deg$) \\ \hline
 S1a & $31.3 \pm 1.0$ & $34.7 \pm 1.4$ & $46.9 \pm 1.2$\\ 
 S1b & $40.2 \pm 1.4$ & $27.0 \pm 1.9$ & $27.1 \pm 1.5$\\
 S2 & $37.1 \pm 2.6$ &  $51.9 \pm 2.0$ & $62.8 \pm 0.9$\\
 S3 & $34.4 \pm 2.8$ &  $48.5 \pm 3.3$ & $61.0 \pm 0.8$\\
 S4 & $53.9 \pm 1.7$ & $49.4 \pm 1.8$ & $51.4 \pm 1.7$\\
 S5 & $50.2 \pm 1.2$ & $54.7 \pm 3.8$ & $51.0 \pm 1.4$\\
 S6 & $41.3\pm0.9$ & $53.2 \pm 2.0$ & $44.9 \pm 0.9$\\ \hline
 \end{tabular}
 \label{Spiral_fits}
\end{table}

Given that the disk is inclined, if the spirals are located in the same plane as the inner edge of the outer disk (i.e.~the bright ring), one has to measure spiral pitch angles in the deprojected image of the disk for a meaningful comparison to the values predicted by different spiral formation mechanisms. We deprojected the image with respect to the center of the disk, i.e.~considering the 52.7 mas shift with respect to the location of the star, and considering the values of inclination and PA of semi-major axis inferred in Section~\ref{sec: Ring}: 40.9$^\circ$ and 74.5$^\circ$, respectively. The pitch angles measured in the deprojected image are provided in the middle column of Table~\ref{Spiral_fits}.
This way of deprojection ignores the vertical characteristics of the spiral feature. For comparison we made another deprojected image with diskmap \citep{Stolker2016} by taking into account a large constant opening angle ($h(r)=0.3r$, where $h$ is the height of the scattering surface) and then conducted the spiral fitting, the results of which are also given at the last column in Table \ref{Spiral_fits}. The difference is significant along the semi-minor axis (S1a, S2, S3, and S6).
We note that our spiral fitting may also be affected by the inclination of HD 34700 A's disk. \cite{Dong2016-scatteredlight} suggested that scattered light of the spiral feature can be distorted by its inclination and image deprojection by $\sim40^\circ$ may not trace the real spiral feature.

The pitch angle values in Table 1 are significantly discrepant with the rough estimates of 20--30$^\circ$ reported in \citet{Monnier2019}. Since they do not mention how the pitch angles were measured nor whether a deprojection was performed for their measurement, it is difficult to discuss the reason for this discrepancy.
From the deprojected pitch angles of Table 1 in a thin-disk case, we notice a significant difference between S1 ($\phi_{\rm deproj} \sim 30^\circ$) and the other spirals ($\phi_{\rm deproj} \sim 50^\circ$), possibly pointing to different origins. In a thick-disk case the pitch angle of S1b is significantly different to other spirals. As S1b is located at a larger separation from the ring and it might have different characteristics than the other spirals.
We note that ADI+SDI reduction could cause distortion of the real shapes because of self-subtraction and follow-up observations with ALMA or high-contrast PDI/RDI reductions with a comparable angular resolution and sensitivity will help to confirm our result of the spiral fitting.

\subsection{Forward Modeling} \label{sec: Forward Modeling}

To investigate the disk's scattering profile, we use forward-modeling to reproduce the observed ring with a synthetic scattered light disk and simultaneously match most of the system's spectral energy distribution.
Here we do not include probable shadowing effects due to inner disk(s).
We followed \cite{Currie2019} with MCMax3D radiative transfer code \citep{Min2009} to model the ring and compared the forward-modeled disk with the CHARIS-RDI result.
We adopted the best-fit parameters in \cite{Monnier2019} as initial parameters and then explored a small range of the model component parameters to reproduce the scattered light image. 

\begin{figure*}
\begin{tabular}{ccc}
\begin{minipage}{0.3\hsize}
    \centering
    \includegraphics[width=\textwidth]{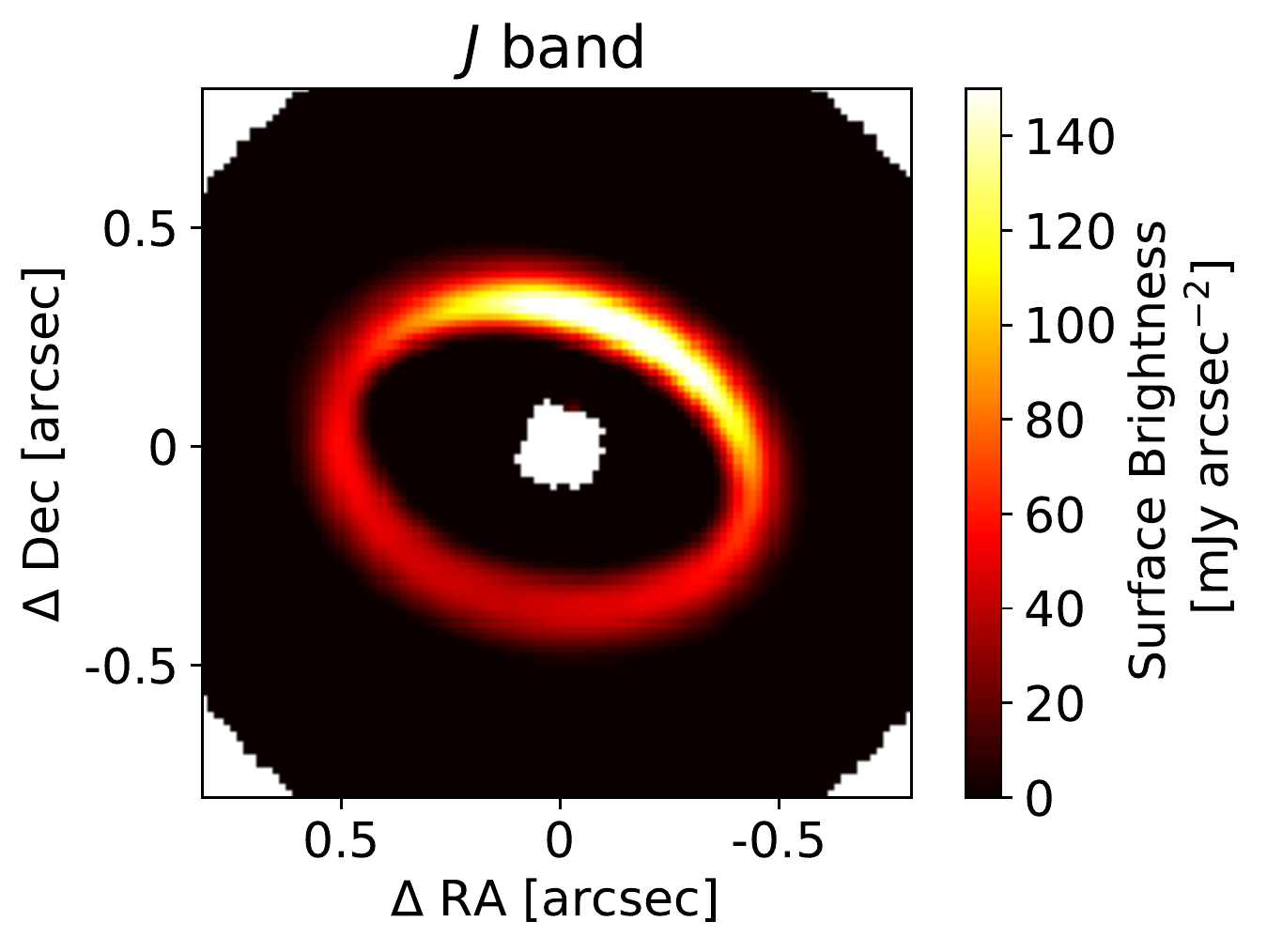}
\end{minipage}
\begin{minipage}{0.3\hsize}
    \centering
    \includegraphics[width=\textwidth]{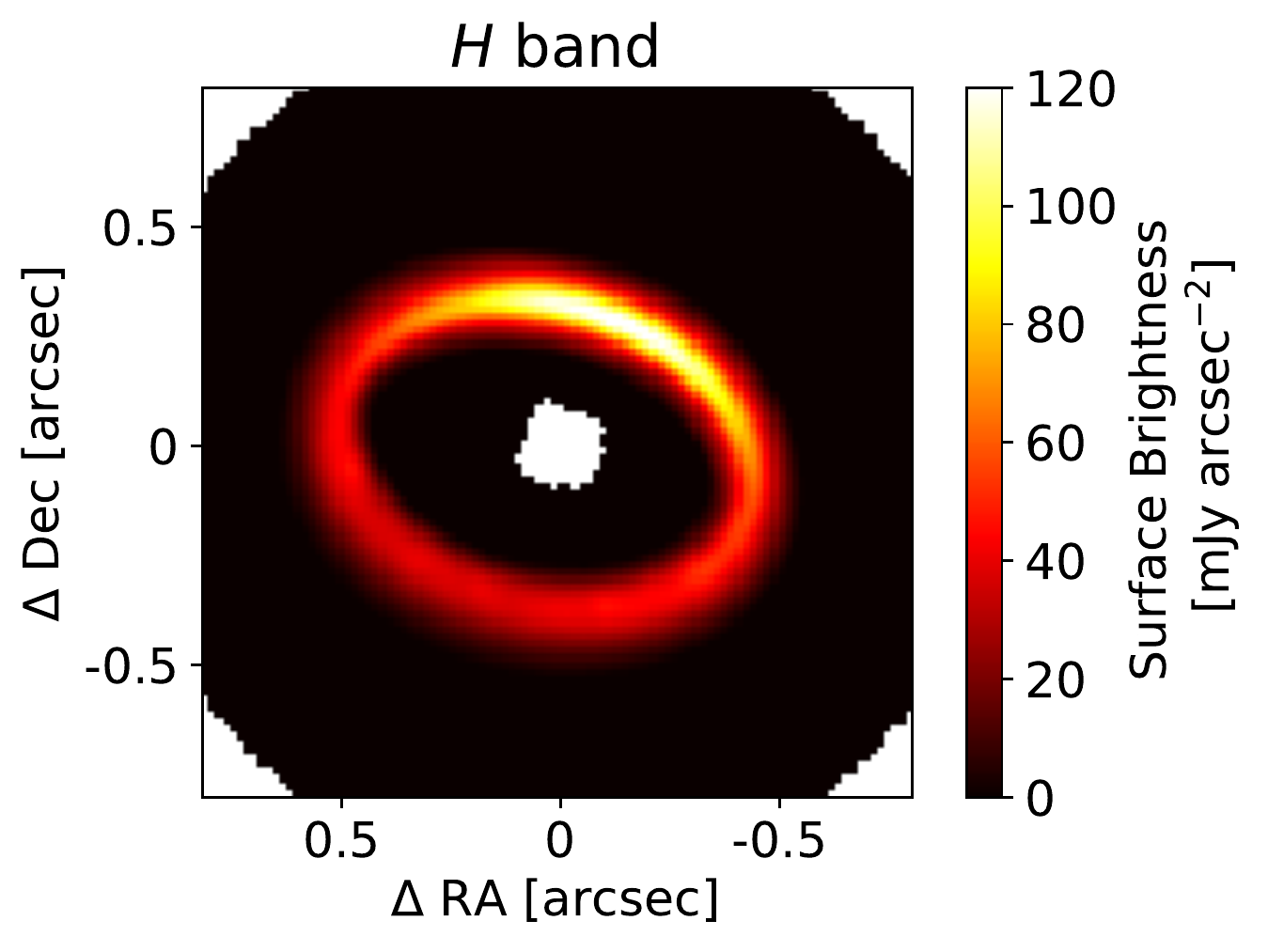}
\end{minipage}
\begin{minipage}{0.3\hsize}
    \centering
    \includegraphics[width=\textwidth]{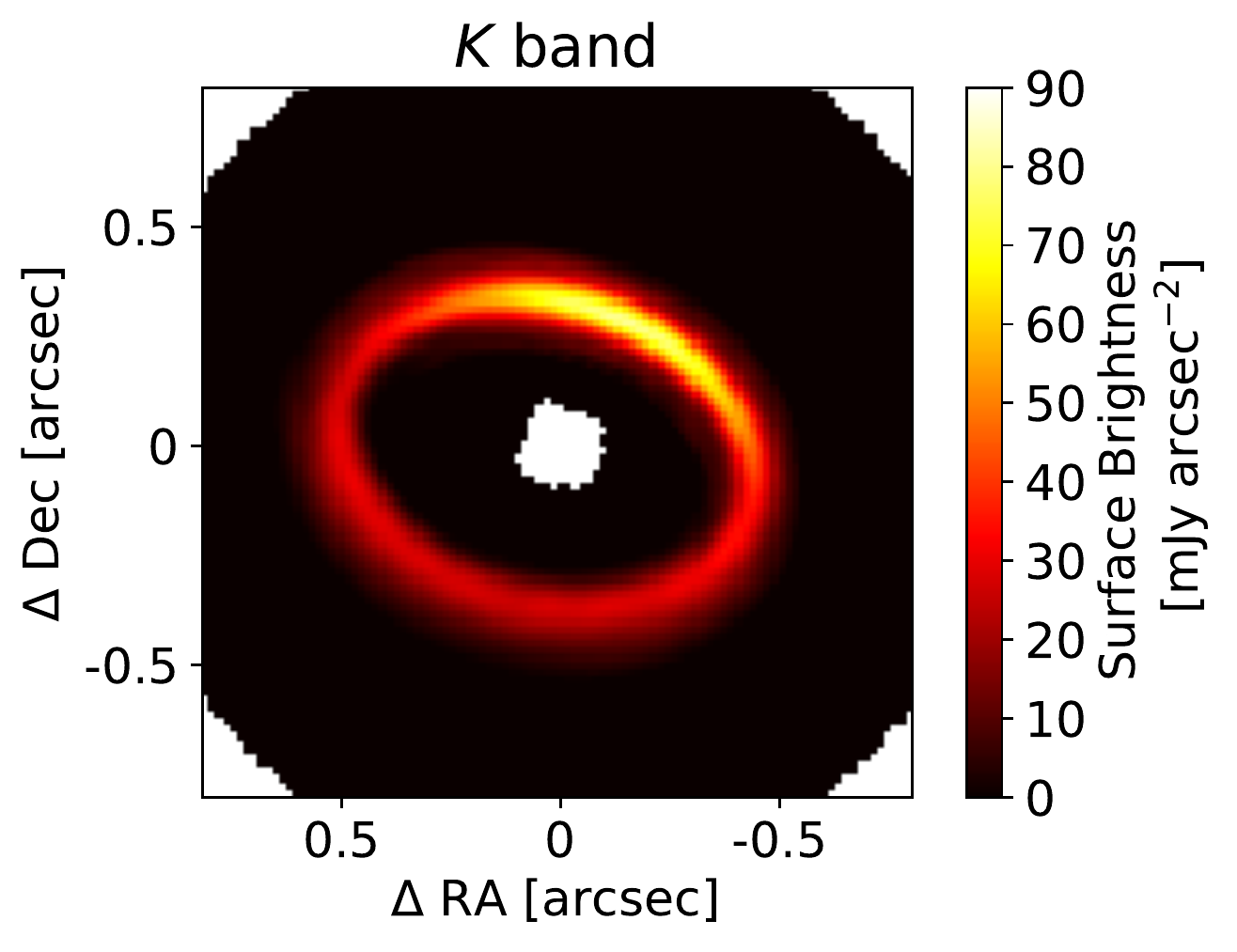}
\end{minipage}
\end{tabular}
\caption{The best-fit forward-modeled disks at $J$ (left), $H$ (center), and $K$ (right) bands. The images are convolved by the instrumental PSF and then reduced by the RDI reduction.}
\label{fig: model disk}
\end{figure*}

Table \ref{mcmaxdisk} summarizes the best-fit model of the ring and Figure \ref{fig: model disk} compares the forward-modeled disks at $JHK$ bands. The disk component surface density follows $\Sigma$ ($R$ $<$ $R_{\rm w}$) $\propto$ $R^{-\epsilon}$$\times$exp(-$(\frac{1-R/R_{\rm exp}}{w})^{3}$)
and $\Sigma$ ($R$ $\ge$ $R_{\rm w}$) $\propto$ $R^{-\epsilon}$.
The scale height in our model is consistent with the best-fit scale height without VSG/PAH in \cite{Monnier2019}. Other best-fit parameters do not have large differences from the \cite{Monnier2019} best-fit parameters but our best-fit model provides a better match to the surface brightness (see Section \ref{sec: Scattering Profiles} for details).
We estimated attenuation factors by comparing the modeled disks before and after the KLIP-RDI reduction, which are used for throughput correction of the RDI reduction ($\sim$10--15\% flux loss at the ring peak). 
The throughput-corrected surface brightness of the ring is shown in Figure \ref{fig: azimuthal profile ring}.

\begin{deluxetable*}{lllll}
\tablecaption{Disk Model Parameters}
\tablewidth{0pt}
\tablehead{\colhead{Parameter} & \colhead{Value} & \colhead{} & \colhead{} & \colhead{}}
\tiny
\centering
\startdata
Disk Parameters \\
\hline
Distance$^*$ (pc) & 365.5 \\
$T_{\rm eff}$$^*$ (for Aa, Ab) (K)   & 5900 , 5800 \\
$L_{\star}$$^*$ (for Aa, Ab) ($L_{\odot}$)   &   13 , 11.5 \\
$R_{\star}$$^*$ (for Aa, Ab) ($R_{\odot}$)   &   3.46 , 3.4 \\
$M_{\star}$$^*$ (for Aa, Ab) ($M_{\odot}$)  &  2.0 , 2.0  \\
Separation between Aab$^*$ (au) & 0.69 \\
$A_{\rm V}$$^*$  & 0.0 \\
  Disk Position Angle ($\theta$) (deg) & 60\\
  Disk inclination ($i$) (deg) & 40.9\\  
  Disk Offset from Star - Major Axis (au) &  -10 \\
  Disk Offset from Star - Minor Axis (au) & 5 \\
  Inner radius, $R_{\rm in}$ (au) & 170   \\
  Outer radius, $R_{\rm out}$ (au) & 400  \\
  Disk wall radius, $R_{\rm w}$ (au) & 200 \\
  Scale height at inner radius, $H_{\rm o, in}$ & 0.1\\
  Scale height power law, $p_{\rm gas}$ & 1.2\\
  Radial surface density power law ($\epsilon$)  &0.5 \\
  Wall shape ($w$) & rounded/0.2\\
  $M_{\rm dust}$ ($M_{\odot}$) & 2.5$\times$10$^{-4}$ \\
  Minimum dust size ($a_{\rm min}$, $\mu m$ [small, large]) & 0.25, 5\\
  Maximum dust size ($a_{\rm max}$, $\mu m$ [small, large])& 5, 1000 \\
  Dust Size Power Law, $p_{a}$  & 3.5\\
  Dust Carbon Mass Fraction & 0.1\\
  Dust Silicate Mass Fraction & 0.9 \\
\hline
\enddata
\vspace{-0.05in}
\tablecomments{We fixed stellar parameters (with $*$ symbol) to those estimated in \cite{Monnier2019}. The dust mass is evenly divided between ``small grain" and ``large grain" components. The wall shape parameter defines the spatial scale over which the disk surface density increases from $R_{\rm in}$ to $R_{\rm w}$.
See \citet{Mulders2010,Mulders2013} and \citet{Thalmann2014} for detailed explanations of the MCMax3D terminology.
}
\label{mcmaxdisk}
\vspace{-0.2in}
\end{deluxetable*}


\subsection{Scattering Profiles} \label{sec: Scattering Profiles}
\begin{figure}
    \centering
    \includegraphics[width=0.5\textwidth]{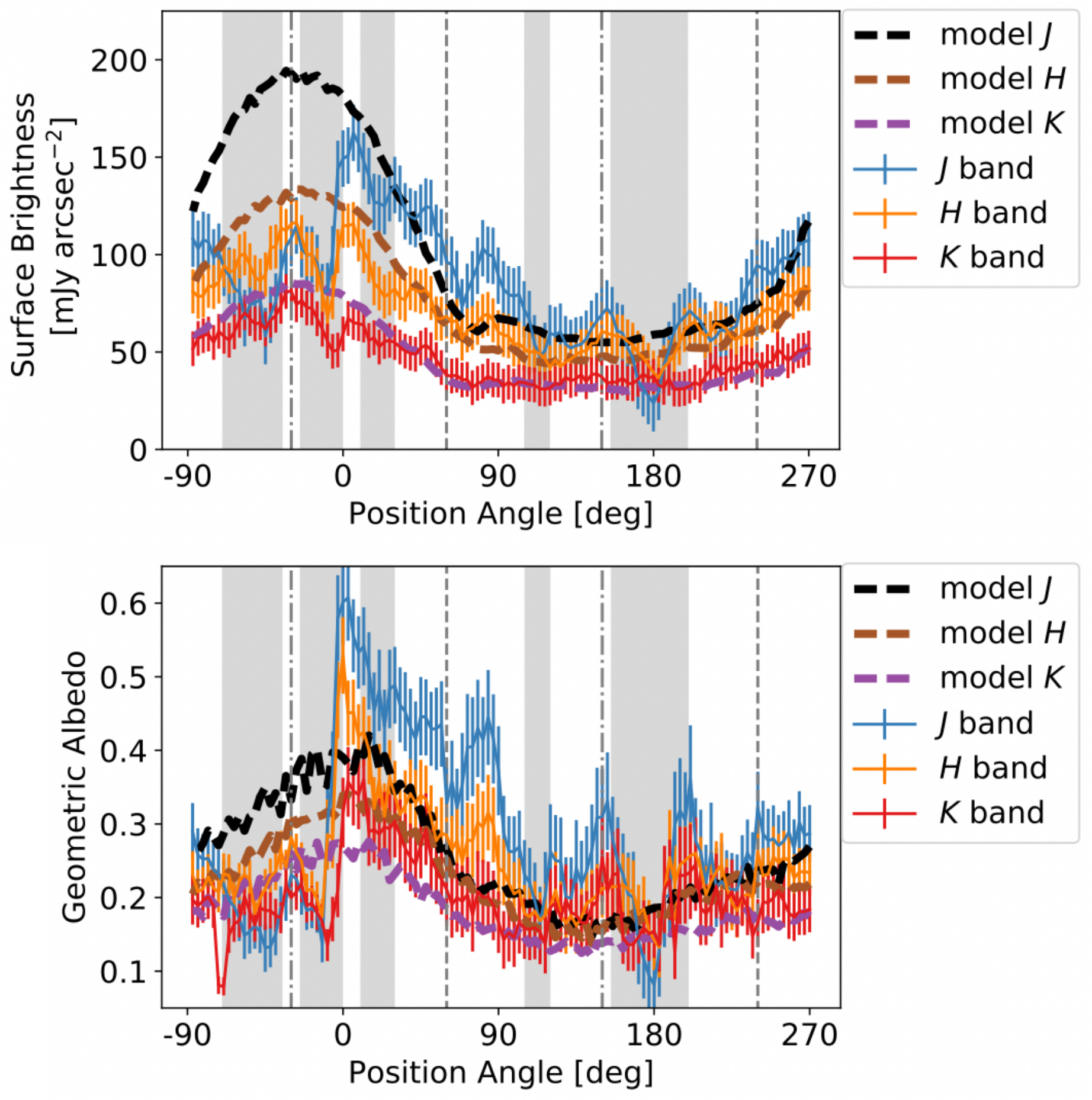}
    \caption{Azimuthal profiles of surface brightness after the throughput correction by tracing the ring peak (top) and geometric albedo converted from the surface brightness (bottom). Those profiles of the modeled disks are overlaid. Gray shaded areas indicate the darkening areas (see Section \ref{sec: Shadows}). Gray vertical dashed and dashed-dotted lines indicate the major and minor axes of the best-fit disk model respectively.
    Error bars in the top image correspond to 14.5, 11.3, and 8.93 mJy arcsec$^{-2}$ at $J$, $H$, and $K$ band, respectively. Note that the model adopted a simple ring without the darkening effects, discontinuity, and spirals seen in the actual disk.}
    \label{fig: azimuthal profile ring}
\end{figure}

Figure \ref{fig: azimuthal profile ring} compares azimuthal profiles of surface brightness and geometric albedo \citep[see Equation (3) of][]{Mulders2013} by tracing the peaks of the ring. Note that this geometric albedo depends on dust albedo and geometry of the disk.
Solid lines with errors correspond to the traced ring peaks from the collapsed $JHK$-band images after the throughput correction. Error bars are extrapolated from the background noise at $0\farcs7$ (see Section \ref{sec: RDI}).
The dashed lines correspond to the ring peaks from the modeled disk before the RDI reduction.

The model matches the general trend in surface brightness very well except for $J$ band at $\sim$-90$^\circ$ to 0$^\circ$, large parts of which are affected by the darkening effects (see also Section \ref{sec: Shadows}), and better reproduces the total intensity of the resolved ring without weighting different passbands. \cite{Monnier2019} needed to multiply the $H$-band model by 2 to match the GPI result.
We note that the model adopts a simple ellipse to approximately reproduce the ring geometry but the actual ring has more complex features such as the darkening features, the discontinuity at PA$\sim0^\circ$, and the spirals.
Compared with the GPI total intensity \citep{Monnier2019} our azimuthal profiles are different in both $J$ and $H$ bands. This difference is mainly due to the difference of data reduction: \cite{Monnier2019} made an approximate reference PSF by assuming a Moffat function and subtracted it from the total intensity (star+disk) image to extract the disk total intensity, while we used the practical star (HR 2466) for a reference PSF and conducted the RDI reduction. As HD 34700 A is a binary the light source onto the disk surface is variable, which can also vary the scattering profile. 

We used photometric results of 2MASS \citep{Cutri2003-2MASS} and deprojected separations (assuming the disk inclination of 40.9$^\circ$ - see Table \ref{mcmaxdisk}) of the traced ring to convert surface brightness into the geometric albedo.
The difference of the geometric albedo profiles between the resolved disk and the modeled disk looks larger than the case of the surface brightness profile because the conversion includes geometric difference between both of the disks (see Appendix \ref{sec: Difference between separations of the traced rings}).
A remarkable feature in the albedo plot is a color tendency at PA between $\sim45^\circ$ and 90$^\circ$. The model-based geometric albedo is comparable at $JH$ band, while the actual $J$-band geometric albedo has a higher value than that in $H$-band.
Such Reyleigh-scattering-like feature appears at higher scale height or where sub-micron dust is prominent and our result suggests either or both of these possibilities at this area.

\subsection{Origin of the Spirals} \label{sec: Origin of the Spirals}

Spiral S1, particularly S1b, appears to be more tightly wound than the other spirals observed in the disk, with measured pitch angles of $\sim35^\circ$-$47^\circ$ and $\sim27^\circ$ for the inner (S1a) and outer (S1b) sections (see Section \ref{sec: Spiral Characterization}), respectively.
In a thin-disk case S1ab can be formed by the same origin and it is also interesting that S1 is the only spiral that is directly crossed by a shadow. Shadowing has been suggested as a possible mechanism to form spiral arms, due to the periodic temperature and hence pressure kick imprinted on material rotating in the disk \citep{Montesinos2016,Montesinos2018}. The morphology of S1a and S1b is roughly compatible with the inner and outer wake of a spiral launched from a shadow, with a larger (resp. smaller) pitch angle for the inner (resp. outer) wake.
In a thick-disk case S1ab might be disconnected and formed via different mechanisms. Future observation may help to investigate whether these spirals are physically connected or not.

The radially-extended feature of S1 is also compatible with being launched by a companion. In that case, it is unlikely to be caused by a yet undetected companion in the cavity, as outer wakes are expected to be tightly wound \citep[$\phi \leq 10^\circ$;][]{Bae2018}.
The large pitch angle of both S1a and S1b would suggest they correspond to an inner spiral wake (with respect to the companion). This could be either a yet undetected protoplanet in the outer disk \citep[e.g.][]{Dong2015,Zhu2015}, or the known K-dwarf outer companion HD~34700~B.

From their flocculent appearance, spirals S2--S6 ($\phi\sim50^\circ$) may resemble those seen in numerical simulations of gravitationally unstable protoplanetary disks \citep[e.g.][]{Rice2003a}. However, \citet{Monnier2019} estimated the Toomre parameter values larger than 25 everywhere in the disk based on their radiative transfer model, which makes this possibility unlikely.

\subsubsection{Stellar Flyby}

Considering the respective proper motion of HD 34700 AB, an interesting possibility is that of a recent flyby.
Hydrodynamical simulations show stellar flybys can induce spirals with a large pitch angle \citep[e.g.][]{Cuello2019,Cuello2020}. In practice arms/spirals in the RW Aur A and UX Tau A disks can be well reproduced by the stellar flyby \citep{Dai2015, Rodriguez2018, Menard2020}.
We checked RA, Dec, and proper motions for HD 34700 A and B \citep{GaiaDR2} and calculated their projected separation over the past. The separation ($r$) between HD 34700 AB is expressed as follows:
\begin{eqnarray*}
r = \sqrt{(\Delta {\rm RA_0}-\Delta pm_{\rm RA}\times t)^2+(\Delta {\rm Dec_0}-\Delta pm_{\rm Dec}\times t)^2},
\end{eqnarray*}
where $\Delta {\rm RA_0}$ and $\Delta {\rm Dec_0}$ are differences of RA and Dec in Gaia DR2, $\Delta pm_{\rm RA}$ and $\Delta pm_{\rm Dec}$ are those of proper motions along RA and Dec, and $t$ corresponds to time [year]. 
We also checked the projected separation with HD 34700 C, which is located at $\sim9\farcs2$ from HD 34700 A.
Current astrometric databases such as Gaia do not have a record of the proper motion of C and we approximately defined the proper motion as difference between Gaia DR2 and \cite{Sterzik2005}, which may include systematic uncertainty of astrometry. The estimation of C's proper motion requires the coordinate of A on 2004 January 30, when the observation of the HD 34700 system was operated by \cite{Sterzik2005}. We used astropy.coordinates libraries to extrapolate the position on this date from Gaia DR2 coordinate and proper motion.
\cite{Monnier2019} suggested another companion candidate HD 34700 D but a proper motion test with HST/STIS coronagraphic data taken in 2018 (PI: Marie Ygouf) revealed that this object is not comoving (Ygouf et al. in prep). Therefore we do not investigate the stellar flyby scenario with D.
We note that these estimations of the separations do not take orbital motions and star-companion interactions into account. Future studies with more inputs of the positions will help to infer their orbits and to discuss the stellar flyby scenario in detail.

\begin{figure*}
\begin{tabular}{cc}
\begin{minipage}{0.5\hsize}
    \centering
    \includegraphics[width=\textwidth]{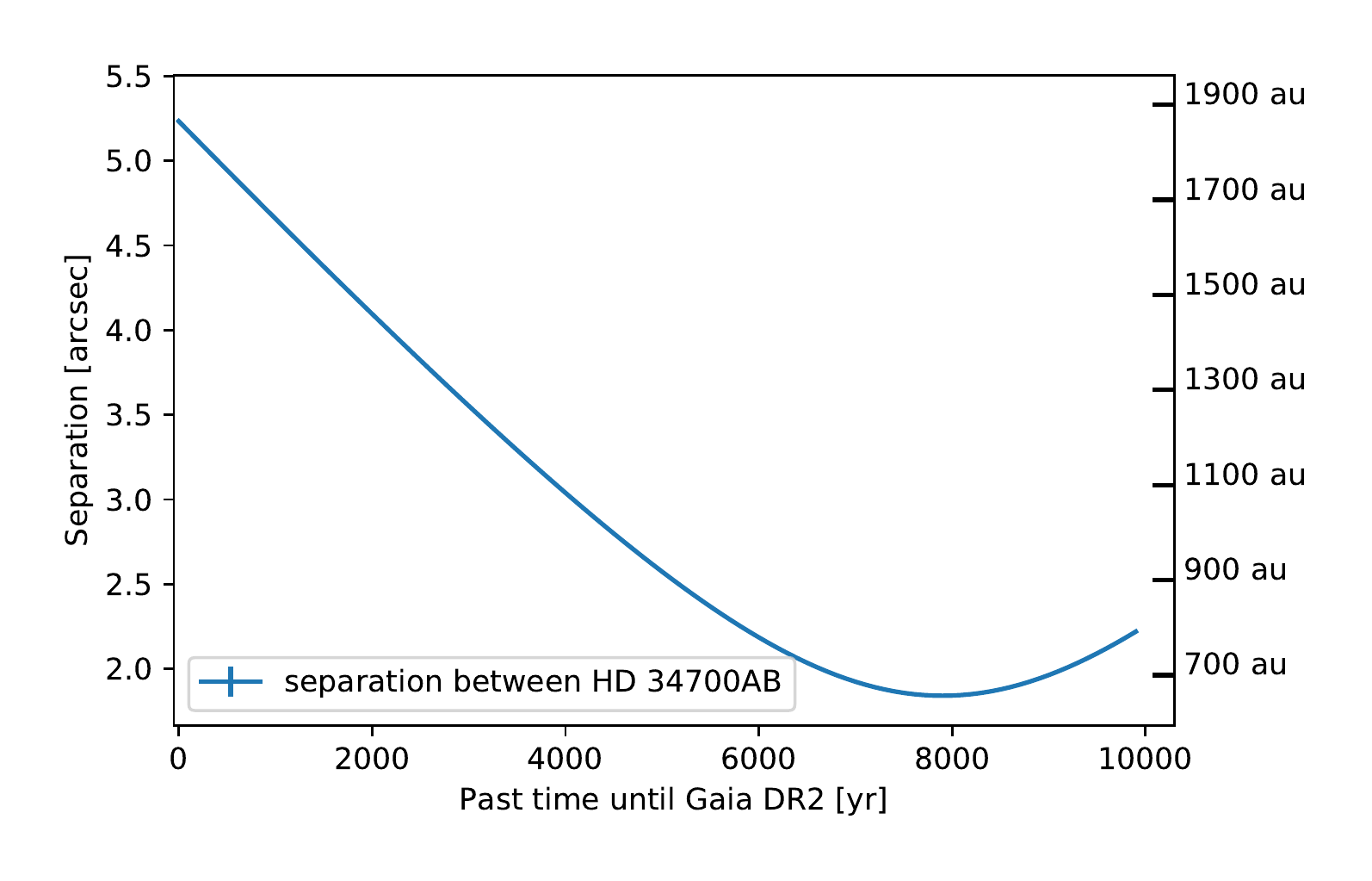}
\end{minipage}
\begin{minipage}{0.5\hsize}
    \centering
    \includegraphics[width=\textwidth]{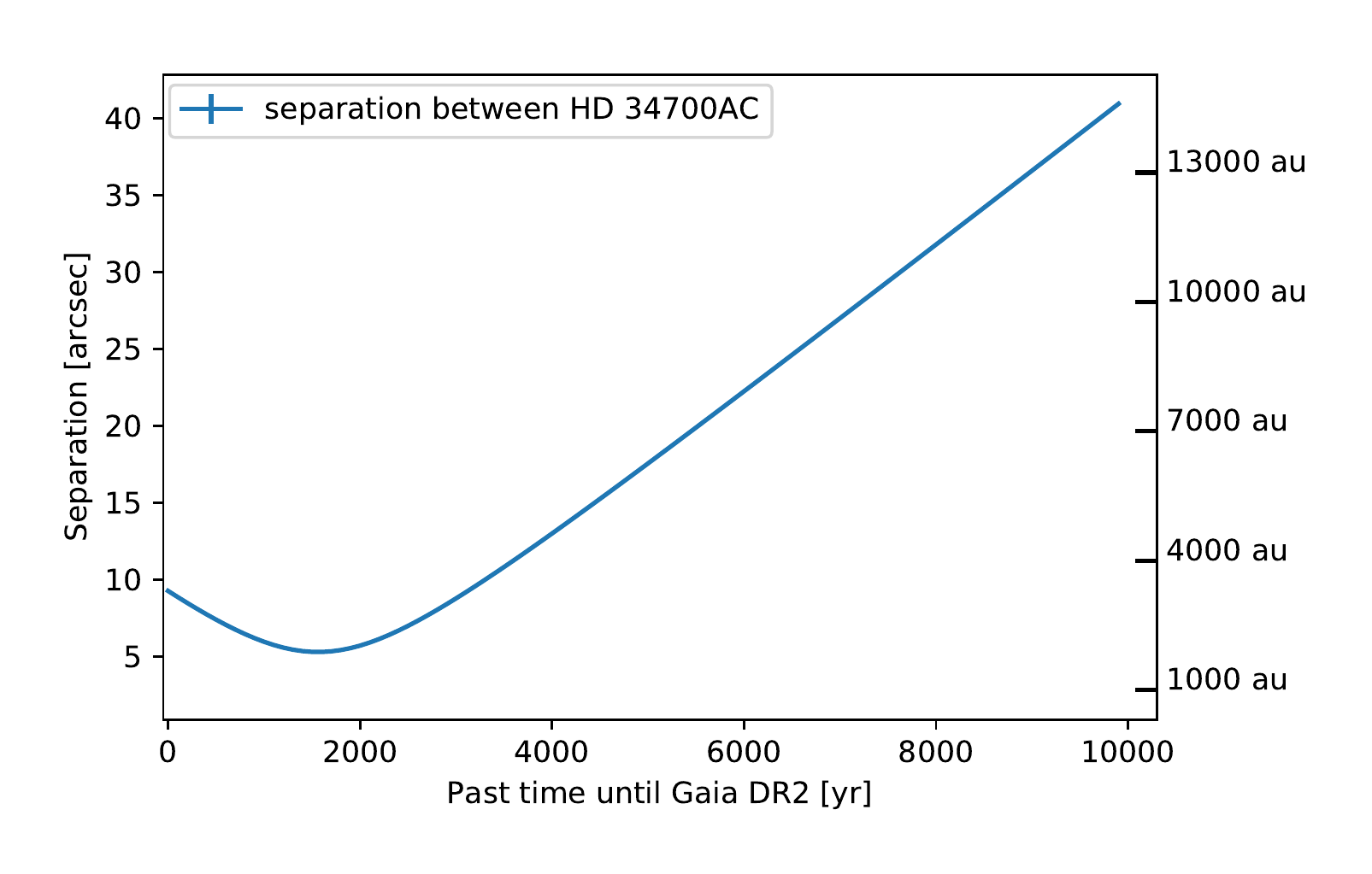}
\end{minipage}
\end{tabular}
\caption{Projected separation between HD 34700 AB (left) and AC (right). Error bars include only measurement errors of RA and Dec (see also Figure \ref{fig: separation w/ errors HD34700AB}).}
\label{fig: separation HD34700AB}
\end{figure*}

Figure \ref{fig: separation HD34700AB} illustrates projected separations between HD 34700 AB (left) and AC (right). The left separation curve suggests a possibility that HD 34700 B could be located $\sim$ 700 au away from HD 34700 A. For the case of C the separation is greater than 1000 au and C may be less responsible for inducing the spirals than B. The larger relative proper motion of C than that of B may be affected by the systematic uncertainty between Gaia DR2 and \cite{Sterzik2005}.
We assume the same distance and do not take into account of line-of-sight motion in these plots because Gaia-based distances are $356.5^{+6.3}_{-6.0}$ pc and $353^{+10}_{-12}$ pc for A and B respectively and are consistent with each other within errors. 
The parallax of C has not been measured and we adopted the same assumption about the distance. 
We note that errors of the separation increase as time increases (see Appendix \ref{sec: Errors of separation}) if we include Gaia measurement errors of the proper motion, and that the error bars in Figure \ref{fig: separation HD34700AB} include only measurement errors of RA and Dec.
\cite{Cuello2019} showed that spirals induced by a stellar flyby can survive for more than 7000 years under some conditions and stellar flyby is perhaps a reasonable scenario if HD 34700 B passed by HD 34700 A $\sim$8000 years ago.
As we have large uncertainties of periastron and we do not set any constraints on an angle between the disk plane and the perturber plane (HD 34700 B's orbit) we do not further simulate the disk feature with the stellar flyby scenario in this study.

\subsubsection{Infall}
An alternative possibility for the origin of the flocculent spiral pattern is infall from a late envelope or a captured cloudlet \citep[e.g.][]{Tang2012, Dullemond2019}. A late-envelope infall was proposed to account for the similar spiral pattern observed in the disk of AB Aur \citep{Fukagawa2004,Tang2012,Tang2017}. Large scale images of the environment of AB Aur show the presence of a large surrounding cloudlet, which led \citet{Dullemond2019} to propose that transitional disks like AB Aur could all be the result of cloudlet capture. In that case, the spirals might be seen in a different plane than that of the inner rim of the outer disk, i.e. the outer disk would be warped, as e.g.~HD~100546 \citep[e.g.][]{Quillen2006a}. This would explain the very large deprojected pitch angle values. 
We note that previous studies and our observation have not yet detected any envelope-like features.
\cite{Monnier2019} implemented SED fitting of HD 34700 A and indicated Av=0. They also presented the large FoV image of HST/NICMOS ($\sim18\farcs9\times18\farcs9$) where one half of its vicinity was explored and there is no significant signal of envelope.
The HST/STIS data cover the whole vicinity (within a radius of $\sim$10$^{\prime\prime}$) and confirmed faint halo extending outside the CHARIS FoV \citep[$\sim$$2-3^{\prime\prime}$ in radius;][and Ygouf et al. in prep]{Ygouf2019}.
We attempted to fit the traced peaks of the spirals with infall but could not set robust constraints on spiral parameters with the infall scenario because of large uncertainties (see Appendix \ref{sec: Fitting of spirals}).
CO rotational line observations with ALMA may help to investigate the kinematics of the outer disk, including the spirals.

Apart from AB~Aur, HD~34700~A also shows a similar spiral pattern to the circumbinary disk HD 142527 \citep[e.g.][]{Fukagawa2006,Christiaens2014,Avenhaus2014}. Both systems harbor a prominent spiral combined with multiple smaller flocculent spiral arms stemming from the edge of the cavity. The hydro-dynamical simulations in \citet{Price2018} suggest that the dynamical interaction between the inner binary and the outer disk can account for the flocculent spiral arms in HD~142527. The prominent spiral might correspond to a secular large-scale spiral density wave \citep[e.g.][]{Demidova2015}. However the separation between the inner binary of HD~34700~Aa and Ab is significantly smaller than HD~142527~AB (0.69 AU versus 25--50 AU) for a similar size cavity ($\sim$ 175 AU versus $\sim$130 AU), so it is unclear whether the inner binary could reproduce all of the spirals of HD~34700~A. Dedicated hydrodynamical simulations are required to pinpoint the origin(s) of the spirals of HD 34700 A.

\section{Summary} \label{sec: Summary}

We have presented Subaru/SCExAO+CHARIS broadband ($JHK$ band) integral field spectroscopy of the HD 34700 A protoplanetary disk.
The observation was conducted under such a good seeing condition that a single frame could resolve the ring without any post-processing.
We then conducted RDI and ADI+SDI reductions to obtain its morphology and to estimate the surface brightness accurately, which resulted in clear detection of both the ring and multiple spirals as shown in \cite{Monnier2019}. Although \cite{Monnier2019} suggested a 50 $M_{\rm Jup}$ companion embedded in the disk, we did not detect any companion candidates.
We calculated contrast limits from the ADI+SDI result and the broadband contrast curve sets a constraint on potential substellar-mass objects down to $\sim12\ M_{\rm Jup}$ at $0\farcs3$ (in the gap) and $\sim5\ M_{\rm Jup}$ at $0\farcs75$ (outside the ring) assuming COND03 model and 5 Myr. We also tested the 50 $M_{\rm Jup}$ companion scenario by injecting a fake source and concluded that our observation could set a robust constraint on this hypothesis.

We used the MCMax3D radiative transfer code to reproduce the ring scattering profile.
By checking the reduced images and comparing surface brightness with the forward-modeled disk we newly confirmed darkening effects on the ring and spiral, large parts of which appear to be shadows cast by possible inner object(s). 
Except at these darkening features our best-fit model provides a better match to the actual surface brightness among $JHK$ bands than \cite{Monnier2019} that showed some discrepancy between their model and surface brightness.
However, part of these features are located by the roots of the spirals and we do not rule out other possibilities such as physical features related to the outer spirals.
Geometric albedo converted from the surface brightness of the ring suggests a higher scale height and/or prominently abundant sub-micron dust at position angles between $\sim45^\circ$ and 90$^\circ$.

We also conducted spiral fitting of S1-S6 and the result suggests very large pitch angles ($\sim30-50^\circ$) that are larger than the estimated pitch angles presented in \cite{Monnier2019}.
A stellar flyby of HD 34700 B or infall from surrounding envelope is perhaps a reasonable scenario to explain the large pitch angles.
We investigated the separation between HD 34700 AB based on Gaia-based coordinates and proper motions and HD 34700 B could be located $\sim$700 au away from HD 34700 A about 8000 years ago. 
Future CO observations with ALMA may investigate the kinematics of the outer disk, including the spirals.   

\acknowledgements
The authors would like to thank the anonymous referees for their constructive comments and suggestions to improve the quality of the paper. We thank John Monnier for authorization to use GPI-PDI images originally presented in \citet{Monnier2019}. The authors are grateful to Gijs Mulders for the helpful comments.
This research is based on data collected at the Subaru Telescope, which is operated by the National Astronomical Observatories of Japan.
This research has made use of NASA's Astrophysics Data System Bibliographic Services.
This research has made use of the SIMBAD database, operated at CDS, Strasbourg, France.
This research made use of Astropy, a community-developed core Python package for Astronomy \citep{astropy:2013, astropy:2018}

TU acknowledges JSPS overseas research fellowship.
TC is funded by a NASA Senior Postdoctoral Fellowship.
JW acknowledges funding support from the NASA XRP program via grants 80NSSC20K0252 and NNX17AF88G.
ST is supported by JSPS KAKENHI Grant-in-Aid for Early-Career Scientists No. 19K14764. 
MT is supported by MEXT/JSPS KAKENHI grant Nos. 18H05442, 15H02063,
and 22000005.
EA is supported by MEXT/JSPS KAKENHI grant No. 17K05399.
The development of SCExAO was supported by JSPS (Grant-in-Aid for Research \#23340051, \#26220704 \& \#23103002), Astrobiology Center of NINS, Japan, the Mt Cuba Foundation, and the director's contingency fund at Subaru Telescope.
CHARIS was developed under the support by the Grant-in-Aid for Scientific Research on Innovative Areas \#2302.

The authors wish to acknowledge the very significant cultural role and reverence that the summit of Mauna Kea has always had within the indigenous Hawaiian community. We are most fortunate to have the opportunity to conduct observations from this mountain.

\newpage
\appendix
\section{Difference between separations of the traced rings} \label{sec: Difference between separations of the traced rings}

\begin{figure}[h]
    \centering
    \includegraphics[width=0.65\textwidth]{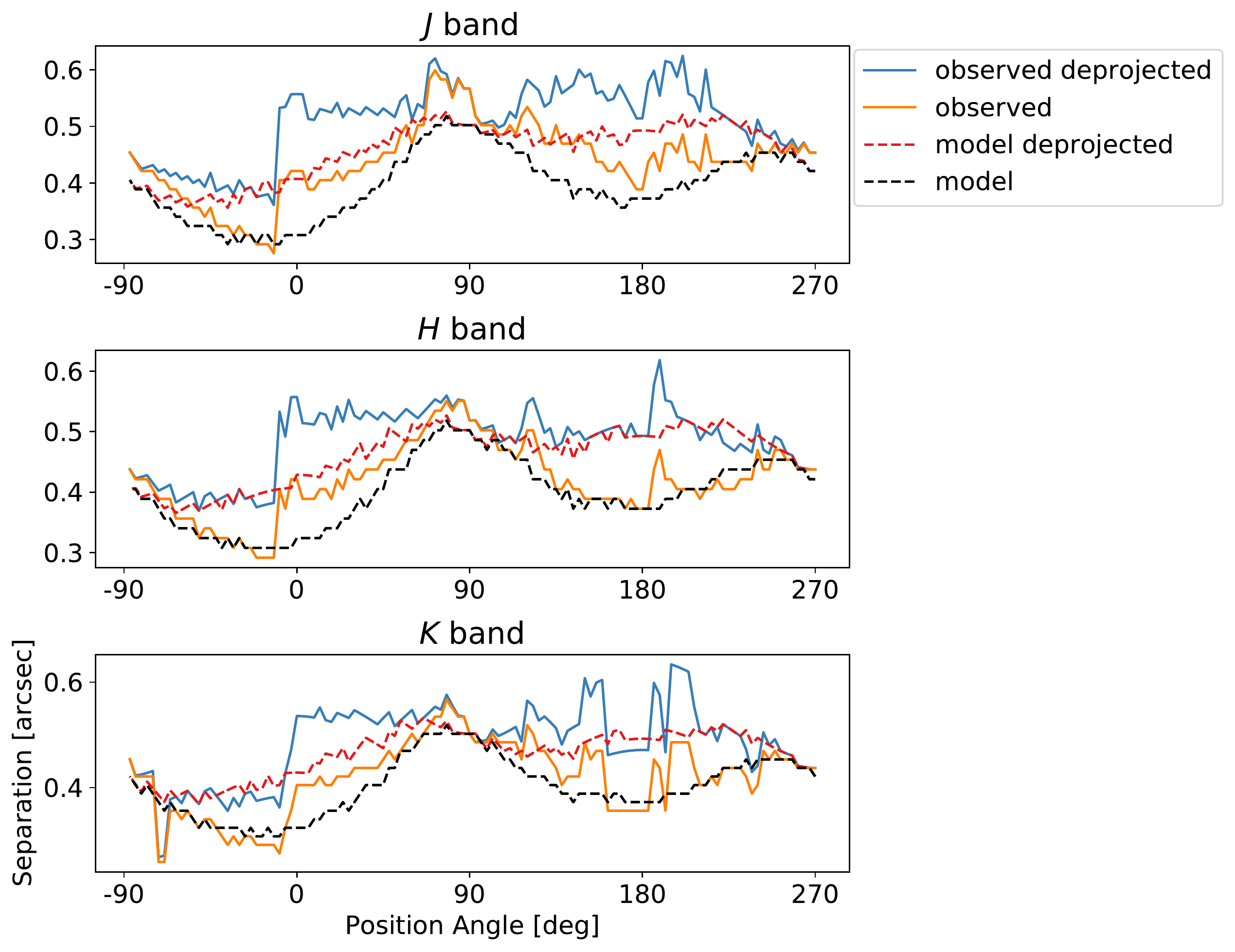}
    \caption{Comparison of peak loci between the resolved ring (`observed' - solid line) and the modeled ring (`model' - dashed line), overlaid with both of deprojected separations (`deprojected'), at $J$, $H$, and $K$ bands, respectively. }
    \label{fig: separation ring}
\end{figure}

Figure \ref{fig: separation ring} shows peak loci of the resolved ring and the modeled ring in each band. The difference of disk geometry affects the conversion from surface brightness into geometric albedo in Section \ref{sec: Scattering Profiles}.

\section{Errors of separation between HD 34700 AB and AC} \label{sec: Errors of separation}
\begin{figure*}
\begin{tabular}{cc}
\begin{minipage}{0.5\hsize}
    \centering
    \includegraphics[width=\textwidth]{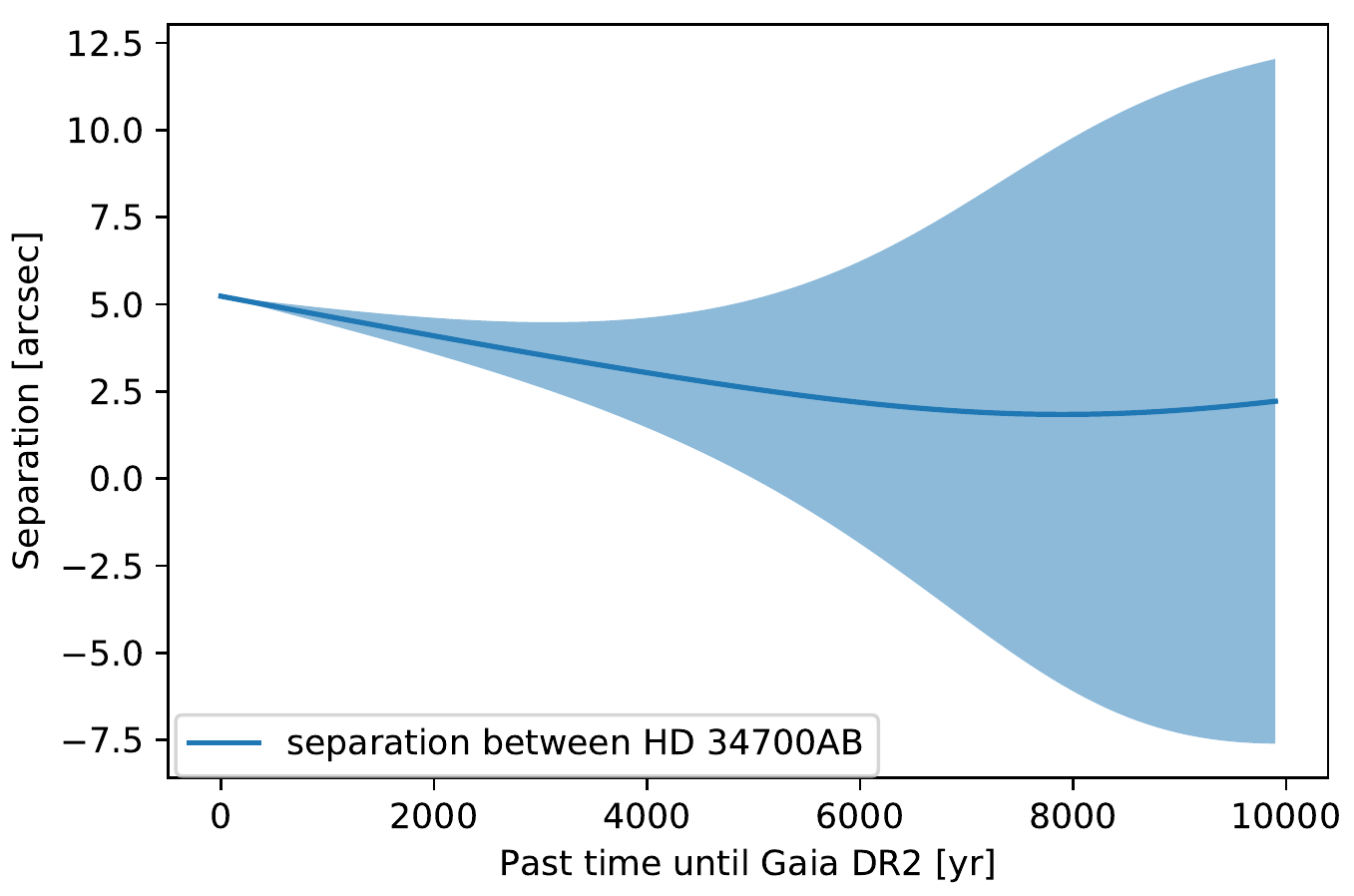}
\end{minipage}
\begin{minipage}{0.5\hsize}
    \centering
    \includegraphics[width=\textwidth]{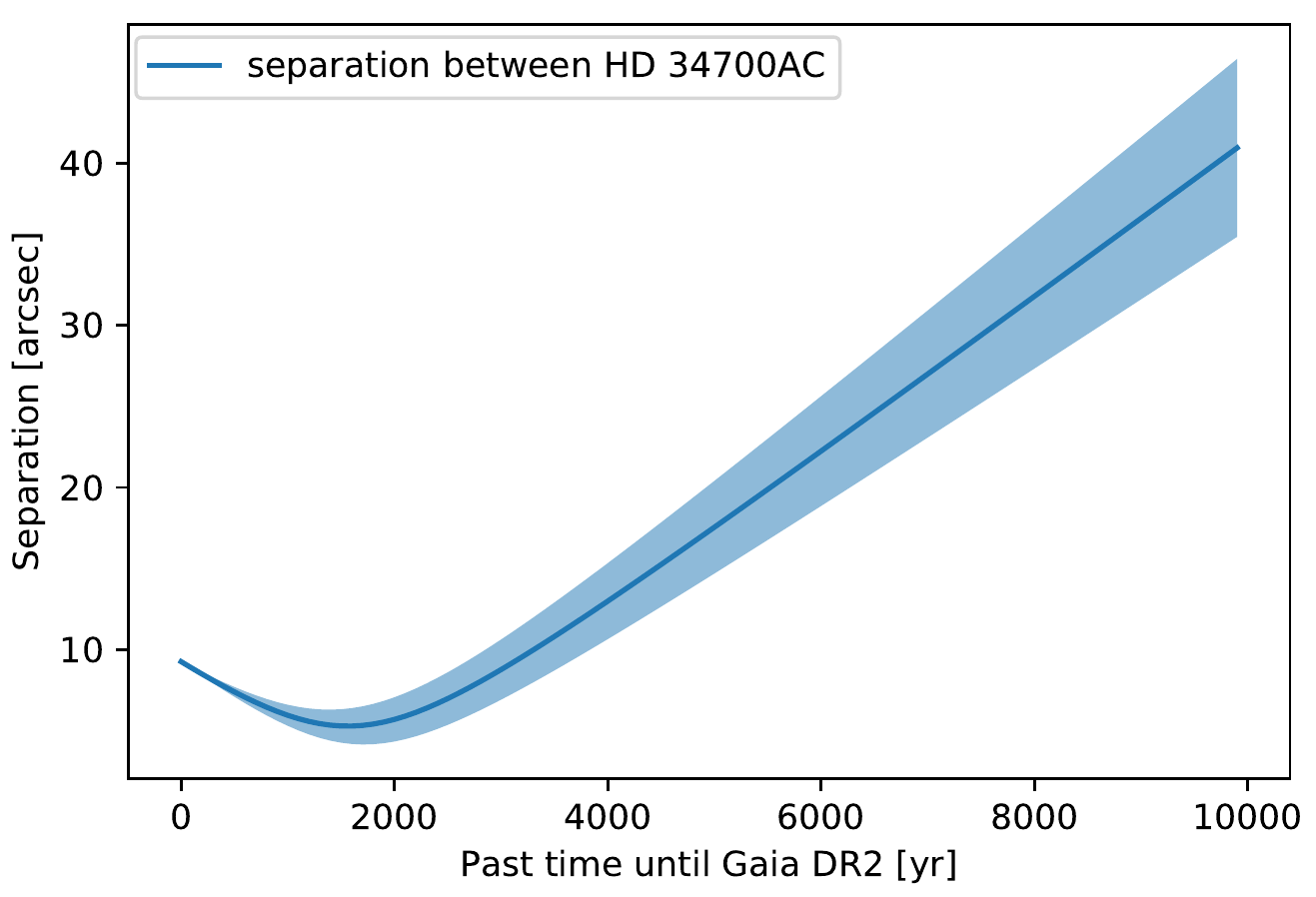}
\end{minipage}
\end{tabular}
\caption{As Figure \ref{fig: separation HD34700AB} with both measurement errors of the coordinates and the proper motions. The solid line corresponds to the separation without errors and the shaded area corresponds to the errors. }
\label{fig: separation w/ errors HD34700AB}
\end{figure*}

The error of separation is estimated according to the law of propagation
\begin{eqnarray*}
\sigma_{r} && = \sqrt{(\frac{\partial r}{\partial {\rm \Delta RA_0}}\sigma_{\Delta RA_0})^2
+(\frac{\partial r}{\partial {\Delta pm_{\rm RA}}}\sigma_{\Delta pm_{\rm RA}})^2
+(\frac{\partial r}{\partial {\rm \Delta Dec_0}}\sigma_{\Delta Dec_0})^2
+(\frac{\partial r}{\partial {\Delta pm_{\rm Dec}}}\sigma_{\Delta pm_{\rm Dec}})^2 }, \\
\end{eqnarray*}
where $\sigma_{\Delta RA_0}$ and other error parameters above are defined as sum of squares of the Gaia DR2 measurement errors. In particular, coefficients of the proper motion errors ($\frac{\partial r}{\partial {\Delta pm_{\rm RA}}}$ and $\frac{\partial r}{\partial {\Delta pm_{\rm Dec}}}$) are expressed as $\frac{\partial r}{\partial \Delta pm_{\rm C}}=-\frac{t(\Delta {\rm C_0}-\Delta pm_{\rm C}\times t)}{r}$, where C is RA or Dec, and have an order of $t$. Therefore the errors of the separation increase according to $t$ if we include the measurement errors of the proper motions (see Figure \ref{fig: separation w/ errors HD34700AB} for the plot with error bars including the proper motion errors).

\section{Fitting of spirals by gas infall model} 
\label{sec: Fitting of spirals}
Infall motion of the envelope gas is written by the parabolic orbit \cite[cf.][]{1981Icar...48..353C}, which is given by
\begin{eqnarray*}
 r'=\frac{a}{1-\cos(\theta'-b) },
\end{eqnarray*}
in the coordinate of the orbital plane ($r',\ \theta'$), where $a$ and $b$ are parameters characterizing the orbit.
The inclination $i$ and the position angle $\Omega$ of the orbital plane are also parameters of the orbit.
We fit the observed spirals by the parabolic orbit by assuming 1) spirals are located foreground and 2) spirals can extends inward the ring and they may not be detected in the CHARIS image.
We summarize the best fit parameters and the standard errors in Table \ref{tab:param}. The errors depend on the traced peaks and the ADI+SDI reduction.

\begin{table}
\begin{center}
    \caption{Best fit parameters and errors obtained from the fitting of gas infall model}
    \label{tab:param}
    \begin{tabular}{cccccccc}
    Parameters  & S1a & S1b &S2 &S3 &S4 &S5 &S6\\
    \hline
    \hline
    Best Fit &&&&&&&\\
    \hline
    $a$ [mas]      & 45.7 & 706 & 290 & 677 & 1.27$\times10^3$ & 62.0 & 876 \\
    $b$ [rad]      & -0.0472 & -0.584 & -0.195 & -0.588 & -0.737 & -0.0997 & -0.727 \\
    $i$ [rad]      & -1.54 & -1.01 & -1.45 & -1.15 & -1.26 & -1.46 & -1.11 \\
    $\Omega$ [rad] & -1.43 & -2.15 & -1.54 & -0.782 & 0.331 & 1.95 & 1.70 \\
    \hline
    Standard Error &&&&&&&\\
    \hline
    $a$ [mas]      & 2.85$\times10^3$ & 1.34$\times10^3$ & 3.26$\times10^3$ & 1.04$\times10^3$ & 313 & 8.78$\times10^3$ & 140 \\
    $b$ [rad]      & 2.96 & 0.732 & 2.17 & 0.764 & 0.222 & 13.9 & 0.0970 \\
    $i$ [rad]      & 1.78 & 1.18 & 1.36 & 0.624 & 0.0404 & 16.2 & 0.0272 \\
    $\Omega$ [rad] & 0.0939 & 0.998 & 0.378 & 0.627 & 0.103 & 1.81 & 0.119 \\
    \hline  
    \end{tabular}
\end{center}
\end{table}

\bibliographystyle{aasjournal}                                                              
\bibliography{library}                                                                

\begin{thebibliography}{}
\expandafter\ifx\csname natexlab\endcsname\relax\def\natexlab#1{#1}\fi
\providecommand{\url}[1]{\href{#1}{#1}}
\providecommand{\dodoi}[1]{doi:~\href{http://doi.org/#1}{\nolinkurl{#1}}}
\providecommand{\doeprint}[1]{\href{http://ascl.net/#1}{\nolinkurl{http://ascl.net/#1}}}
\providecommand{\doarXiv}[1]{\href{https://arxiv.org/abs/#1}{\nolinkurl{https://arxiv.org/abs/#1}}}

\bibitem[{{Andrews} {et~al.}(2018){Andrews}, {Huang}, {P{\'e}rez}, {Isella},
  {Dullemond}, {Kurtovic}, {Guzm{\'a}n}, {Carpenter}, {Wilner}, {Zhang}, {Zhu},
  {Birnstiel}, {Bai}, {Benisty}, {Hughes}, {{\"O}berg}, \&
  {Ricci}}]{Andrews2018}
{Andrews}, S.~M., {Huang}, J., {P{\'e}rez}, L.~M., {et~al.} 2018, \apjl, 869,
  L41, \dodoi{10.3847/2041-8213/aaf741}

\bibitem[{{Astropy Collaboration} {et~al.}(2013){Astropy Collaboration},
  {Robitaille}, {Tollerud}, {Greenfield}, {Droettboom}, {Bray}, {Aldcroft},
  {Davis}, {Ginsburg}, {Price-Whelan}, {Kerzendorf}, {Conley}, {Crighton},
  {Barbary}, {Muna}, {Ferguson}, {Grollier}, {Parikh}, {Nair}, {Unther},
  {Deil}, {Woillez}, {Conseil}, {Kramer}, {Turner}, {Singer}, {Fox}, {Weaver},
  {Zabalza}, {Edwards}, {Azalee Bostroem}, {Burke}, {Casey}, {Crawford},
  {Dencheva}, {Ely}, {Jenness}, {Labrie}, {Lim}, {Pierfederici}, {Pontzen},
  {Ptak}, {Refsdal}, {Servillat}, \& {Streicher}}]{astropy:2013}
{Astropy Collaboration}, {Robitaille}, T.~P., {Tollerud}, E.~J., {et~al.} 2013,
  \aap, 558, A33, \dodoi{10.1051/0004-6361/201322068}

\bibitem[{{Astropy Collaboration} {et~al.}(2018){Astropy Collaboration},
  {Price-Whelan}, {Sip{\H{o}}cz}, {G{\"u}nther}, {Lim}, {Crawford}, {Conseil},
  {Shupe}, {Craig}, {Dencheva}, {Ginsburg}, {Vand erPlas}, {Bradley},
  {P{\'e}rez-Su{\'a}rez}, {de Val-Borro}, {Aldcroft}, {Cruz}, {Robitaille},
  {Tollerud}, {Ardelean}, {Babej}, {Bach}, {Bachetti}, {Bakanov}, {Bamford},
  {Barentsen}, {Barmby}, {Baumbach}, {Berry}, {Biscani}, {Boquien}, {Bostroem},
  {Bouma}, {Brammer}, {Bray}, {Breytenbach}, {Buddelmeijer}, {Burke},
  {Calderone}, {Cano Rodr{\'\i}guez}, {Cara}, {Cardoso}, {Cheedella}, {Copin},
  {Corrales}, {Crichton}, {D'Avella}, {Deil}, {Depagne}, {Dietrich}, {Donath},
  {Droettboom}, {Earl}, {Erben}, {Fabbro}, {Ferreira}, {Finethy}, {Fox},
  {Garrison}, {Gibbons}, {Goldstein}, {Gommers}, {Greco}, {Greenfield},
  {Groener}, {Grollier}, {Hagen}, {Hirst}, {Homeier}, {Horton}, {Hosseinzadeh},
  {Hu}, {Hunkeler}, {Ivezi{\'c}}, {Jain}, {Jenness}, {Kanarek}, {Kendrew},
  {Kern}, {Kerzendorf}, {Khvalko}, {King}, {Kirkby}, {Kulkarni}, {Kumar},
  {Lee}, {Lenz}, {Littlefair}, {Ma}, {Macleod}, {Mastropietro}, {McCully},
  {Montagnac}, {Morris}, {Mueller}, {Mumford}, {Muna}, {Murphy}, {Nelson},
  {Nguyen}, {Ninan}, {N{\"o}the}, {Ogaz}, {Oh}, {Parejko}, {Parley}, {Pascual},
  {Patil}, {Patil}, {Plunkett}, {Prochaska}, {Rastogi}, {Reddy Janga},
  {Sabater}, {Sakurikar}, {Seifert}, {Sherbert}, {Sherwood-Taylor}, {Shih},
  {Sick}, {Silbiger}, {Singanamalla}, {Singer}, {Sladen}, {Sooley},
  {Sornarajah}, {Streicher}, {Teuben}, {Thomas}, {Tremblay}, {Turner},
  {Terr{\'o}n}, {van Kerkwijk}, {de la Vega}, {Watkins}, {Weaver}, {Whitmore},
  {Woillez}, {Zabalza}, \& {Astropy Contributors}}]{astropy:2018}
{Astropy Collaboration}, {Price-Whelan}, A.~M., {Sip{\H{o}}cz}, B.~M., {et~al.}
  2018, \aj, 156, 123, \dodoi{10.3847/1538-3881/aabc4f}

\bibitem[{{Avenhaus} {et~al.}(2014){Avenhaus}, {Quanz}, {Schmid}, {Meyer},
  {Garufi}, {Wolf}, \& {Dominik}}]{Avenhaus2014}
{Avenhaus}, H., {Quanz}, S.~P., {Schmid}, H.~M., {et~al.} 2014, \apj, 781, 87,
  \dodoi{10.1088/0004-637X/781/2/87}

\bibitem[{{Avenhaus} {et~al.}(2018){Avenhaus}, {Quanz}, {Garufi}, {Perez},
  {Casassus}, {Pinte}, {Bertrang}, {Caceres}, {Benisty}, \&
  {Dominik}}]{Avenhaus2018}
{Avenhaus}, H., {Quanz}, S.~P., {Garufi}, A., {et~al.} 2018, \apj, 863, 44,
  \dodoi{10.3847/1538-4357/aab846}

\bibitem[{{Bae} \& {Zhu}(2018)}]{Bae2018}
{Bae}, J., \& {Zhu}, Z. 2018, \apj, 859, 119, \dodoi{10.3847/1538-4357/aabf93}

\bibitem[{{Baraffe} {et~al.}(2003){Baraffe}, {Chabrier}, {Barman}, {Allard}, \&
  {Hauschildt}}]{Baraffe2003-COND03}
{Baraffe}, I., {Chabrier}, G., {Barman}, T.~S., {Allard}, F., \& {Hauschildt},
  P.~H. 2003, \aap, 402, 701, \dodoi{10.1051/0004-6361:20030252}

\bibitem[{{Benisty} {et~al.}(2017){Benisty}, {Stolker}, {Pohl}, {de Boer},
  {Lesur}, {Dominik}, {Dullemond}, {Langlois}, {Min}, {Wagner}, {Henning},
  {Juhasz}, {Pinilla}, {Facchini}, {Apai}, {van Boekel}, {Garufi}, {Ginski},
  {M{\'e}nard}, {Pinte}, {Quanz}, {Zurlo}, {Boccaletti}, {Bonnefoy}, {Beuzit},
  {Chauvin}, {Cudel}, {Desidera}, {Feldt}, {Fontanive}, {Gratton}, {Kasper},
  {Lagrange}, {LeCoroller}, {Mouillet}, {Mesa}, {Sissa}, {Vigan}, {Antichi},
  {Buey}, {Fusco}, {Gisler}, {Llored}, {Magnard}, {Moeller-Nilsson}, {Pragt},
  {Roelfsema}, {Sauvage}, \& {Wildi}}]{Benisty2017}
{Benisty}, M., {Stolker}, T., {Pohl}, A., {et~al.} 2017, \aap, 597, A42,
  \dodoi{10.1051/0004-6361/201629798}

\bibitem[{{Bertrang} {et~al.}(2018){Bertrang}, {Avenhaus}, {Casassus},
  {Montesinos}, {Kirchschlager}, {Perez}, {Cieza}, \& {Wolf}}]{Bertrang2018}
{Bertrang}, G.~H.~M., {Avenhaus}, H., {Casassus}, S., {et~al.} 2018, \mnras,
  474, 5105, \dodoi{10.1093/mnras/stx3052}

\bibitem[{{Bonnefoy} {et~al.}(2014){Bonnefoy}, {Chauvin}, {Lagrange}, {Rojo},
  {Allard}, {Pinte}, {Dumas}, \& {Homeier}}]{Bonnefoy2014}
{Bonnefoy}, M., {Chauvin}, G., {Lagrange}, A.~M., {et~al.} 2014, \aap, 562,
  A127, \dodoi{10.1051/0004-6361/201118270}

\bibitem[{{Bowler}(2016)}]{Bowler2016}
{Bowler}, B.~P. 2016, \pasp, 128, 102001,
  \dodoi{10.1088/1538-3873/128/968/102001}

\bibitem[{{Brandt} {et~al.}(2017){Brandt}, {Rizzo}, {Groff}, {Chilcote},
  {Greco}, {Kasdin}, {Limbach}, {Galvin}, {Loomis}, {Knapp}, {McElwain},
  {Jovanovic}, {Currie}, {Mede}, {Tamura}, {Takato}, \& {Hayashi}}]{Brandt2017}
{Brandt}, T.~D., {Rizzo}, M., {Groff}, T., {et~al.} 2017, Journal of
  Astronomical Telescopes, Instruments, and Systems, 3, 048002,
  \dodoi{10.1117/1.JATIS.3.4.048002}

\bibitem[{{Brauer} {et~al.}(2019){Brauer}, {Pantin}, {Di Folco}, {Habart},
  {Dutrey}, \& {Guilloteau}}]{Baruer2019}
{Brauer}, R., {Pantin}, E., {Di Folco}, E., {et~al.} 2019, \aap, 628, A88,
  \dodoi{10.1051/0004-6361/201935966}

\bibitem[{{Canovas} {et~al.}(2017){Canovas}, {Hardy}, {Zurlo}, {Wahhaj},
  {Schreiber}, {Vigan}, {Villaver}, {Olofsson}, {Meeus}, {M{\'e}nard},
  {Caceres}, {Cieza}, \& {Garufi}}]{Canovas2017}
{Canovas}, H., {Hardy}, A., {Zurlo}, A., {et~al.} 2017, \aap, 598, A43,
  \dodoi{10.1051/0004-6361/201629145}

\bibitem[{{Casassus} {et~al.}(2018){Casassus}, {Avenhaus}, {P{\'e}rez},
  {Navarro}, {C{\'a}rcamo}, {Marino}, {Cieza}, {Quanz}, {Alarc{\'o}n}, {Zurlo},
  {Osses}, {Rannou}, {Rom{\'a}n}, \& {Barraza}}]{Casassus2018}
{Casassus}, S., {Avenhaus}, H., {P{\'e}rez}, S., {et~al.} 2018, \mnras, 477,
  5104, \dodoi{10.1093/mnras/sty894}

\bibitem[{{Cassen} \& {Moosman}(1981)}]{1981Icar...48..353C}
{Cassen}, P., \& {Moosman}, A. 1981, \icarus, 48, 353,
  \dodoi{10.1016/0019-1035(81)90051-8}

\bibitem[{{Castelli} \& {Kurucz}(2003)}]{Castelli2003}
{Castelli}, F., \& {Kurucz}, R.~L. 2003, in IAU Symposium, Vol. 210, Modelling
  of Stellar Atmospheres, ed. N.~{Piskunov}, W.~W. {Weiss}, \& D.~F. {Gray},
  A20.
\newblock \doarXiv{astro-ph/0405087}

\bibitem[{{Christiaens} {et~al.}(2014){Christiaens}, {Casassus}, {Perez}, {van
  der Plas}, \& {M{\'e}nard}}]{Christiaens2014}
{Christiaens}, V., {Casassus}, S., {Perez}, S., {van der Plas}, G., \&
  {M{\'e}nard}, F. 2014, \apjl, 785, L12, \dodoi{10.1088/2041-8205/785/1/L12}

\bibitem[{{Cuello} {et~al.}(2019){Cuello}, {Dipierro}, {Mentiplay}, {Price},
  {Pinte}, {Cuadra}, {Laibe}, {M{\'e}nard}, {Poblete}, \&
  {Montesinos}}]{Cuello2019}
{Cuello}, N., {Dipierro}, G., {Mentiplay}, D., {et~al.} 2019, \mnras, 483,
  4114, \dodoi{10.1093/mnras/sty3325}

\bibitem[{{Cuello} {et~al.}(2020){Cuello}, {Louvet}, {Mentiplay}, {Pinte},
  {Price}, {Winter}, {Nealon}, {M{\'e}nard}, {Lodato}, {Dipierro},
  {Christiaens}, {Montesinos}, {Cuadra}, {Laibe}, {Cieza}, {Dong}, \&
  {Alexander}}]{Cuello2020}
{Cuello}, N., {Louvet}, F., {Mentiplay}, D., {et~al.} 2020, \mnras, 491, 504,
  \dodoi{10.1093/mnras/stz2938}

\bibitem[{{Currie} {et~al.}(2015){Currie}, {Cloutier}, {Brittain}, {Grady},
  {Burrows}, {Muto}, {Kenyon}, \& {Kuchner}}]{Currie2015}
{Currie}, T., {Cloutier}, R., {Brittain}, S., {et~al.} 2015, \apjl, 814, L27,
  \dodoi{10.1088/2041-8205/814/2/L27}

\bibitem[{{Currie} {et~al.}(2011){Currie}, {Burrows}, {Itoh}, {Matsumura},
  {Fukagawa}, {Apai}, {Madhusudhan}, {Hinz}, {Rodigas}, {Kasper}, {Pyo}, \&
  {Ogino}}]{Currie2011}
{Currie}, T., {Burrows}, A., {Itoh}, Y., {et~al.} 2011, \apj, 729, 128,
  \dodoi{10.1088/0004-637X/729/2/128}

\bibitem[{{Currie} {et~al.}(2012){Currie}, {Debes}, {Rodigas}, {Burrows},
  {Itoh}, {Fukagawa}, {Kenyon}, {Kuchner}, \& {Matsumura}}]{Currie2012}
{Currie}, T., {Debes}, J., {Rodigas}, T.~J., {et~al.} 2012, \apjl, 760, L32,
  \dodoi{10.1088/2041-8205/760/2/L32}

\bibitem[{{Currie} {et~al.}(2013){Currie}, {Burrows}, {Madhusudhan},
  {Fukagawa}, {Girard}, {Dawson}, {Murray-Clay}, {Kenyon}, {Kuchner},
  {Matsumura}, {Jayawardhana}, {Chambers}, \& {Bromley}}]{Currie2013}
{Currie}, T., {Burrows}, A., {Madhusudhan}, N., {et~al.} 2013, \apj, 776, 15,
  \dodoi{10.1088/0004-637X/776/1/15}

\bibitem[{{Currie} {et~al.}(2018){Currie}, {Brandt}, {Uyama}, {Nielsen},
  {Blunt}, {Guyon}, {Tamura}, {Marois}, {Mede}, {Kuzuhara}, {Groff},
  {Jovanovic}, {Kasdin}, {Lozi}, {Hodapp}, {Chilcote}, {Carson}, {Martinache},
  {Goebel}, {Grady}, {McElwain}, {Akiyama}, {Asensio-Torres}, {Hayashi},
  {Janson}, {Knapp}, {Kwon}, {Nishikawa}, {Oh}, {Schlieder}, {Serabyn},
  {Sitko}, \& {Skaf}}]{Currie2018}
{Currie}, T., {Brandt}, T.~D., {Uyama}, T., {et~al.} 2018, \aj, 156, 291,
  \dodoi{10.3847/1538-3881/aae9ea}

\bibitem[{{Currie} {et~al.}(2019){Currie}, {Marois}, {Cieza}, {Mulders},
  {Lawson}, {Caceres}, {Rodriguez-Ruiz}, {Wisniewski}, {Guyon}, {Brandt},
  {Kasdin}, {Groff}, {Lozi}, {Chilcote}, {Hodapp}, {Jovanovic}, {Martinache},
  {Skaf}, {Lyra}, {Tamura}, {Asensio-Torres}, {Dong}, {Grady}, {Gerard},
  {Fukagawa}, {Hand}, {Hayashi}, {Henning}, {Kudo}, {Kuzuhara}, {Kwon},
  {McElwain}, \& {Uyama}}]{Currie2019}
{Currie}, T., {Marois}, C., {Cieza}, L., {et~al.} 2019, \apjl, 877, L3,
  \dodoi{10.3847/2041-8213/ab1b42}

\bibitem[{{Cutri} {et~al.}(2003){Cutri}, {Skrutskie}, {van Dyk}, {Beichman},
  {Carpenter}, {Chester}, {Cambresy}, {Evans}, {Fowler}, {Gizis}, {Howard},
  {Huchra}, {Jarrett}, {Kopan}, {Kirkpatrick}, {Light}, {Marsh}, {McCallon},
  {Schneider}, {Stiening}, {Sykes}, {Weinberg}, {Wheaton}, {Wheelock}, \&
  {Zacarias}}]{Cutri2003-2MASS}
{Cutri}, R.~M., {Skrutskie}, M.~F., {van Dyk}, S., {et~al.} 2003, VizieR Online
  Data Catalog, II/246

\bibitem[{{Dai} {et~al.}(2015){Dai}, {Facchini}, {Clarke}, \&
  {Haworth}}]{Dai2015}
{Dai}, F., {Facchini}, S., {Clarke}, C.~J., \& {Haworth}, T.~J. 2015, \mnras,
  449, 1996, \dodoi{10.1093/mnras/stv403}

\bibitem[{{Debes} {et~al.}(2017){Debes}, {Poteet}, {Jang-Condell}, {Gaspar},
  {Hines}, {Kastner}, {Pueyo}, {Rapson}, {Roberge}, {Schneider}, \&
  {Weinberger}}]{Debes2017}
{Debes}, J.~H., {Poteet}, C.~A., {Jang-Condell}, H., {et~al.} 2017, \apj, 835,
  205, \dodoi{10.3847/1538-4357/835/2/205}

\bibitem[{{Demidova} \& {Shevchenko}(2015)}]{Demidova2015}
{Demidova}, T.~V., \& {Shevchenko}, I.~I. 2015, \apj, 805, 38,
  \dodoi{10.1088/0004-637X/805/1/38}

\bibitem[{{Dodson-Robinson} \& {Salyk}(2011)}]{Dodson-Robinson2011}
{Dodson-Robinson}, S.~E., \& {Salyk}, C. 2011, \apj, 738, 131,
  \dodoi{10.1088/0004-637X/738/2/131}

\bibitem[{{Dong} {et~al.}(2016){Dong}, {Fung}, \&
  {Chiang}}]{Dong2016-scatteredlight}
{Dong}, R., {Fung}, J., \& {Chiang}, E. 2016, \apj, 826, 75,
  \dodoi{10.3847/0004-637X/826/1/75}

\bibitem[{{Dong} {et~al.}(2018{\natexlab{a}}){Dong}, {Najita}, \&
  {Brittain}}]{Dong2018-spiral}
{Dong}, R., {Najita}, J.~R., \& {Brittain}, S. 2018{\natexlab{a}}, \apj, 862,
  103, \dodoi{10.3847/1538-4357/aaccfc}

\bibitem[{{Dong} {et~al.}(2015){Dong}, {Zhu}, {Rafikov}, \& {Stone}}]{Dong2015}
{Dong}, R., {Zhu}, Z., {Rafikov}, R.~R., \& {Stone}, J.~M. 2015, \apjl, 809,
  L5, \dodoi{10.1088/2041-8205/809/1/L5}

\bibitem[{{Dong} {et~al.}(2018{\natexlab{b}}){Dong}, {Liu}, {Eisner},
  {Andrews}, {Fung}, {Zhu}, {Chiang}, {Hashimoto}, {Liu}, {Casassus},
  {Esposito}, {Hasegawa}, {Muto}, {Pavlyuchenkov}, {Wilner}, {Akiyama},
  {Tamura}, \& {Wisniewski}}]{Dong2018}
{Dong}, R., {Liu}, S.-y., {Eisner}, J., {et~al.} 2018{\natexlab{b}}, \apj, 860,
  124, \dodoi{10.3847/1538-4357/aac6cb}

\bibitem[{{Dullemond} {et~al.}(2019){Dullemond}, {K{\"u}ffmeier}, {Goicovic},
  {Fukagawa}, {Oehl}, \& {Kramer}}]{Dullemond2019}
{Dullemond}, C.~P., {K{\"u}ffmeier}, M., {Goicovic}, F., {et~al.} 2019, \aap,
  628, A20, \dodoi{10.1051/0004-6361/201832632}

\bibitem[{{Fukagawa} {et~al.}(2006){Fukagawa}, {Tamura}, {Itoh}, {Kudo},
  {Imaeda}, {Oasa}, {Hayashi}, \& {Hayashi}}]{Fukagawa2006}
{Fukagawa}, M., {Tamura}, M., {Itoh}, Y., {et~al.} 2006, \apjl, 636, L153,
  \dodoi{10.1086/500128}

\bibitem[{{Fukagawa} {et~al.}(2004){Fukagawa}, {Hayashi}, {Tamura}, {Itoh},
  {Hayashi}, {Oasa}, {Takeuchi}, {Morino}, {Murakawa}, {Oya}, {Yamashita},
  {Suto}, {Mayama}, {Naoi}, {Ishii}, {Pyo}, {Nishikawa}, {Takato}, {Usuda},
  {Ando}, {Iye}, {Miyama}, \& {Kaifu}}]{Fukagawa2004}
{Fukagawa}, M., {Hayashi}, M., {Tamura}, M., {et~al.} 2004, \apjl, 605, L53,
  \dodoi{10.1086/420699}

\bibitem[{{Gaia Collaboration} {et~al.}(2018){Gaia Collaboration}, {Brown},
  {Vallenari}, {Prusti}, {de Bruijne}, {Babusiaux}, {Bailer-Jones}, {Biermann},
  {Evans}, {Eyer}, {Jansen}, {Jordi}, {Klioner}, {Lammers}, {Lindegren},
  {Luri}, {Mignard}, {Panem}, {Pourbaix}, {Randich}, {Sartoretti}, {Siddiqui},
  {Soubiran}, {van Leeuwen}, {Walton}, {Arenou}, {Bastian}, {Cropper},
  {Drimmel}, {Katz}, {Lattanzi}, {Bakker}, {Cacciari}, {Casta{\~n}eda},
  {Chaoul}, {Cheek}, {De Angeli}, {Fabricius}, {Guerra}, {Holl}, {Masana},
  {Messineo}, {Mowlavi}, {Nienartowicz}, {Panuzzo}, {Portell}, {Riello},
  {Seabroke}, {Tanga}, {Th{\'e}venin}, {Gracia-Abril}, {Comoretto},
  {Garcia-Reinaldos}, {Teyssier}, {Altmann}, {Andrae}, {Audard},
  {Bellas-Velidis}, {Benson}, {Berthier}, {Blomme}, {Burgess}, {Busso},
  {Carry}, {Cellino}, {Clementini}, {Clotet}, {Creevey}, {Davidson}, {De
  Ridder}, {Delchambre}, {Dell'Oro}, {Ducourant},
  {Fern{\'a}ndez-Hern{\'a}ndez}, {Fouesneau}, {Fr{\'e}mat}, {Galluccio},
  {Garc{\'\i}a-Torres}, {Gonz{\'a}lez-N{\'u}{\~n}ez}, {Gonz{\'a}lez-Vidal},
  {Gosset}, {Guy}, {Halbwachs}, {Hambly}, {Harrison}, {Hern{\'a}ndez},
  {Hestroffer}, {Hodgkin}, {Hutton}, {Jasniewicz}, {Jean-Antoine-Piccolo},
  {Jordan}, {Korn}, {Krone-Martins}, {Lanzafame}, {Lebzelter}, {L{\"o}ffler},
  {Manteiga}, {Marrese}, {Mart{\'\i}n-Fleitas}, {Moitinho}, {Mora}, {Muinonen},
  {Osinde}, {Pancino}, {Pauwels}, {Petit}, {Recio-Blanco}, {Richards},
  {Rimoldini}, {Robin}, {Sarro}, {Siopis}, {Smith}, {Sozzetti}, {S{\"u}veges},
  {Torra}, {van Reeven}, {Abbas}, {Abreu Aramburu}, {Accart}, {Aerts},
  {Altavilla}, {{\'A}lvarez}, {Alvarez}, {Alves}, {Anderson}, {Andrei},
  {Anglada Varela}, {Antiche}, {Antoja}, {Arcay}, {Astraatmadja}, {Bach},
  {Baker}, {Balaguer-N{\'u}{\~n}ez}, {Balm}, {Barache}, {Barata}, {Barbato},
  {Barblan}, {Barklem}, {Barrado}, {Barros}, {Barstow}, {Bartholom{\'e}
  Mu{\~n}oz}, {Bassilana}, {Becciani}, {Bellazzini}, {Berihuete}, {Bertone},
  {Bianchi}, {Bienaym{\'e}}, {Blanco-Cuaresma}, {Boch}, {Boeche}, {Bombrun},
  {Borrachero}, {Bossini}, {Bouquillon}, {Bourda}, {Bragaglia}, {Bramante},
  {Breddels}, {Bressan}, {Brouillet}, {Br{\"u}semeister}, {Brugaletta},
  {Bucciarelli}, {Burlacu}, {Busonero}, {Butkevich}, {Buzzi}, {Caffau},
  {Cancelliere}, {Cannizzaro}, {Cantat-Gaudin}, {Carballo}, {Carlucci},
  {Carrasco}, {Casamiquela}, {Castellani}, {Castro-Ginard}, {Charlot},
  {Chemin}, {Chiavassa}, {Cocozza}, {Costigan}, {Cowell}, {Crifo}, {Crosta},
  {Crowley}, {Cuypers}, {Dafonte}, {Damerdji}, {Dapergolas}, {David}, {David},
  {de Laverny}, {De Luise}, {De March}, {de Martino}, {de Souza}, {de Torres},
  {Debosscher}, {del Pozo}, {Delbo}, {Delgado}, {Delgado}, {Di Matteo},
  {Diakite}, {Diener}, {Distefano}, {Dolding}, {Drazinos}, {Dur{\'a}n},
  {Edvardsson}, {Enke}, {Eriksson}, {Esquej}, {Eynard Bontemps}, {Fabre},
  {Fabrizio}, {Faigler}, {Falc{\~a}o}, {Farr{\`a}s Casas}, {Federici},
  {Fedorets}, {Fernique}, {Figueras}, {Filippi}, {Findeisen}, {Fonti},
  {Fraile}, {Fraser}, {Fr{\'e}zouls}, {Gai}, {Galleti}, {Garabato},
  {Garc{\'\i}a-Sedano}, {Garofalo}, {Garralda}, {Gavel}, {Gavras}, {Gerssen},
  {Geyer}, {Giacobbe}, {Gilmore}, {Girona}, {Giuffrida}, {Glass}, {Gomes},
  {Granvik}, {Gueguen}, {Guerrier}, {Guiraud}, {Guti{\'e}rrez-S{\'a}nchez},
  {Haigron}, {Hatzidimitriou}, {Hauser}, {Haywood}, {Heiter}, {Helmi}, {Heu},
  {Hilger}, {Hobbs}, {Hofmann}, {Holland}, {Huckle}, {Hypki}, {Icardi},
  {Jan{\ss}en}, {Jevardat de Fombelle}, {Jonker}, {Juh{\'a}sz}, {Julbe},
  {Karampelas}, {Kewley}, {Klar}, {Kochoska}, {Kohley}, {Kolenberg},
  {Kontizas}, {Kontizas}, {Koposov}, {Kordopatis}, {Kostrzewa-Rutkowska},
  {Koubsky}, {Lambert}, {Lanza}, {Lasne}, {Lavigne}, {Le Fustec}, {Le
  Poncin-Lafitte}, {Lebreton}, {Leccia}, {Leclerc}, {Lecoeur-Taibi},
  {Lenhardt}, {Leroux}, {Liao}, {Licata}, {Lindstr{\o}m}, {Lister}, {Livanou},
  {Lobel}, {L{\'o}pez}, {Managau}, {Mann}, {Mantelet}, {Marchal}, {Marchant},
  {Marconi}, {Marinoni}, {Marschalk{\'o}}, {Marshall}, {Martino}, {Marton},
  {Mary}, {Massari}, {Matijevi{\v{c}}}, {Mazeh}, {McMillan}, {Messina},
  {Michalik}, {Millar}, {Molina}, {Molinaro}, {Moln{\'a}r}, {Montegriffo},
  {Mor}, {Morbidelli}, {Morel}, {Morris}, {Mulone}, {Muraveva}, {Musella},
  {Nelemans}, {Nicastro}, {Noval}, {O'Mullane}, {Ord{\'e}novic},
  {Ord{\'o}{\~n}ez-Blanco}, {Osborne}, {Pagani}, {Pagano}, {Pailler},
  {Palacin}, {Palaversa}, {Panahi}, {Pawlak}, {Piersimoni}, {Pineau}, {Plachy},
  {Plum}, {Poggio}, {Poujoulet}, {Pr{\v{s}}a}, {Pulone}, {Racero}, {Ragaini},
  {Rambaux}, {Ramos-Lerate}, {Regibo}, {Reyl{\'e}}, {Riclet}, {Ripepi}, {Riva},
  {Rivard}, {Rixon}, {Roegiers}, {Roelens}, {Romero-G{\'o}mez}, {Rowell},
  {Royer}, {Ruiz-Dern}, {Sadowski}, {Sagrist{\`a} Sell{\'e}s}, {Sahlmann},
  {Salgado}, {Salguero}, {Sanna}, {Santana-Ros}, {Sarasso}, {Savietto},
  {Schultheis}, {Sciacca}, {Segol}, {Segovia}, {S{\'e}gransan}, {Shih},
  {Siltala}, {Silva}, {Smart}, {Smith}, {Solano}, {Solitro}, {Sordo}, {Soria
  Nieto}, {Souchay}, {Spagna}, {Spoto}, {Stampa}, {Steele},
  {Steidelm{\"u}ller}, {Stephenson}, {Stoev}, {Suess}, {Surdej}, {Szabados},
  {Szegedi-Elek}, {Tapiador}, {Taris}, {Tauran}, {Taylor}, {Teixeira},
  {Terrett}, {Teyssand ier}, {Thuillot}, {Titarenko}, {Torra Clotet}, {Turon},
  {Ulla}, {Utrilla}, {Uzzi}, {Vaillant}, {Valentini}, {Valette}, {van Elteren},
  {Van Hemelryck}, {van Leeuwen}, {Vaschetto}, {Vecchiato}, {Veljanoski},
  {Viala}, {Vicente}, {Vogt}, {von Essen}, {Voss}, {Votruba}, {Voutsinas},
  {Walmsley}, {Weiler}, {Wertz}, {Wevers}, {Wyrzykowski}, {Yoldas},
  {{\v{Z}}erjal}, {Ziaeepour}, {Zorec}, {Zschocke}, {Zucker}, {Zurbach}, \&
  {Zwitter}}]{GaiaDR2}
{Gaia Collaboration}, {Brown}, A.~G.~A., {Vallenari}, A., {et~al.} 2018, \aap,
  616, A1, \dodoi{10.1051/0004-6361/201833051}

\bibitem[{{Grady} {et~al.}(2013){Grady}, {Muto}, {Hashimoto}, {Fukagawa},
  {Currie}, {Biller}, {Thalmann}, {Sitko}, {Russell}, {Wisniewski}, {Dong},
  {Kwon}, {Sai}, {Hornbeck}, {Schneider}, {Hines}, {Moro Mart{\'\i}n}, {Feldt},
  {Henning}, {Pott}, {Bonnefoy}, {Bouwman}, {Lacour}, {Mueller}, {Juh{\'a}sz},
  {Crida}, {Chauvin}, {Andrews}, {Wilner}, {Kraus}, {Dahm}, {Robitaille},
  {Jang-Condell}, {Abe}, {Akiyama}, {Brandner}, {Brandt}, {Carson}, {Egner},
  {Follette}, {Goto}, {Guyon}, {Hayano}, {Hayashi}, {Hayashi}, {Hodapp},
  {Ishii}, {Iye}, {Janson}, {Kandori}, {Knapp}, {Kudo}, {Kusakabe}, {Kuzuhara},
  {Mayama}, {McElwain}, {Matsuo}, {Miyama}, {Morino}, {Nishimura}, {Pyo},
  {Serabyn}, {Suto}, {Suzuki}, {Takami}, {Takato}, {Terada}, {Tomono},
  {Turner}, {Watanabe}, {Yamada}, {Takami}, {Usuda}, \& {Tamura}}]{Grady2013}
{Grady}, C.~A., {Muto}, T., {Hashimoto}, J., {et~al.} 2013, \apj, 762, 48,
  \dodoi{10.1088/0004-637X/762/1/48}

\bibitem[{{Haffert} {et~al.}(2019){Haffert}, {Bohn}, {de Boer}, {Snellen},
  {Brinchmann}, {Girard}, {Keller}, \& {Bacon}}]{Haffert2019}
{Haffert}, S.~Y., {Bohn}, A.~J., {de Boer}, J., {et~al.} 2019, Nature
  Astronomy, 3, 749, \dodoi{10.1038/s41550-019-0780-5}

\bibitem[{Hammel \& Sullivan-Molina(2020)}]{Hammel2020}
Hammel, B., \& Sullivan-Molina, N. 2020,
  bdhammel/least-squares-ellipse-fitting: v2.0.0, v2.0.0,  Zenodo,
  \dodoi{10.5281/zenodo.3723294}

\bibitem[{{Hashimoto} {et~al.}(2011){Hashimoto}, {Tamura}, {Muto}, {Kudo},
  {Fukagawa}, {Fukue}, {Goto}, {Grady}, {Henning}, {Hodapp}, {Honda},
  {Inutsuka}, {Kokubo}, {Knapp}, {McElwain}, {Momose}, {Ohashi}, {Okamoto},
  {Takami}, {Turner}, {Wisniewski}, {Janson}, {Abe}, {Brandner}, {Carson},
  {Egner}, {Feldt}, {Golota}, {Guyon}, {Hayano}, {Hayashi}, {Hayashi}, {Ishii},
  {Kandori}, {Kusakabe}, {Matsuo}, {Mayama}, {Miyama}, {Morino}, {Moro-Martin},
  {Nishimura}, {Pyo}, {Suto}, {Suzuki}, {Takato}, {Terada}, {Thalmann},
  {Tomono}, {Watanabe}, {Yamada}, {Takami}, \& {Usuda}}]{Hashimoto2011}
{Hashimoto}, J., {Tamura}, M., {Muto}, T., {et~al.} 2011, \apjl, 729, L17,
  \dodoi{10.1088/2041-8205/729/2/L17}

\bibitem[{{Itoh} {et~al.}(2002){Itoh}, {Tamura}, {Hayashi}, {Oasa}, {Fukagawa},
  {Kaifu}, {Suto}, {Murakawa}, {Doi}, {Ebizuka}, {Naoi}, {Takami}, {Takato},
  {Gaessler}, {Kanzawa}, {Hayano}, {Kamata}, {Saint-Jacques}, \&
  {Iye}}]{Itoh2002}
{Itoh}, Y., {Tamura}, M., {Hayashi}, S.~S., {et~al.} 2002, \pasj, 54, 963,
  \dodoi{10.1093/pasj/54.6.963}

\bibitem[{{Itoh} {et~al.}(2014){Itoh}, {Oasa}, {Kudo}, {Kusakabe}, {Hashimoto},
  {Abe}, {Brand ner}, {Brandt}, {Carson}, {Egner}, {Feldt}, {Grady}, {Guyon},
  {Hayano}, {Hayashi}, {Hayashi}, {Henning}, {Hodapp}, {Ishii}, {Iye},
  {Janson}, {Kand ori}, {Knapp}, {Kuzuhara}, {Kwon}, {Matsuo}, {McElwain},
  {Miyama}, {Morino}, {Moro-Martin}, {Nishimura}, {Pyo}, {Serabyn}, {Suenaga},
  {Suto}, {Suzuki}, {Takahashi}, {Takato}, {Terada}, {Thalmann}, {Tomono},
  {Turner}, {Watanabe}, {Wisniewski}, {Yamada}, {Mayama}, {Currie}, {Takami},
  {Usuda}, \& {Tamura}}]{Itoh2014}
{Itoh}, Y., {Oasa}, Y., {Kudo}, T., {et~al.} 2014, Research in Astronomy and
  Astrophysics, 14, 1438, \dodoi{10.1088/1674-4527/14/11/007}

\bibitem[{{Jovanovic} {et~al.}(2015){Jovanovic}, {Guyon}, {Martinache},
  {Pathak}, {Hagelberg}, \& {Kudo}}]{Jovanovic2015-astrogrids}
{Jovanovic}, N., {Guyon}, O., {Martinache}, F., {et~al.} 2015, \apjl, 813, L24,
  \dodoi{10.1088/2041-8205/813/2/L24}

\bibitem[{{Keppler} {et~al.}(2018){Keppler}, {Benisty}, {M{\"u}ller},
  {Henning}, {van Boekel}, {Cantalloube}, {Ginski}, {van Holstein}, {Maire},
  {Pohl}, {Samland }, {Avenhaus}, {Baudino}, {Boccaletti}, {de Boer},
  {Bonnefoy}, {Chauvin}, {Desidera}, {Langlois}, {Lazzoni}, {Marleau},
  {Mordasini}, {Pawellek}, {Stolker}, {Vigan}, {Zurlo}, {Birnstiel},
  {Brandner}, {Feldt}, {Flock}, {Girard}, {Gratton}, {Hagelberg}, {Isella},
  {Janson}, {Juhasz}, {Kemmer}, {Kral}, {Lagrange}, {Launhardt}, {Matter},
  {M{\'e}nard}, {Milli}, {Molli{\`e}re}, {Olofsson}, {P{\'e}rez}, {Pinilla},
  {Pinte}, {Quanz}, {Schmidt}, {Udry}, {Wahhaj}, {Williams}, {Buenzli},
  {Cudel}, {Dominik}, {Galicher}, {Kasper}, {Lannier}, {Mesa}, {Mouillet},
  {Peretti}, {Perrot}, {Salter}, {Sissa}, {Wildi}, {Abe}, {Antichi},
  {Augereau}, {Baruffolo}, {Baudoz}, {Bazzon}, {Beuzit}, {Blanchard}, {Brems},
  {Buey}, {De Caprio}, {Carbillet}, {Carle}, {Cascone}, {Cheetham}, {Claudi},
  {Costille}, {Delboulb{\'e}}, {Dohlen}, {Fantinel}, {Feautrier}, {Fusco},
  {Giro}, {Gluck}, {Gry}, {Hubin}, {Hugot}, {Jaquet}, {Le Mignant}, {Llored},
  {Madec}, {Magnard}, {Martinez}, {Maurel}, {Meyer}, {M{\"o}ller-Nilsson},
  {Moulin}, {Mugnier}, {Orign{\'e}}, {Pavlov}, {Perret}, {Petit}, {Pragt},
  {Puget}, {Rabou}, {Ramos}, {Rigal}, {Rochat}, {Roelfsema}, {Rousset}, {Roux},
  {Salasnich}, {Sauvage}, {Sevin}, {Soenke}, {Stadler}, {Suarez}, {Turatto}, \&
  {Weber}}]{Keppler2018}
{Keppler}, M., {Benisty}, M., {M{\"u}ller}, A., {et~al.} 2018, \aap, 617, A44,
  \dodoi{10.1051/0004-6361/201832957}

\bibitem[{{Keppler} {et~al.}(2020){Keppler}, {Penzlin}, {Benisty}, {van
  Boekel}, {Henning}, {van Holstein}, {Kley}, {Garufi}, {Ginski}, {Brandner},
  {Bertrang}, {Boccaletti}, {de Boer}, {Bonavita}, {Brown Sevilla}, {Chauvin},
  {Dominik}, {Janson}, {Langlois}, {Lodato}, {Maire}, {M{\'e}nard}, {Pantin},
  {Pinte}, {Stolker}, {Szul{\'a}gyi}, {Thebault}, {Villenave}, {Zurlo},
  {Rabou}, {Feautrier}, {Feldt}, {Madec}, \& {Wildi}}]{Keppler2020}
{Keppler}, M., {Penzlin}, A., {Benisty}, M., {et~al.} 2020, \aap, 639, A62,
  \dodoi{10.1051/0004-6361/202038032}

\bibitem[{{Krist} {et~al.}(2002){Krist}, {Stapelfeldt}, \&
  {Watson}}]{Krist2002}
{Krist}, J.~E., {Stapelfeldt}, K.~R., \& {Watson}, A.~M. 2002, \apj, 570, 785,
  \dodoi{10.1086/339777}

\bibitem[{{Lafreni{\`e}re} {et~al.}(2009){Lafreni{\`e}re}, {Marois}, {Doyon},
  \& {Barman}}]{Lafreniere2009}
{Lafreni{\`e}re}, D., {Marois}, C., {Doyon}, R., \& {Barman}, T. 2009, \apjl,
  694, L148, \dodoi{10.1088/0004-637X/694/2/L148}

\bibitem[{{Lafreni{\`e}re} {et~al.}(2007){Lafreni{\`e}re}, {Marois}, {Doyon},
  {Nadeau}, \& {Artigau}}]{Lafreniere2007}
{Lafreni{\`e}re}, D., {Marois}, C., {Doyon}, R., {Nadeau}, D., \& {Artigau},
  {\'E}. 2007, \apj, 660, 770, \dodoi{10.1086/513180}

\bibitem[{{Laws} {et~al.}(2020){Laws}, {Harries}, {Setterholm}, {Monnier},
  {Rich}, {Aarnio}, {Adams}, {Andrews}, {Bae}, {Calvet}, {Espaillat},
  {Hartmann}, {Hinkley}, {Isella}, {Kraus}, {Wilner}, \& {Zhu}}]{Lass2020}
{Laws}, A. S.~E., {Harries}, T.~J., {Setterholm}, B.~R., {et~al.} 2020, \apj,
  888, 7, \dodoi{10.3847/1538-4357/ab59e2}

\bibitem[{{Marino} {et~al.}(2015){Marino}, {Perez}, \&
  {Casassus}}]{Marino2015a}
{Marino}, S., {Perez}, S., \& {Casassus}, S. 2015, \apjl, 798, L44,
  \dodoi{10.1088/2041-8205/798/2/L44}

\bibitem[{{Marois} {et~al.}(2006){Marois}, {Lafreni{\`e}re}, {Doyon},
  {Macintosh}, \& {Nadeau}}]{Marois2006}
{Marois}, C., {Lafreni{\`e}re}, D., {Doyon}, R., {Macintosh}, B., \& {Nadeau},
  D. 2006, \apj, 641, 556, \dodoi{10.1086/500401}

\bibitem[{{Mawet} {et~al.}(2014){Mawet}, {Milli}, {Wahhaj}, {Pelat}, {Absil},
  {Delacroix}, {Boccaletti}, {Kasper}, {Kenworthy}, {Marois}, {Mennesson}, \&
  {Pueyo}}]{Mawet2014}
{Mawet}, D., {Milli}, J., {Wahhaj}, Z., {et~al.} 2014, \apj, 792, 97,
  \dodoi{10.1088/0004-637X/792/2/97}

\bibitem[{{M{\'e}nard} {et~al.}(2020){M{\'e}nard}, {Cuello}, {Ginski}, {van der
  Plas}, {Villenave}, {Gonzalez}, {Pinte}, {Benisty}, {Boccaletti}, {Price},
  {Boehler}, {Chripko}, {de Boer}, {Dominik}, {Garufi}, {Gratton}, {Hagelberg},
  {Henning}, {Langlois}, {Maire}, {Pinilla}, {Ruane}, {Schmid}, {van Holstein},
  {Vigan}, {Zurlo}, {Hubin}, {Pavlov}, {Rochat}, {Sauvage}, \&
  {Stadler}}]{Menard2020}
{M{\'e}nard}, F., {Cuello}, N., {Ginski}, C., {et~al.} 2020, \aap, 639, L1,
  \dodoi{10.1051/0004-6361/202038356}

\bibitem[{{Min} {et~al.}(2009){Min}, {Dullemond}, {Dominik}, {de Koter}, \&
  {Hovenier}}]{Min2009}
{Min}, M., {Dullemond}, C.~P., {Dominik}, C., {de Koter}, A., \& {Hovenier},
  J.~W. 2009, \aap, 497, 155, \dodoi{10.1051/0004-6361/200811470}

\bibitem[{{Monnier} {et~al.}(2019){Monnier}, {Harries}, {Bae}, {Setterholm},
  {Laws}, {Aarnio}, {Adams}, {Andrews}, {Calvet}, {Espaillat}, {Hartmann},
  {Kraus}, {McClure}, {Miller}, {Oppenheimer}, {Wilner}, \&
  {Zhu}}]{Monnier2019}
{Monnier}, J.~D., {Harries}, T.~J., {Bae}, J., {et~al.} 2019, \apj, 872, 122,
  \dodoi{10.3847/1538-4357/aafe87}

\bibitem[{{Montesinos} \& {Cuello}(2018)}]{Montesinos2018}
{Montesinos}, M., \& {Cuello}, N. 2018, \mnras, 475, L35,
  \dodoi{10.1093/mnrasl/sly001}

\bibitem[{{Montesinos} {et~al.}(2016){Montesinos}, {Perez}, {Casassus},
  {Marino}, {Cuadra}, \& {Christiaens}}]{Montesinos2016}
{Montesinos}, M., {Perez}, S., {Casassus}, S., {et~al.} 2016, \apjl, 823, L8,
  \dodoi{10.3847/2041-8205/823/1/L8}

\bibitem[{{Mulders} {et~al.}(2010){Mulders}, {Dominik}, \& {Min}}]{Mulders2010}
{Mulders}, G.~D., {Dominik}, C., \& {Min}, M. 2010, \aap, 512, A11,
  \dodoi{10.1051/0004-6361/200912743}

\bibitem[{{Mulders} {et~al.}(2013){Mulders}, {Min}, {Dominik}, {Debes}, \&
  {Schneider}}]{Mulders2013}
{Mulders}, G.~D., {Min}, M., {Dominik}, C., {Debes}, J.~H., \& {Schneider}, G.
  2013, \aap, 549, A112, \dodoi{10.1051/0004-6361/201219522}

\bibitem[{{Muto} {et~al.}(2012){Muto}, {Grady}, {Hashimoto}, {Fukagawa},
  {Hornbeck}, {Sitko}, {Russell}, {Werren}, {Cur{\'e}}, {Currie}, {Ohashi},
  {Okamoto}, {Momose}, {Honda}, {Inutsuka}, {Takeuchi}, {Dong}, {Abe},
  {Brandner}, {Brandt}, {Carson}, {Egner}, {Feldt}, {Fukue}, {Goto}, {Guyon},
  {Hayano}, {Hayashi}, {Hayashi}, {Henning}, {Hodapp}, {Ishii}, {Iye},
  {Janson}, {Kandori}, {Knapp}, {Kudo}, {Kusakabe}, {Kuzuhara}, {Matsuo},
  {Mayama}, {McElwain}, {Miyama}, {Morino}, {Moro-Martin}, {Nishimura}, {Pyo},
  {Serabyn}, {Suto}, {Suzuki}, {Takami}, {Takato}, {Terada}, {Thalmann},
  {Tomono}, {Turner}, {Watanabe}, {Wisniewski}, {Yamada}, {Takami}, {Usuda}, \&
  {Tamura}}]{Muto2012}
{Muto}, T., {Grady}, C.~A., {Hashimoto}, J., {et~al.} 2012, \apjl, 748, L22,
  \dodoi{10.1088/2041-8205/748/2/L22}

\bibitem[{{Nielsen} {et~al.}(2019){Nielsen}, {De Rosa}, {Macintosh}, {Wang},
  {Ruffio}, {Chiang}, {Marley}, {Saumon}, {Savransky}, {Ammons}, {Bailey},
  {Barman}, {Blain}, {Bulger}, {Burrows}, {Chilcote}, {Cotten}, {Czekala},
  {Doyon}, {Duch{\^e}ne}, {Esposito}, {Fabrycky}, {Fitzgerald}, {Follette},
  {Fortney}, {Gerard}, {Goodsell}, {Graham}, {Greenbaum}, {Hibon}, {Hinkley},
  {Hirsch}, {Hom}, {Hung}, {Dawson}, {Ingraham}, {Kalas}, {Konopacky},
  {Larkin}, {Lee}, {Lin}, {Maire}, {Marchis}, {Marois}, {Metchev},
  {Millar-Blanchaer}, {Morzinski}, {Oppenheimer}, {Palmer}, {Patience},
  {Perrin}, {Poyneer}, {Pueyo}, {Rafikov}, {Rajan}, {Rameau}, {Rantakyr{\"o}},
  {Ren}, {Schneider}, {Sivaramakrishnan}, {Song}, {Soummer}, {Tallis},
  {Thomas}, {Ward-Duong}, \& {Wolff}}]{Nielsen2019}
{Nielsen}, E.~L., {De Rosa}, R.~J., {Macintosh}, B., {et~al.} 2019, \aj, 158,
  13, \dodoi{10.3847/1538-3881/ab16e9}

\bibitem[{{P{\'e}rez} {et~al.}(2016){P{\'e}rez}, {Carpenter}, {Andrews},
  {Ricci}, {Isella}, {Linz}, {Sargent}, {Wilner}, {Henning}, {Deller},
  {Chandler}, {Dullemond}, {Lazio}, {Menten}, {Corder}, {Storm}, {Testi},
  {Tazzari}, {Kwon}, {Calvet}, {Greaves}, {Harris}, \& {Mundy}}]{Perez2016}
{P{\'e}rez}, L.~M., {Carpenter}, J.~M., {Andrews}, S.~M., {et~al.} 2016,
  Science, 353, 1519, \dodoi{10.1126/science.aaf8296}

\bibitem[{{Price} {et~al.}(2018){Price}, {Cuello}, {Pinte}, {Mentiplay},
  {Casassus}, {Christiaens}, {Kennedy}, {Cuadra}, {Sebastian Perez}, {Marino},
  {Armitage}, {Zurlo}, {Juhasz}, {Ragusa}, {Laibe}, \& {Lodato}}]{Price2018}
{Price}, D.~J., {Cuello}, N., {Pinte}, C., {et~al.} 2018, \mnras, 477, 1270,
  \dodoi{10.1093/mnras/sty647}

\bibitem[{{Quanz} {et~al.}(2013){Quanz}, {Amara}, {Meyer}, {Kenworthy},
  {Kasper}, \& {Girard}}]{Quanz2013}
{Quanz}, S.~P., {Amara}, A., {Meyer}, M.~R., {et~al.} 2013, \apjl, 766, L1,
  \dodoi{10.1088/2041-8205/766/1/L1}

\bibitem[{{Quillen}(2006)}]{Quillen2006a}
{Quillen}, A.~C. 2006, \apj, 640, 1078, \dodoi{10.1086/500165}

\bibitem[{{Reggiani} {et~al.}(2018){Reggiani}, {Christiaens}, {Absil}, {Mawet},
  {Huby}, {Choquet}, {Gomez Gonzalez}, {Ruane}, {Femenia}, {Serabyn},
  {Matthews}, {Barraza}, {Carlomagno}, {Defr{\`e}re}, {Delacroix}, {Habraken},
  {Jolivet}, {Karlsson}, {Orban de Xivry}, {Piron}, {Surdej}, {Vargas Catalan},
  \& {Wertz}}]{Reggiani2018}
{Reggiani}, M., {Christiaens}, V., {Absil}, O., {et~al.} 2018, \aap, 611, A74,
  \dodoi{10.1051/0004-6361/201732016}

\bibitem[{{Rice} {et~al.}(2003){Rice}, {Armitage}, {Bate}, \&
  {Bonnell}}]{Rice2003a}
{Rice}, W.~K.~M., {Armitage}, P.~J., {Bate}, M.~R., \& {Bonnell}, I.~A. 2003,
  \mnras, 339, 1025, \dodoi{10.1046/j.1365-8711.2003.06253.x}

\bibitem[{{Rich} {et~al.}(2019){Rich}, {Wisniewski}, {Currie}, {Fukagawa},
  {Grady}, {Sitko}, {Pikhartova}, {Hashimoto}, {Abe}, {Brand ner}, {Brandt},
  {Carson}, {Chilcote}, {Dong}, {Feldt}, {Goto}, {Groff}, {Guyon}, {Hayano},
  {Hayashi}, {Hayashi}, {Henning}, {Hodapp}, {Ishii}, {Iye}, {Janson},
  {Jovanovic}, {Kand ori}, {Kasdin}, {Knapp}, {Kudo}, {Kusakabe}, {Kuzuhara},
  {Kwon}, {Lozi}, {Martinache}, {Matsuo}, {Mayama}, {McElwain}, {Miyama},
  {Morino}, {Moro-Martin}, {Nakagawa}, {Nishimura}, {Pyo}, {Serabyn}, {Suto},
  {Russel}, {Suzuki}, {Takami}, {Takato}, {Terada}, {Thalmann}, {Turner},
  {Uyama}, {Wagner}, {Watanabe}, {Yamada}, {Takami}, {Usuda}, \&
  {Tamura}}]{Rich2019}
{Rich}, E.~A., {Wisniewski}, J.~P., {Currie}, T., {et~al.} 2019, \apj, 875, 38,
  \dodoi{10.3847/1538-4357/ab0f3b}

\bibitem[{{Rodriguez} {et~al.}(2018){Rodriguez}, {Loomis}, {Cabrit}, {Haworth},
  {Facchini}, {Dougados}, {Booth}, {Jensen}, {Clarke}, {Stassun}, {Dent}, \&
  {Pety}}]{Rodriguez2018}
{Rodriguez}, J.~E., {Loomis}, R., {Cabrit}, S., {et~al.} 2018, \apj, 859, 150,
  \dodoi{10.3847/1538-4357/aac08f}

\bibitem[{{Sahoo} {et~al.}(2020){Sahoo}, {Guyon}, {Lozi}, {Chilcote},
  {Jovanovic}, {Brandt}, {Groff}, \& {Martinache}}]{Sahoo2020}
{Sahoo}, A., {Guyon}, O., {Lozi}, J., {et~al.} 2020, \aj, 159, 250,
  \dodoi{10.3847/1538-3881/ab88cd}

\bibitem[{{Soummer} {et~al.}(2012){Soummer}, {Pueyo}, \&
  {Larkin}}]{Soummer2012-KLIP}
{Soummer}, R., {Pueyo}, L., \& {Larkin}, J. 2012, \apjl, 755, L28,
  \dodoi{10.1088/2041-8205/755/2/L28}

\bibitem[{{Spiegel} \& {Burrows}(2012)}]{Spiegel2012}
{Spiegel}, D.~S., \& {Burrows}, A. 2012, \apj, 745, 174,
  \dodoi{10.1088/0004-637X/745/2/174}

\bibitem[{{Sterzik} {et~al.}(2005){Sterzik}, {Melo}, {Tokovinin}, \& {van der
  Bliek}}]{Sterzik2005}
{Sterzik}, M.~F., {Melo}, C.~H.~F., {Tokovinin}, A.~A., \& {van der Bliek}, N.
  2005, \aap, 434, 671, \dodoi{10.1051/0004-6361:20042302}

\bibitem[{{Stolker} {et~al.}(2016){Stolker}, {Dominik}, {Min}, {Garufi},
  {Mulders}, \& {Avenhaus}}]{Stolker2016}
{Stolker}, T., {Dominik}, C., {Min}, M., {et~al.} 2016, \aap, 596, A70,
  \dodoi{10.1051/0004-6361/201629098}

\bibitem[{{Stolker} {et~al.}(2017){Stolker}, {Sitko}, {Lazareff}, {Benisty},
  {Dominik}, {Waters}, {Min}, {Perez}, {Milli}, {Garufi}, {de Boer}, {Ginski},
  {Kraus}, {Berger}, \& {Avenhaus}}]{Stolker2017}
{Stolker}, T., {Sitko}, M., {Lazareff}, B., {et~al.} 2017, \apj, 849, 143,
  \dodoi{10.3847/1538-4357/aa886a}

\bibitem[{{Takami} {et~al.}(2013){Takami}, {Karr}, {Hashimoto}, {Kim},
  {Wisniewski}, {Henning}, {Grady}, {Kandori}, {Hodapp}, {Kudo}, {Kusakabe},
  {Chou}, {Itoh}, {Momose}, {Mayama}, {Currie}, {Follette}, {Kwon}, {Abe},
  {Brandner}, {Brandt}, {Carson}, {Egner}, {Feldt}, {Guyon}, {Hayano},
  {Hayashi}, {Hayashi}, {Ishii}, {Iye}, {Janson}, {Knapp}, {Kuzuhara},
  {McElwain}, {Matsuo}, {Miyama}, {Morino}, {Moro-Martin}, {Nishimura}, {Pyo},
  {Serabyn}, {Suto}, {Suzuki}, {Takato}, {Terada}, {Thalmann}, {Tomono},
  {Turner}, {Watanabe}, {Yamada}, {Takami}, {Usuda}, \& {Tamura}}]{Takami2013}
{Takami}, M., {Karr}, J.~L., {Hashimoto}, J., {et~al.} 2013, \apj, 772, 145,
  \dodoi{10.1088/0004-637X/772/2/145}

\bibitem[{{Tang} {et~al.}(2012){Tang}, {Guilloteau}, {Pi{\'e}tu}, {Dutrey},
  {Ohashi}, \& {Ho}}]{Tang2012}
{Tang}, Y.~W., {Guilloteau}, S., {Pi{\'e}tu}, V., {et~al.} 2012, \aap, 547,
  A84, \dodoi{10.1051/0004-6361/201219414}

\bibitem[{{Tang} {et~al.}(2017){Tang}, {Guilloteau}, {Dutrey}, {Muto}, {Shen},
  {Gu}, {Inutsuka}, {Momose}, {Pietu}, {Fukagawa}, {Chapillon}, {Ho}, {di
  Folco}, {Corder}, {Ohashi}, \& {Hashimoto}}]{Tang2017}
{Tang}, Y.-W., {Guilloteau}, S., {Dutrey}, A., {et~al.} 2017, \apj, 840, 32,
  \dodoi{10.3847/1538-4357/aa6af7}

\bibitem[{{Thalmann} {et~al.}(2014){Thalmann}, {Mulders}, {Hodapp}, {Janson},
  {Grady}, {Min}, {de Juan Ovelar}, {Carson}, {Brandt}, {Bonnefoy}, {McElwain},
  {Leisenring}, {Dominik}, {Henning}, \& {Tamura}}]{Thalmann2014}
{Thalmann}, C., {Mulders}, G.~D., {Hodapp}, K., {et~al.} 2014, \aap, 566, A51,
  \dodoi{10.1051/0004-6361/201322915}

\bibitem[{{Torres}(2004)}]{Torres2004}
{Torres}, G. 2004, \aj, 127, 1187, \dodoi{10.1086/381066}

\bibitem[{{Uyama} {et~al.}(2018){Uyama}, {Hashimoto}, {Muto}, {Akiyama},
  {Dong}, {de Leon}, {Sakon}, {Kudo}, {Kusakabe}, {Kuzuhara}, {Bonnefoy},
  {Abe}, {Brand ner}, {Brandt}, {Carson}, {Currie}, {Egner}, {Feldt}, {Fung},
  {Goto}, {Grady}, {Guyon}, {Hayano}, {Hayashi}, {Hayashi}, {Henning},
  {Hodapp}, {Ishii}, {Iye}, {Janson}, {Kand ori}, {Knapp}, {Kwon}, {Matsuo},
  {Mayama}, {Mcelwain}, {Miyama}, {Morino}, {Moro-Martin}, {Nishimura}, {Pyo},
  {Serabyn}, {Sitko}, {Suenaga}, {Suto}, {Suzuki}, {Takahashi}, {Takami},
  {Takato}, {Terada}, {Thalmann}, {Turner}, {Watanabe}, {Wisniewski}, {Yamada},
  {Yang}, {Takami}, {Usuda}, \& {Tamura}}]{Uyama2018}
{Uyama}, T., {Hashimoto}, J., {Muto}, T., {et~al.} 2018, \aj, 156, 63,
  \dodoi{10.3847/1538-3881/aacbd1}

\bibitem[{{Uyama} {et~al.}(2020){Uyama}, {Muto}, {Mawet}, {Christiaens},
  {Hashimoto}, {Kudo}, {Kuzuhara}, {Ruane}, {Beichman}, {Absil}, {Akiyama},
  {Bae}, {Bottom}, {Choquet}, {Currie}, {Dong}, {Follette}, {Fukagawa},
  {Guidi}, {Huby}, {Kwon}, {Mayama}, {Meshkat}, {Reggiani}, {Ricci}, {Serabyn},
  {Tamura}, {Testi}, {Wallack}, {Williams}, \& {Zhu}}]{Uyama2020}
{Uyama}, T., {Muto}, T., {Mawet}, D., {et~al.} 2020, \aj, 159, 118,
  \dodoi{10.3847/1538-3881/ab7006}

\bibitem[{{Vigan} {et~al.}(2015){Vigan}, {Gry}, {Salter}, {Mesa}, {Homeier},
  {Moutou}, \& {Allard}}]{Vigan2015}
{Vigan}, A., {Gry}, C., {Salter}, G., {et~al.} 2015, \mnras, 454, 129,
  \dodoi{10.1093/mnras/stv1928}

\bibitem[{{Wagner} {et~al.}(2015){Wagner}, {Apai}, {Kasper}, \&
  {Robberto}}]{Wagner2015}
{Wagner}, K., {Apai}, D., {Kasper}, M., \& {Robberto}, M. 2015, \apjl, 813, L2,
  \dodoi{10.1088/2041-8205/813/1/L2}

\bibitem[{{Wagner} {et~al.}(2018){Wagner}, {Follete}, {Close}, {Apai}, {Gibbs},
  {Keppler}, {M{\"u}ller}, {Henning}, {Kasper}, {Wu}, {Long}, {Males},
  {Morzinski}, \& {McClure}}]{Wagner2018}
{Wagner}, K., {Follete}, K.~B., {Close}, L.~M., {et~al.} 2018, \apjl, 863, L8,
  \dodoi{10.3847/2041-8213/aad695}

\bibitem[{{Wisniewski} {et~al.}(2008){Wisniewski}, {Clampin}, {Grady},
  {Ardila}, {Ford}, {Golimowski}, {Illingworth}, \& {Krist}}]{Wisniewski2008}
{Wisniewski}, J.~P., {Clampin}, M., {Grady}, C.~A., {et~al.} 2008, \apj, 682,
  548, \dodoi{10.1086/589629}

\bibitem[{{Ygouf} {et~al.}(2019){Ygouf}, {Patel}, {Debes}, {Beichman},
  {Duchene}, {Weinberger}, {Pueyo}, {Choquet}, {Meshkat}, {Lubow}, {Sahlmann},
  {Mawet}, {Girard}, {Akeson}, {Dong}, {Perrin}, \& {de Boer}}]{Ygouf2019}
{Ygouf}, M., {Patel}, R., {Debes}, J., {et~al.} 2019, in American Astronomical
  Society Meeting Abstracts, Vol. 233, American Astronomical Society Meeting
  Abstracts \#233, 436.02

\bibitem[{{Zhu} {et~al.}(2015){Zhu}, {Dong}, {Stone}, \& {Rafikov}}]{Zhu2015}
{Zhu}, Z., {Dong}, R., {Stone}, J.~M., \& {Rafikov}, R.~R. 2015, \apj, 813, 88,
  \dodoi{10.1088/0004-637X/813/2/88}

\bibitem[{{Zhu} {et~al.}(2011){Zhu}, {Nelson}, {Hartmann}, {Espaillat}, \&
  {Calvet}}]{Zhu2011}
{Zhu}, Z., {Nelson}, R.~P., {Hartmann}, L., {Espaillat}, C., \& {Calvet}, N.
  2011, \apj, 729, 47, \dodoi{10.1088/0004-637X/729/1/47}

\end{thebibliography}

\end{document}